\documentclass[fleqn,usenatbib]{mnras}

\usepackage{newtxtext,newtxmath}

\usepackage[T1]{fontenc}

\DeclareRobustCommand{\VAN}[3]{#2}
\let\VANthebibliography\thebibliography
\def\thebibliography{\DeclareRobustCommand{\VAN}[3]{##3}\VANthebibliography}

\usepackage{graphicx}	
\usepackage{amsmath}	
\usepackage{fix-cm}
\usepackage{subfig}
\usepackage{float}
\usepackage{placeins}
\usepackage{MnSymbol}
\usepackage{enumerate}
\usepackage{threeparttable}
\usepackage{booktabs,caption}
\usepackage{xspace}
\usepackage{adjustbox}  

\newcommand\orcid[1]{\href{http://orcid.org/#1}{\adjustbox{trim={-.15\width} {0\height} {-.15\width} {0\height},clip}{\includegraphics[height=10pt]{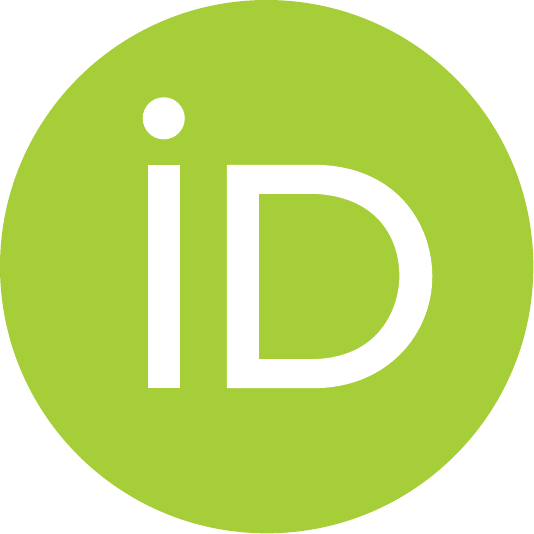}}}}

\title[Merger rate out to Cosmic Dawn]{Constraining the major merger history of $z \sim 3-9$ galaxies using JADES: dominant in-situ star formation}

\author[Puskás et al.]{Dávid Puskás\orcid{0000-0001-8630-2031},$^{1,2}$\thanks{E-mail: dp670@cam.ac.uk}
Sandro Tacchella\orcid{0000-0002-8224-4505},$^{1,2}$
Charlotte Simmonds\orcid{0000-0003-4770-7516},$^{1,2}$
Kevin Hainline\orcid{0000-0003-4565-8239},$^{3}$\newauthor
Francesco D'Eugenio\orcid{0000-0003-2388-8172},$^{1,2}$
Stacey Alberts\orcid{0000-0002-8909-8782},$^{3}$
Santiago Arribas\orcid{0000-0001-7997-1640},$^{4}$
William M. Baker\orcid{0000-0003-0215-1104},$^{5}$\newauthor
Andrew J.\ Bunker\orcid{0000-0002-8651-9879},$^{6}$
Stefano Carniani\orcid{0000-0002-6719-380X},$^{7}$
Stéphane Charlot\orcid{0000-0003-3458-2275},$^{8}$
Qiao Duan\orcid{0009-0009-8105-4564},$^{1,2}$
Daniel J.\ Eisenstein\orcid{0000-0002-2929-3121},$^{9}$\newauthor
Zhiyuan Ji\orcid{0000-0001-7673-2257},$^{3}$
Benjamin D.\ Johnson\orcid{0000-0002-9280-7594},$^{9}$
Gareth C. Jones\orcid{0000-0002-0267-9024},$^{1,2}$
Roberto Maiolino\orcid{0000-0002-4985-3819},$^{1,2}$\newauthor
William McClymont\orcid{0009-0009-5565-3790},$^{1,2}$
Marcia Rieke\orcid{0000-0002-7893-6170},$^{3}$
Pierluigi Rinaldi\orcid{0000-0002-5104-8245},$^{3}$
Brant Robertson\orcid{0000-0002-4271-0364},$^{10}$
Hannah \"Ubler\orcid{0000-0003-4891-0794},$^{11}$\newauthor
Christina C. Williams\orcid{0000-0003-2919-7495},$^{12}$
Christopher N. A. Willmer\orcid{0000-0001-9262-9997},$^{3}$
Chris Willott\orcid{0000-0002-4201-7367}$^{13}$ and
Joris Witstok\orcid{0000-0002-7595-121X}$^{14,15}$
\\
\\
$^{1}$Cavendish Laboratory, University of Cambridge, 19 JJ Thomson Avenue, Cambridge, CB3 OHE, UK\\
$^{2}$Kavli Institute for Cosmology, Madingley Road, Cambridge, CB3 0HA, UK\\
$^{3}$Steward Observatory, University of Arizona, 933 N. Cherry Avenue, Tucson, AZ 85721, USA\\
$^{4}$Centro de Astrobiolog\'ia (CAB), CSIC–INTA, Cra. de Ajalvir Km.~4, 28850- Torrej\'on de Ardoz, Madrid, Spain\\
$^{5}$DARK, Niels Bohr Institute, University of Copenhagen, Jagtvej 128, DK-2200 Copenhagen, Denmark\\
$^{6}$Department of Physics, University of Oxford, Denys Wilkinson Building, Keble Road, Oxford OX1 3RH, UK\\
$^{7}$Scuola Normale Superiore, Piazza dei Cavalieri 7, I-56126 Pisa, Italy\\
$^{8}$Sorbonne Universit\'e, CNRS, UMR 7095, Institut d'Astrophysique de Paris, 98 bis bd Arago, 75014 Paris, France\\
$^{9}$Center for Astrophysics $|$ Harvard \& Smithsonian, 60 Garden St., Cambridge MA 02138 USA\\
$^{10}$Department of Astronomy and Astrophysics University of California, Santa Cruz, 1156 High Street, Santa Cruz CA 96054, USA\\
$^{11}$Max-Planck-Institut f\"ur extraterrestrische Physik (MPE), Gie{\ss}enbachstra{\ss}e 1, 85748 Garching, Germany\\
$^{12}$NSF National Optical-Infrared Astronomy Research Laboratory, 950 North Cherry Avenue, Tucson, AZ 85719, USA\\
$^{13}$NRC Herzberg, 5071 West Saanich Rd, Victoria, BC V9E 2E7, Canada\\
$^{14}$Cosmic Dawn Center (DAWN), Copenhagen, Denmark\\
$^{15}$Niels Bohr Institute, University of Copenhagen, Jagtvej 128, DK-2200, Copenhagen, Denmark
}

\pubyear{2025}

\begin{document}
\label{firstpage}
\pagerange{\pageref{firstpage}--\pageref{lastpage}}
\maketitle

\begin{abstract}
We present a comprehensive analysis of galaxy close-pair fractions and major merger rates to evaluate the importance of mergers in the hierarchical growth of galaxies over cosmic time. This study focuses on the previously poorly understood redshift range of $z \approx 3-9$ using JADES observations. Our mass-complete sample includes primary galaxies with stellar masses of ${\rm log}(M_\star/{\rm M_\odot}) = [8, 10]$, having major companions (mass ratio $\geq 1/4$) selected by $5-30$ pkpc projected separation and redshift proximity criteria. Pair fractions are measured using a statistically robust method incorporating photometric redshift posteriors and available spectroscopic data. The pair fraction evolves with redshift and shows dependence on the stellar mass: at ${\rm log}(M_\star/{\rm M_\odot}) = [8.0, 8.5]$ there is an increase up to $z\sim5-6$, followed by a turnover; while at higher stellar masses there is a flattening and weak decline with increasing redshift. Similarly, the derived galaxy major merger rate increases and flattens beyond $z \sim 6$ to $2-8~{\rm Gyr^{-1}}$ per galaxy, showing a weak scaling with stellar mass, driven by the evolution of the galaxy stellar mass function. A comparison between the cumulative mass accretion from major mergers and the mass assembled through star formation indicates that major mergers contribute approximately $3-13\%$ to the total mass growth over the studied redshift range, which is in agreement with the ex-situ mass fraction estimated from our simple numerical model. These results highlight that major mergers contribute little to the direct stellar mass growth compared to in-situ star formation but could still play an indirect role by driving star formation itself.
\end{abstract}




\section{Introduction}

Galaxy mergers have long been predicted to play a significant role in the evolution and overall mass build-up of the galaxy population through the hierarchical growth of dark matter haloes in the $\Lambda$CDM cosmological model \citep{White_1978}. A fundamental question from an evolutionary perspective is how massive galaxies have accumulated their stellar mass. It is believed that galaxies grow through two main channels: the formation of new stars via the accretion of cold gas from their surrounding intergalactic medium and mergers with nearby galaxies, resulting in a single, more massive galaxy. The relative contribution of these channels to galaxy mass growth, particularly at early epochs, remains poorly constrained. While galaxy star formation rates (SFRs) have been reliably measured in the past, determining the galaxy merger rate over cosmic time and assessing the relative importance of these growth mechanisms is considerably more challenging. In this paper, we present the most extensive study to date about the importance of mergers in the mass growth of galaxies in the first 2 billion years of cosmic history.

In addition to understanding the mass build-up of galaxies, mergers were first proposed to explain observed trends in the physical properties of galaxies. In their seminal paper, \citet{Toomre72} identified two spiral galaxies in the process of merging by observing morphological disturbances and tidal tails. Since then, mergers have been considered crucial in the structural evolution and morphological transformation of massive elliptical galaxies \citep{Barnes96, Bell06, Naab_2006, Bournaud_2011}. Gas-rich mergers have been found to trigger starburst events (exceptionally high rates of star formation), and merging galaxies exhibit enhanced SFRs compared to isolated ones \citep{Mihos94, Patton11, Torrey_2012, Patton13, Lanz13, Moreno_2015, Pearson_2019, Moreno_2019, Patton_2020, Garduno_2021, Ellison_2022, Thorp_2022, Montenegro-Taborda_2023, Duan_2024b, Reeves_2024, Yuan_2024}. Mergers can also trigger active galactic nuclei (AGN) activity \citep{Silk98, Hopkins08, Ellison11, Satyapal_2014, Gao_2020, Li_2023, Bickley_2023, Sharma_2024, Duan_2024b}, and the most luminous AGN, ultra-luminous infrared galaxies, and extreme emission line galaxies are often associated with mergers \citep{Kartaltepe10, Ellison13, Gupta_2023, Marshall_2023, Perna_2025}. Moreover, \citet{Witten_2024} recently discovered several Ly$\alpha$ emitters with close companions at $z > 7$, concluding that mergers may drive Ly$\alpha$ emission in these early systems and facilitate the escape of Ly$\alpha$ photons from ionized bubbles \citep[see also][]{Saxena_2023, Witstok_2024}. This intense star formation activity in mergers is subsequently quenched \citep{Hopkins08, Toft14, Ellison_2022}, leading to quiescent galaxies characterized by low or suppressed star formation rates. This sequence of events ultimately contributes to bulge formation, resulting in the massive elliptical galaxies observed in the local Universe. Recently, it has been found that mergers can also induce bar formation in spiral galaxies at high redshifts \citep[see e.g.,][]{Fragkoudi_2025}. Galaxy mergers are directly related to the mergers of supermassive black holes (SMBHs), which produce a low-frequency gravitational-wave background that has been measured by the North American Nanohertz Observatory for Gravitational Waves (NANOGrav) Collaboration \citep{Agazie_2023}. Therefore, it is crucial to accurately measure and constrain the galaxy merger history throughout cosmic time to understand these physical processes better.

There are two main methods to empirically study the fraction of galaxies that are undergoing mergers: \((i)\) counting the galaxies that are in close pairs on the projected plane of the sky and from this estimating the abundance of mergers -- \textit{close-pair method} \citep[e.g.,][]{Zepf_1989, LeFevre_2000, Patton_2002,Lopez-Sanjuan15, Man16, Mundy17, Mantha18, Duncan19, Conselice_2022, Duan_2024}, \((ii)\) finding systems that are at the late stages or have recently completed merging based on disturbed morphologies -- \textit{morphological method} -- that can be further broken down to \textit{quantitative} techniques, i.e. using morphological parameters \citep[e.g.,][]{Conselice_2003, Lotz_2008a, Jogee_2009, Desmons_2023, Rose_2023, Dalmasso_2024}, and more recently, using deep learning to identify mergers \citep[e.g.,][]{Pearson_2019, Ferreira_2020, Pearson_2022, Bickley_2022, Margalef-Bentabol_2024}, or \textit{qualitative} methods, for example, visual classifications \citep[e.g.,][]{Kartaltepe_2015}, and studies which utilise the identification and presence of tidal signatures \citep[e.g.,][]{Kado-Fong_2018, Mantha_2019}. These two methods complement each other as they probe different phases of galaxy mergers with different observability timescales. However, one of the main sources of uncertainty in the derived merger rates is the timescale itself, which makes comparisons between the results obtained by the two different methods difficult and ambiguous. In addition, several other methods exist based on the internal kinematics from spectroscopy (e.g., velocity fields) to identify galaxies involved in mergers or recent merger remnants \citep[e.g.,][]{Jesseit_2007, Shapiro_2008}. Several JWST/NIRSpec IFU works, mostly as part of the Galaxy Assembly with NIRSpec IFS (GA-NIFS) survey, have found and studied mergers and multiple companion groups, both in star-forming galaxies \citep[SFGs e.g.,][]{Arribas_2024, Jones_2024,  Lamperti_2024, Marconcini_2024, Rodriguez_Del_Pino_2024, Scholtz_2025} and in AGN \citep[e.g.,][]{Perna_2023, Marshall_2024, Perna_2025}. These are studies on individual systems with spatially resolved spectroscopy in 2D regions of typically $r < 10~{\rm kpc}$, using a methodology rather different from this work. In the following, we will only focus on the close-pair methodology to constrain the galaxy merger rate.

When determining merger rates by the close-pair method, the main galaxy and its companion(s) have to satisfy some selection criteria. The typical criteria are that galaxies have to be at the same (or similar) redshift, which translates to being at a close radial distance from each other and within some projected 2D physical separation (usually between $20-50$ kpc) on the sky to be close-pairs. In observational studies, the challenge lies in constraining these potential pairs in the third spatial coordinate, i.e. requiring them to be close in redshift. In spectroscopic surveys this corresponds to a maximum velocity offset of $\sim 500$ km/s. However, few large spectroscopic surveys exist that probe deep enough to measure redshifts for a mass-complete sample at a meaningful scale, especially at high redshifts. Instead, most works resort to photometric redshifts from large deep photometric surveys and define a similar criterion in redshift space (although less restrictive due to the uncertainty in photometric redshifts), usually defined as $\delta z / (1+z) \simeq 0.01$ \citep[e.g.,][]{Molino_2014, Man16}, which translates into a velocity uncertainty of $3000~{\rm km/s}$. However, only using the peaks of the photometric redshift posterior distributions and their 1$\sigma$ uncertainties \citep[e.g.,][]{Man16, Mantha18} leads to information loss about the exact shape of the probability distribution function (PDF) and could lead to unreliable results. The method developed by \citet{Lopez-Sanjuan15} therefore incorporates the full posterior distribution of the photometric redshifts and propagates the associated uncertainties throughout the full close-pair analysis \citep[see also ][]{Mundy17, Duncan19, Conselice_2022}. In this work we use the same method to obtain close-pair fractions.

The first observational studies of galaxy close-pair fractions relied on robust spectroscopic datasets but could only study bright and massive galaxies at low redshifts up to $z \sim 1.2$ \citep[e.g.,][]{Patton_2002, Lin_2004, Kartaltepe_2007, Lopez-Sanjuan_2012, Xu_2012}. All of these results agree that the major galaxy pair fraction increases with redshift, evolving as a power law. To probe merger fractions at higher redshifts, studies had to rely on large photometric surveys, either by identifying morphologically disturbed systems \citep[using CAS and M$_{20}$/Gini parameters e.g., as in][]{Conselice_2008} or by looking for close-pairs using photometric redshifts derived from fluxes measured in multiple bands \citep[e.g.,][]{Bluck_2009, Williams_2011, Man16, Mantha18}. These latter works mainly utilise deep photometric data using Hubble Space Telescope (HST) from large surveys such as the Cosmic Assembly Near-infrared Deep Extragalactic Legacy Survey \citep[CANDELS, ][]{Koekemoer_2011}, and study close-pair fractions in the range of $z \sim 0-3$. Whereas at low redshifts, all studies agreed that the pair fraction rises, at these intermediate redshifts of $z \sim 2-3$, some find a further increase, a flattening evolution, or in some cases, even a turn-over. This shows that there was already a significant uncertainty in close-pair fraction trends found by HST at $z \sim 3$. It is important to note, that all these works only use the peak values of the derived photometric redshifts, which could be highly affected by uncertainties. Therefore, works propagating the entire probability distributions of photometric redshifts throughout their analysis by the method described in \citet{Lopez-Sanjuan15} are generally more robust. Although these studies provide more reliable results \citep{Mundy17, Conselice_2022} and probe to higher redshifts \citep[up to $z \sim 6$, see][]{Duncan19}, they still disagree on the trends of the pair fraction and merger rate evolution. Other instruments and surveys also investigated the number of close-pairs at higher redshifts (up to $z \sim 6-7$), such as the Multi Unit Spectroscopic Explorer (MUSE) deep fields \citep{Ventou_2017, Ventou_2019}, Subaru Hyper Suprime-Cam (HSC) imaging surveys \citep{Shibuya_2022}, or the ALMA [CII] surveys \citep[e.g., ALPINE;][]{Romano_2021}. Similar to previous studies, these results do not resolve the disagreement between a steep increase or turn-over in pair fractions at high redshifts. It is also worth noting, that most of these studies mentioned above use significantly different selection functions and employ different projected separation limits (see Table~\ref{tab:merger_studies} for a list of a few relevant studies), which makes comparison difficult and might amplify the lack of agreement.

The advent of the James Webb Space Telescope \citep[JWST, ][]{Gardner_2006, Gardner_2023} opens a new window upon the study of galaxy mergers at high redshifts using deep photometry by the NIRCam instrument \citep{Rieke_2023a}, complemented by high-resolution spectra from NIRSpec \citep{Jakobsen_2022, Boker_2023}. A high number of recent studies present discoveries at high redshifts about early galaxy formation, and in particular related to galaxy mergers \citep[e.g.,][]{Claeyssens_2023, Hashimoto_2023, Hsiao_2023, Jin_2023, Suess_2023, Tacchella_2023a, Treu_2023, Alberts_2024, Carnall_2024, Decarli_2024, Duan_2024, Duan_2024b, deGraaff_2024, Hsiao_2024}. A recent study by \citet{Duan_2024} examines close-pair fractions and major merger rates in the redshift range $4.5 \leq z \leq 11.5$ using the probabilistic method mentioned previously, where they find rising merger rates up to $z\sim6$ followed by a flattening. In the discussion section, we review the differences and improvements compared to their study.

The aim of this study is to investigate the redshift evolution of close-pair fractions and major galaxy merger rates at the ambiguous and previously under-explored redshift range of $z \sim 3-9$. We use deep photometric and spectroscopic observations of the GOODS-South and GOODS-North fields by the JWST Advanced Deep Extragalactic Survey (JADES) collaboration \citep{Eisenstein_2023}. We adopt the statistical method developed by \citet{Lopez-Sanjuan15}, but instead of analysing a luminosity-selected galaxy sample, we perform a sample selection based on stellar masses as in \citet{Mundy17}. We find that the close-pair fraction rises and peaks at $z \approx 5-6$, followed by a turn-over at higher redshifts that also scales with the stellar mass of the primary galaxies. In the case of merger rates, we find that there is an increase and a subsequent flattening beyond $z \approx 6$. The resulting mass accretion rate is significantly (by a factor of ${\sim}5$) lower than the star-forming main sequence (SFMS) SFR at these high redshifts. This suggests that mergers are not the dominant channel for the mass growth of galaxies and contribute to about 3-13\% of the ex-situ mass fraction of an average galaxy.

The paper is structured as follows. Section~\ref{sec:data} provides an overview of the catalogues and data products used in this study, starting with an overview of the JADES survey. We detail the methodology for obtaining photometric redshifts and compare them with existing spectroscopic surveys, as well as discussing the derived stellar masses and their completeness. Section~\ref{sec:close-pair methodology} outlines the process for identifying close pairs, including selection criteria and sample selection based on redshift and stellar mass. We also describe the pair probability function and correct for selection effects. In Section~\ref{sec:pair fraction}, we present results on the pair fraction and its evolution with redshift. Section~\ref{sec:major merger rate} covers the conversion of observational pair fractions into a physically meaningful major merger rate by assuming a merger observability timescale. Section~\ref{sec:discussion} discusses our findings, including evolutionary trends, comparisons with cosmological simulations, and the relationship between star formation and merger rates. Finally, Section~\ref{sec:conclusion} summarises our work and highlights the main conclusions.

Throughout this paper, we adopt the AB magnitude system \citep{Oke_1983}. We use a standard cosmology with $\Omega_{m} = 0.310$,  $\Omega_{\Lambda} = 0.689$, and $H_0 = 67.66 \mathrm{\ km \ s^{-1} \ Mpc^{-1}}$ \citep{Planck_2020}, and we refrain from using the little $h$ notation for the Hubble parameter \citep{Croton_2013}. Throughout our analysis, we use the \texttt{Astropy} python package \citep{astropy:2022}, and its subpackage \texttt{astropy.cosmology}, where we assume a flat $\Lambda$CDM cosmology with parameters from \citet{Planck_2020}.

\section{Data}
\label{sec:data}

In this section, we present the various data sets utilised to measure the close-pair fractions of galaxies, which we will later convert into major merger rates. We start by discussing the data and footprints from the JADES survey, followed by a description of the estimated photometric redshifts and their quality assessment, as well as the available spectroscopic data. Finally, we explain how stellar masses are calculated for each galaxy using spectral energy distribution (SED) fitting, as well as the overall stellar mass completeness of the different survey regions.

\subsection{JADES survey}
\label{sec:JADES survey}

\begin{figure*}
 \centering
 \subfloat{\includegraphics[width=0.5\linewidth]{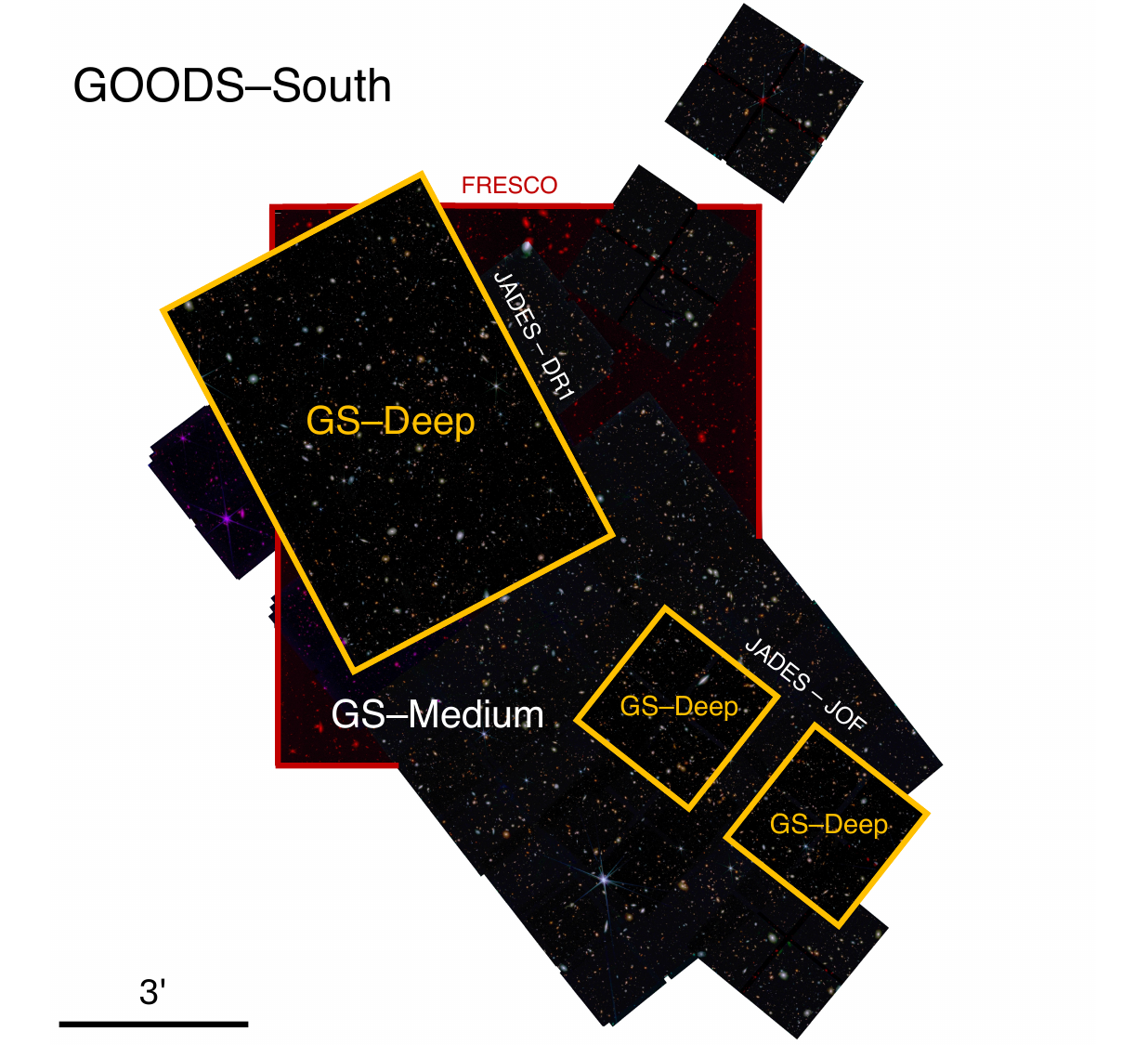}}
 \subfloat{\includegraphics[width=0.5\linewidth]{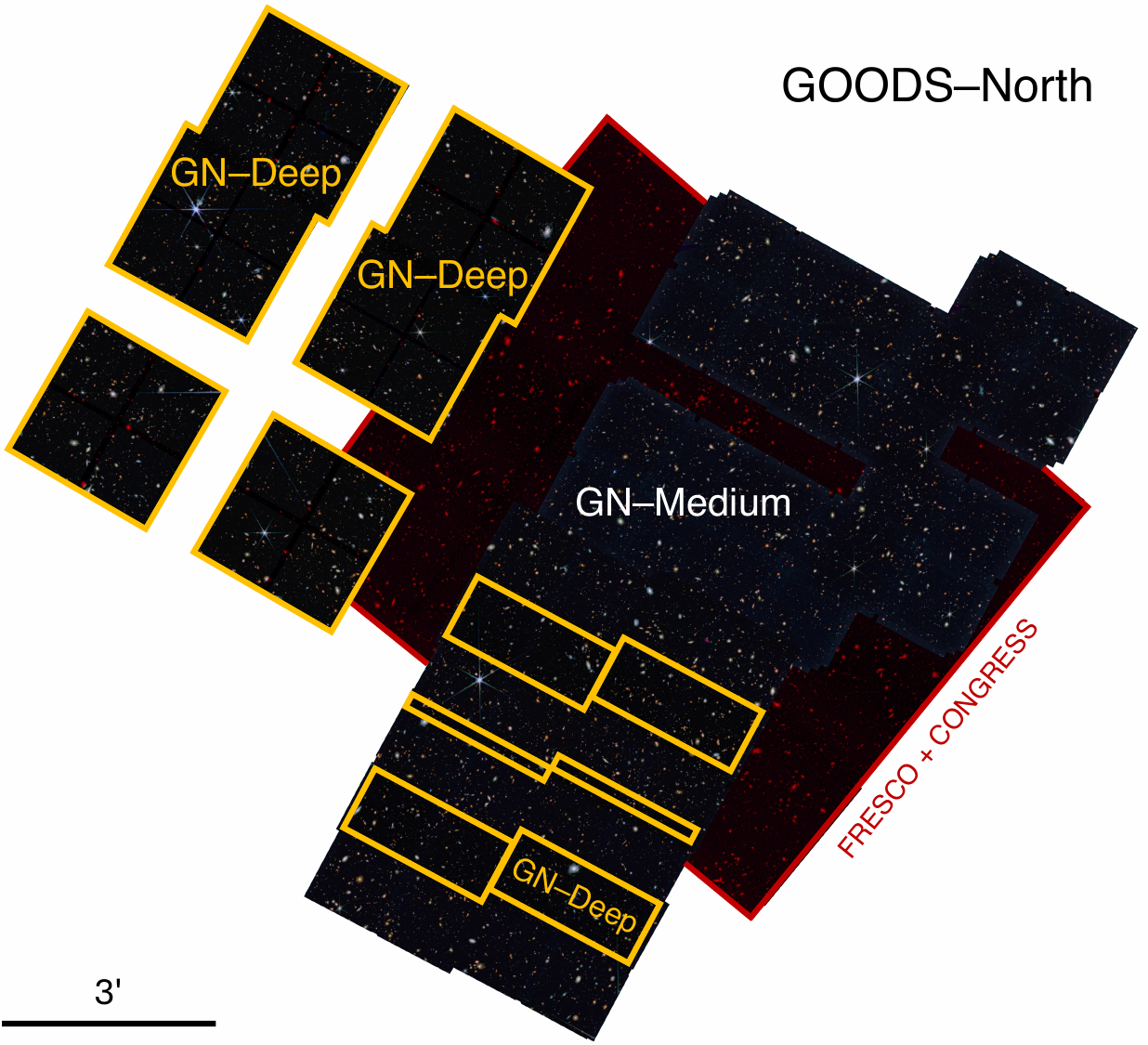}}
 \caption{Footprint of the JADES GOODS-South and GOODS-North fields from the mosaics as RGB images (using the F090W, F200W, and F444W bands). Both the GOODS-South and GOODS-North field is further divided into deep and medium tiers based on the exposure time in the F444W band, with the deep tier indicated by the orange area with exposure time above the threshold of $\mathrm{T_{exp}^{GS} = 32.5 \/\ ks}$, and $\mathrm{T_{exp}^{GN} = 9.5 \/\ ks}$, respectively. The GS-Deep tier corresponds to the JADES DR1 and JOF footprints.}
 \label{fig:tiers}
\end{figure*}

The JWST Advanced Deep Extragalactic Survey \citep[JADES, ][]{Eisenstein_2023} is the deepest and most extensive extragalactic survey to date, probing the areas of the  Great Observatories Origins Deep Survey \citep[GOODS, ][]{Giavalisco_2004}, covering the Hubble Ultra Deep Field \citep[HUDF, ][]{Beckwith_2006} in the GOODS-South region, and the GOODS-North region. JADES is a joint GTO program between the NIRCam and NIRSpec GTO teams that consists of NIRCam imaging, NIRSpec spectroscopy, and MIRI imaging. With DR3 and previous data releases \citep{Rieke_2023, Eisenstein_2023b, Bunker_2024, D'Eugenio_2025}, complemented by JWST Extragalactic Medium-band Survey \citep[JEMS, ][]{Williams_2023} and First Reionization Epoch Spectroscopically Complete Observations \citep[FRESCO, ][]{Oesch_2023} data, JADES has more than ${\sim} 175 \/\ \mathrm{arcmin^2}$ area covered by 8-10 photometric filters (${\sim}200 \/\ \mathrm{arcmin^2}$ in F444W) and over 6000 spectra. It probes down to an unprecedented photometric depth of 30.8 AB magnitude in F444W.

We divide the JADES photometric sample into four different tiers based on the area and exposure time. Both the GOODS-South and GOODS-North fields have different flux depths due to multiple observations and exposure times. Therefore, we divide the footprints by imposing an exposure threshold of $\mathrm{T_{exp}^{GS} = 32.5 \/\ ks}$ for GOODS-South, and $\mathrm{T_{exp}^{GN} = 9.5 \/\ ks}$ for GOODS-North. In the case of GOODS-South, the area that was observed with longer exposure time than $\mathrm{T_{exp}^{GS}}$ is defined as \textit{GS-Deep} and corresponds to the JADES footprint from DR1 \citep{Rieke_2023} and the JADES Origins Field \citep[JOF, ][]{Eisenstein_2023b} and it is indicated on Figure~\ref{fig:tiers}. The remaining area in the GOODS-South field is defined as \textit{GS-Medium}, which corresponds to an exposure time lower than $\mathrm{T_{exp}^{GS}}$ and was released as part of DR2 \citep{Eisenstein_2023b}. Similarly, GOODS-North is divided into a deep and medium region based on the integration time.

Here, we list the coverage by different photometric bands that are later used in photometric redshift estimates and SED-fitting to obtain stellar masses (see Sections~\ref{sec:photometric redshifts} and ~\ref{sec:Stellar masses}). We use JADES NIRCam photometry in bands F090W, F115W, F150W, F162M, F200W, F250M, F277W, F300M, F335M, F356W, F410M, and F444W, which are available in both GOODS-South and -North Deep and Medium tiers. This is supplemented by deep medium-band photometry in F162M, F250M, and F300M as part of the JOF observations \citep{Eisenstein_2023b}, which is contained in our GS-Deep tier. In addition, we use JEMS \citep{Williams_2023} NIRCam photometry in the F182M, F210M, F430M, F460M, and F480M bands as part of GOODS-South. Both GOODS fields are further complemented by FRESCO \citep{Oesch_2023} and CONGRESS (GOODS-North) F444W NIRCam photometry, obtaining the most extensive coverage in this band. Furthermore, both GOODS fields have ancillary HST photometric observations using the ACS bands F435W, F606W, F775W, F814W, F850LP, and from the IR bands F105W, F125W, F140W, and F160W.

For the photometric catalogues, we are using internal versions v0.9.3 for GOODS-South and v0.9.1 for GOODS-North, most of which were released as part of DR2 \citep{Eisenstein_2023b} and DR3 \citep{D'Eugenio_2025} and is publicly available on Mikulski Archive for Space Telescopes (MAST; \url{https://archive.stsci.edu/hlsp/jades})\footnote{DOI: \href{https://archive.stsci.edu/doi/resolve/resolve.html?doi=10.17909/8tdj-8n28}{10.17909/8tdj-8n28}}. A detailed description about the exact catalogue construction can be found in \citet{Rieke_2023}, and here we give a brief summary of the main steps as these could significantly affect the number of sources and satellites detected, and hence the pair fractions derived. A detection catalogue of blended sources is first generated using \texttt{Photutils} \citep{Bradley_2024}, adopting a threshold of ${\rm SNR} \geq 1.5$. Deblending is performed on the segmentation map utilising a logarithmically scaled F200W image. For larger segments, further deblending is applied with the \texttt{deblend\_sources} function from \texttt{Photutils}. Satellite sources are identified by running \texttt{detect\_sources} on the high-pass filtered outer light profiles of extended sources. Compact sources missed or excluded in previous steps are recovered by reapplying source detection to blank regions, using a threshold of ${\rm SNR} = 3.5$. The final segmentation map serves as the basis for constructing the photometric catalogue. Object centroids are determined using the windowed position algorithm provided in \texttt{Photutils}, with implemented \texttt{Source Extractor} \citep{Bertin_1996} methodology, applied to the NIRCam long-wavelength signal image. Elliptical sizes, orientations, and Kron radii (with $K = 2.5$) for each source are derived from the signal image. The photometric catalogues provide flux measurements in multiple apertures for each source, with forced photometry performed at object positions identified during detection. In this analysis, we use Kron apertures applied to images convolved (\texttt{KRON\_CONV}) to a consistent resolution across photometric bands.

\subsection{Photometric redshifts}
\label{sec:photometric redshifts}

\subsubsection{\texttt{EAZY}}

We obtain photometric redshifts for every galaxy in the GS and GN fields by using the photometric redshift code \texttt{EAZY} \citep{Brammer_2008}. \texttt{EAZY} is a template fitting code that combines galaxy spectra to fit the observed photometric data by searching for the best fit on a redshift grid. The templates and assumptions made for our fits are detailed in Section 3.1 of \citet{Hailine_2024}. We take $z_a$ as the photometric redshift, which corresponds to the minimum in the $\chi^2$ of the fit. The grid search is performed in the redshift range $z = 0.01 - 22$ with a step size of $\Delta z = 0.01$.

Here, we briefly summarise the templates used in the \texttt{EAZY} fits. We started with the modified version of the original seven templates from \citet{Brammer_2008} to include line emission, and complemented them with a dusty and a high-equivalent width template \citep{Erb_2010}. We further supplemented these by seven new JADES templates\footnote{Publicly available on Zenodo: \href{https://zenodo.org/records/7996500}{https://zenodo.org/records/7996500}} that were optimised for better redshift estimates using mock-galaxy observations from the JAGUAR simulation \citep{Williams_2018}. These templates better cover the colour space of JAGUAR galaxies, including blue UV-bright and red dusty galaxies, and were generated by using Flexible Stellar Population Synthesis \citep[\texttt{fsps};][]{Conroy_2010}. For a comprehensive description of the \texttt{EAZY} templates used for the fits, we refer the reader to \citet{Hailine_2024}.

\subsubsection{Photometric redshift probability distributions}

We obtain the photometric redshift posterior distributions from \texttt{EAZY} by using the $\chi^2(z)$ outputs of the fits. Assuming a constant prior for the redshift, we calculate the posterior distribution by $P(z) = \textrm{exp}[-\chi^2(z) / 2]$ with a normalisation of $\int P(z) dz = 1.0$. We note here that generally, $z_a$ does not coincide exactly with $z_\mathrm{peak}$, which is internally calculated by \texttt{EAZY} as a probability-weighted average redshift. As described in \citet{Hailine_2024}, we use $z_a$ as the best photometric redshift in the following analysis. We note here that recent studies found that photometric redshift estimation could be affected by the Eddington bias \citep[see e.g.,][]{Serjeant_2023, Donnan_2024} that leads to higher estimated values than the true population, therefore we compare our sample with a robust spectroscopic redshift catalogue (see Section~\ref{sec:Spectroscopic redshift catalogue}).

\subsubsection{Odds quality parameter}
\label{sec:odds quality parameter}

The \textit{odds} quality parameter $\mathcal{O}$ is a proxy for the reliability of the photometric redshift fit, and it is a useful measure to select a robust sample of galaxies with accurate photometric redshifts and a low rate of catastrophic outliers. The odds parameter is defined as the redshift probability distribution function (PDF) integrated over a small region $\pm K (1+z_a)$ \citep{Benitez_2000, Molino_2014} around the best photometric redshift $z_a$,
\begin{equation}
 \mathcal{O} = \int^{+K (1 + z_a)}_{-K(1+z_a)} P(z-z_a) dz,
\label{odds_eq}
\end{equation}
where $K$ is a survey-specific constant. Previous studies have empirically chosen values for K ranging between $0.0125-0.05$ \citep[e.g.,][]{Molino_2014, Lopez-Sanjuan_2014, Lopez-Sanjuan15}, that typically depend on the photometric redshift accuracy of the data. For example, \citet{Molino_2014} adopts $K = 0.0125$ for medium-band filters, while \citet{Conselice_2022} uses a larger value of $K=0.05$ in the case of broadband filters. In this work, we choose $K = 0.03$ since we have photometric redshifts that have relatively high accuracy obtained from multiple wide- and medium-band JWST filters and additional HST photometry.

In our initial sample selection, we require that all objects must have an odds parameter $\mathcal{O} \geq 0.3$. This choice of odds cut is consistent with previous studies \citep[e.g.,][]{Lopez-Sanjuan15, Mundy17, Conselice_2022}, and ensures that our initial sample has accurate and robust photometric redshifts. This odds quality cut significantly reduces the initial sample (discarding ${\sim}40\%$ of the sources in both GS and GN).


\subsection{Spectroscopic redshift catalogue}
\label{sec:Spectroscopic redshift catalogue}

\begin{figure}
 \includegraphics[width=\columnwidth]{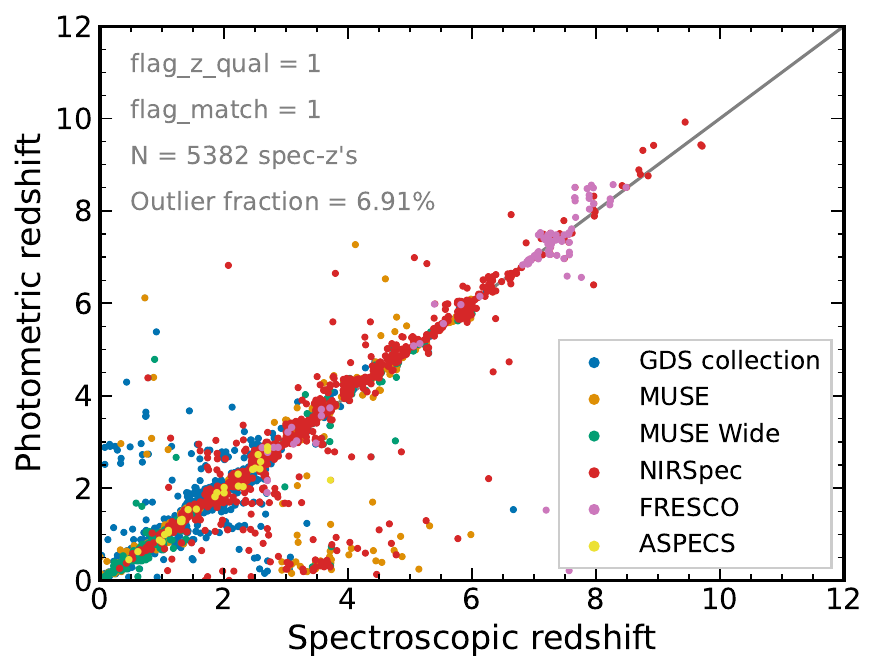}
 \caption{One-to-one comparison between the estimated photometric redshifts from \texttt{EAZY} and all the available spectroscopic redshifts from JADES and other surveys as indicated by the colours. This comparison is for objects in the GOODS-South field, including only the highest-quality redshifts and the best coordinate matches between the photometric and spectroscopic sources. There is a good agreement between the two types of redshifts with an outlier fraction of $6.91\%$. The cloud of outlier point at $z_{\mathrm{spec}} \sim 3-4$ below the one-to-one line might be present due to \texttt{EAZY} confusing the Lyman-break with a Balmer-break, and hence estimating the photometric redshifts to be much lower.}
 \label{fig:specz_vs_photz}
\end{figure}

To assess the accuracy of the estimated photometric redshifts from \texttt{EAZY}, we compare them to spectroscopic redshifts. We compile a catalogue containing all the available spectroscopic redshifts of objects falling in the JADES footprints. We include spectroscopic redshifts from the JADES NIRSpec observations \citep{Bunker_2024, D'Eugenio_2025}, the FRESCO and CONGRESS surveys (F. Sun, private communication; see also \citealt{Meyer_2024} for H$\alpha$ and \citealt{Covelo-Paz_2025} for H$\beta$+[O {\sc III}] catalogues, and survey paper by \citealt{Oesch_2023}), the MUSE DR2 \citep{Bacon_2023}, the MUSE-Wide DR1 \citep{Urrutia_2019}, the ASPECS (ALMA SPECtroscopic Survey in the UDF) Large Program \citep{Decarli_2019}, and a collection of all publicly available spectroscopic redshifts from legacy observations falling in the CANDELS fields (GDS collection on Figure~\ref{fig:specz_vs_photz}; N. Hathi, private communication). The catalogue is built by coordinate matching and correcting for any systematic coordinate offsets between surveys. The classification of reliability and quality of the spectroscopic redshifts from these different surveys is converted to a unified system, categorising them as secure/best (1), less confident (2), and unreliable (3).

Our final assembled catalogue of spectroscopic redshifts consists of 5382 sources in the GOODS-South and 2591 in the GOODS-North field, falling within the best category in quality and coordinate match. The assembled spectroscopic redshift catalogue is compared to the matched photometric redshift catalogue from \texttt{EAZY}. We plot the highest quality matched sources for GOODS-South in Figure~\ref{fig:specz_vs_photz}. We defined the outlier fraction as $|z_{\mathrm{spec}} - z_{\mathrm{phot}}| / (1 + z_{\mathrm{spec}}) > 0.15$. Overall, there is a good agreement between the two redshifts, with an outlier fraction of $6.91\%$ for GS and $13.62\%$ for GN. The higher outlier fraction in case of GN is likely due to the shallower depth compared to GS and to the fewer number of medium photometric bands available, making the \texttt{EAZY} fits less constrained. The average difference between the spectroscopic and photometric redshifts is $\langle z_{\mathrm{spec}} - z_{\mathrm{phot}} \rangle = 0.026$ for GS. The scatter around the one-to-one relation is measured by the Normalised Median Absolute Deviation (NMAD), which is defined as
\begin{equation}
 \sigma_{\mathrm{NMAD}} = 1.48 \times \mathrm{median} \left( \left| \frac{\delta z - \mathrm{median}(\delta z)}{1 + z_{\mathrm{spec}}} \right| \right),
\label{odds}
\end{equation}
where $\delta z = z_{\mathrm{spec}} - z_{\mathrm{phot}}$. For the GS sample, this quantity is $\sigma_{\mathrm{NMAD}} = 0.065$, which reflects a good agreement between the spectroscopic and photometric redshifts and the robustness of the catalogue. We note that at high redshift ($z > 8$), damped Lyman-alpha (DLA) absorption might bias photometric redshifts upwards \citep{Fujimoto_2023, Helton_2024, Hailine_2024, Finkelstein_2024, Willott_2024}. However, our sample of interest ($3 \leq z \leq 9$) is barely affected by this bias.

Beyond this comparison being a useful and necessary check on the photometric redshifts estimated by \texttt{EAZY}, we also compile a catalogue of the best available redshifts, including the spectroscopic redshifts instead of the photometric ones if available. If multiple $z_{\mathrm{spec}}$ exists for an individual object, we select the best match and highest quality for our catalogue. If multiple $z_{\mathrm{spec}}$ of the same highest quality and match exist for a single source, we take the average of the spectroscopic redshifts if they are in good agreement ($\delta z_{\mathrm{spec}} < 0.1$). In very few cases, there is a disagreement between the highest quality spectroscopic redshifts, and we do not include these values in our final catalogue of best available redshifts.

\subsection{Stellar masses}
\label{sec:Stellar masses}

For our close-pair selection to find major mergers, a crucial parameter for the galaxies considered is their stellar mass. We obtain stellar masses by two different codes, which allows us to perform two independent analyses and compare the results, providing a further check on the robustness of this study.

\subsubsection{\texttt{Prospector}}

To obtain a robust catalogue of stellar masses, we run \texttt{Prospector} \citep{Leja_2017, Johnson_2021} for a subsample that is relevant to this study. \texttt{Prospector} is an SED-fitting code that uses Bayesian inference to estimate galaxy parameters using different priors, employing a flexible stellar population synthesis (FSPS) framework \citep{Conroy_2009, Conroy_2010} through \texttt{python-fsps} \citep{Johnson_2024}, using Markov Chain Monte Carlo (MCMC) methods through \texttt{emcee} \citep{Foreman-Mackey_2013} to explore the parameter space and determine the probability distributions of various galaxy properties employing dynamic nested sampling by \texttt{dynesty}. The code incorporates models for stellar populations, star formation histories, including non-parametric SFH \citep{Leja_2019}, dust attenuation, nebular emission \citep{Byler_2017}, and other physical processes, which allows for robust estimates and uncertainties for parameters such as stellar mass, SFR, metallicity, and dust content \citep[see e.g.,][]{Tacchella_2022, Robertson_2023}. On the other hand, it is computationally resource-intensive and significantly slower than \texttt{eazy-py} but provides more accurate and robust results.

Therefore, we only fit a subsample of galaxies using \texttt{Prospector} from our initial catalogue ($57,863$ fits in total for GS and GN), corresponding to the redshift range of $3 \leq z \leq 9$ above a signal-to-noise ratio (SNR) threshold in the flux of $\mathrm{SNR} > 3$ measured in the F444W band, which significantly reduces the number of objects to be fitted. The SED fitting used in this work is detailed in \citet{Simmonds_2024}, and we briefly summarise the priors used and their allowed ranges. We assume a \citet{Chabrier_2003} initial mass function (IMF) with mass cut-off at 0.1 and 100 $\rm M_\odot$, a top-hat distribution (between $\log(M_\star/{\rm M_\odot}) = 6-12$) for the stellar mass prior, a non-parametric SFH prior \citep{Leja_2019} with eight time bins, a free ionisation parameter with an upper limit of $\log \langle U \rangle_{\rm max} = -1.0$, and we allow the stellar metallicity to be in the range $0.01-1~{\rm Z_\odot}$. We assume a two-component dust model \citep{Charlot_2000, Conroy_2009}, which accounts for the differential effect of dust on young stars ($<10$ Myr) and nebular emission lines through different optical depths and a variable dust index \citep{Kriek_2013}. Finally, we assume a constant photometric redshift prior based on the \texttt{EAZY} photo-$z$ estimates, or fixing the redshift at the spec-$z$ if available in our previously compiled catalogue.

We note here that the estimated stellar masses by \texttt{Prospector} are robust; nevertheless, they can be prone to uncertainties due to the different assumptions in the priors. One such source of uncertainty is the choice of the prior SFH \citep{Tacchella_2022, Whitler_2023b}, for which we choose a non-parametric model \citep[continuity SFH;][]{Leja_2019}. In this model, the SFH is divided into eight different SFR bins, with the first and last bins fixed 5 Myr and $z=20$, respectively, and the remaining time bins divided into equal intervals of $\log_{10}(t_{\rm lookback})$. The ratios between adjacent time bins are determined by the bursty-continuity prior \citep{Tacchella_2022}. In this model, we adopt the Student’s t-distribution (with width $\sigma = 0.3$) for fitting for the $\Delta \log({\rm SFR})$ between the neighbouring bins, which prevents sharp transitions \citep{Leja_2017}. Furthermore, this model is also supported by \citet{Lower_2020}, who demonstrate that non-parametric SFHs outperform traditional parametric models and lead to significantly improved and robust stellar masses. Another source of uncertainty can be the choice of IMF that can introduce variations in the resulting stellar masses \citep{Conroy_2013}. These uncertainties caused by the different model assumptions would likely introduce systematic offsets in the estimated stellar masses, and since we estimate the completeness and define the mass bins self-consistently, the resulting pair fractions would likely be similar in the offset stellar mass bins.

\subsubsection{\texttt{eazy-py}}

We also obtain the stellar mass of each photometric source by \texttt{eazy-py} \citep{Brammer_2008, Brammer_2021}, which is a set of \texttt{python} photometric redshift tools based on \texttt{EAZY}. It is a fast template-fitting SED code that outputs different galaxy parameters beyond only estimating the photometric redshift, such as the stellar mass, SFR, age, luminosity, and UBVJ colours. Although these parameter estimates are often crude and uncertain, the main advantage of this code is its speed compared to other more sophisticated SED-fitting codes, which makes it ideal for our large dataset (over $300,000$ photometric sources in total). 

These stellar masses are used for an independent analysis of pair fractions using a data set that is less robust, which is discussed in Appendix~\ref{sec:appendix_eazy}.


\subsection{Stellar mass completeness}

It is important to accurately assess the completeness of our survey and different sub-tiers in order to perform a robust close-pair analysis at high redshift, where the incompleteness in stellar mass could potentially significantly affect the trends observed in the evolution of close-pair fractions and major merger rates.

\subsubsection{$5\sigma$ point-source flux depth estimates}
\label{sec:5sigma point-source flux depth estimates}

\begin{table}
\centering
\caption{Flux limits derived from the 5$\sigma$ point-source depths measured in the different tiers for GOODS-North and GOODS-South fields, and the area of each footprint calculated from the non-zero pixels in the F444W band.}
\resizebox{\columnwidth}{!}{%
\begin{tabular}{lccc}
\hline
Tier        & 5$\sigma$ flux limit [nJy] & Flux limit [AB mag] & Area [arcmin$^2$] \\
\hline
GN-Deep     & 4.34             & 29.81             & 36.97               \\
GN-Medium   & 6.99             & 29.29             & 68.29               \\
GN          & 5.75             & 29.50             & 105.26              \\
GS-Deep     & 1.97             & 30.66             & 33.59               \\
GS-Medium   & 4.36             & 29.80             & 65.37               \\
GS          & 3.37             & 30.08             & 98.96               \\
\hline
\end{tabular}%
}
\label{tab:5-sigma}
\end{table}

The first step to assess the stellar mass completeness is to estimate the $5\sigma$ point-source flux depths of our different subfields. For this purpose, we use the full mosaics of the GS and GN fields in the F444W band. We perform aperture photometry using the \texttt{photutils} python package \citep{Bradley_2024} for a fixed $0.3 "$ diameter aperture. We use the segmentation map obtained from \texttt{Source Extractor} \citep{Bertin_1996}, perform a binary dilation (expands the regions flagged by the segmentation map to exclude any contamination caused by extended sources), and create a mask for any source pixels based on the segmentation map or any non-observed area. After sigma-clipping, we obtain the final mask to ensure that we only put apertures on the background, which are not affected by any source pixels. We randomly place apertures with a constant diameter of $0.3 "$ where we measure the flux and use the $80\%$ enclosed energy radius to do aperture correction (resulting in an aperture correction factor of 1.52). Finally, we calculate the $5\sigma$ flux limit, which corresponds to the point source flux depth of the survey considered. Additionally, while we measure the depth for fixed $0.3" $ apertures, the Kron-apertures for the selected initial sample are similar on average, with a median value of ${\sim}0.23" $, and therefore, the overall effect is small and the choice of $0.3" $ is a good approximation for the depth estimates.

We perform this measurement for the four different tiers defined in Section~\ref{sec:JADES survey}. We obtain $5\sigma$ flux limits of 1.97 nJy (30.66 AB mag), 4.36 nJy (29.80 AB mag), 4.34 nJy (29.81 AB mag), and 6.99 nJy (29.29 AB mag) for the GS-Deep, GS-Medium, GN-Deep, and GN-Medium subfields respectively (see Table~\ref{tab:5-sigma}). These $5\sigma$ limits are lower (fainter) than the values cited in previous JADES release papers \citep{Rieke_2023, Eisenstein_2023b, Hailine_2024, D'Eugenio_2025}, which is partly due to using different aperture sizes and using more data that was observed after the data releases which increases the flux depths.

\subsubsection{Completeness analysis}
\label{sec:completeness analysis}

\begin{figure}
 \includegraphics[width=\columnwidth]{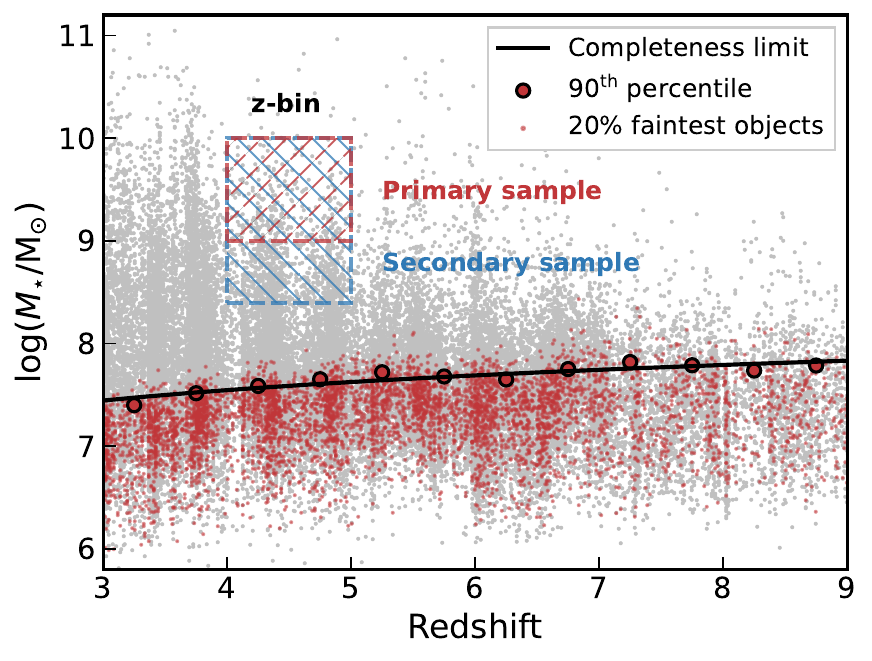}
 \caption{Stellar mass distribution of our sample in the GOODS-South field, with grey dots representing masses estimated with \texttt{Prospector}. The stellar mass completeness limit is calculated by the method of \citet{Pozetti_2010}. The red dots represent the $M_{\mathrm{lim}}$ values of the faintest $20\%$ galaxies, while the red circles with black edges are the $M_{\mathrm{min}}$ values in each redshift bin (width of $\Delta z = 0.5$), obtained by taking the $90^{\mathrm{th}}$ percentile of the $M_{\mathrm{lim}}$ distribution. The black solid curve is the redshift-dependent completeness limit $M_\star^{\mathrm{comp}}(z)$ obtained by fitting a logarithmic curve to the $M_{\mathrm{min}}$ values. Throughout this work we focus on the redshift range of $z = 3-9$, and the stellar mass range of $\log_{10}(M_\star / {\rm M_\odot}) = 8-10$. The red hatched box represents a fixed arbitrary stellar mass bin centred around $\mathrm{log}(M_\star / \mathrm{M_\odot}) = 9.5 \pm 0.5$, which determines the primary galaxy sample (in the example above ${N\rm _{primary} = 308}$) around which we search for major companions. The blue hatched box is the corresponding secondary sample (where ${N\rm _{secondary} = 1120}$), which represents a larger pool of galaxies that contains potential satellites with up to 1/4 lower mass than the primary galaxies.}
 \label{fig:sample}
\end{figure}

We perform the stellar mass completeness analysis by adopting the method described in \citet{Pozetti_2010}. We aim to find a redshift-dependent stellar mass completeness limit $M_\star^{\mathrm{comp}}(z)$, above which we essentially have a complete sample in stellar mass and potentially can observe all types of galaxies. In our flux-limited sample, the limiting stellar mass we can observe depends on the redshift and the mass-to-light ratio $M_\star/L$. Therefore, we calculate a minimum mass $M_{\mathrm{min}}$ for individual redshift bins of width $\Delta z = 0.5$, where we assume that all galaxies can be observed above a typical $M_\star/L$.

In each redshift bin, we calculate $M_{\mathrm{min}}$ by first calculating the limiting stellar mass $M_{\mathrm{lim}}$ of each galaxy. The limiting stellar mass is the mass a galaxy would have at its spectroscopic (or photometric) redshift if its apparent magnitude ($m_{\rm AB}$ measured in F444W) were equal to the limiting magnitude of the survey or subfield considered, which we obtain from the $5\sigma$ flux limits. The limiting stellar mass can be obtained by
\begin{equation}
 \mathrm{log}(M_{\mathrm{lim}}) = \mathrm{log}(M_{\star}) + 0.4(m_{\mathrm{AB}} - m_{\mathrm{AB}}^{\mathrm{lim}}),
\end{equation}
where $m_{\mathrm{AB}}^{\mathrm{lim}}$ is the limiting magnitude of the survey considered (measured in the F444W mosaic; see Section~\ref{sec:5sigma point-source flux depth estimates}). We then look at the resulting distribution of $M_{\mathrm{lim}}$ and only include the faintest $20\%$ galaxies in each redshift bin. To obtain the $90\%$ completeness limit $M_{\mathrm{min}}$ in each redshift bin, we take the $90^{\mathrm{th}}$ percentile of the $M_{\mathrm{lim}}$ distribution. Finally, we obtain the redshift-dependent stellar mass completeness limit $M_\star^{\mathrm{comp}}(z)$ by fitting a logarithmic curve to the $M_{\mathrm{min}}$ values at different redshifts, as shown on Figure~\ref{fig:sample}. These completeness limits are different for each subfield, but in general, are in the range $\log_{10}(M_\star / {\rm M_\odot}) \approx 7-8$ depending on redshift, with GS-Deep having the most complete and GN-Medium being the least complete out of our four subfields.

\section{Close-pair methodology}
\label{sec:close-pair methodology}

This section discusses our methodology for finding close pairs of major galaxy mergers at a wide redshift range. We adopt the close-pair selection method developed by \citet{Lopez-Sanjuan15}, where we make use of the full photometric redshift probability distribution function to propagate the associated uncertainties, as opposed to only using the peaks of these distributions. However, instead of selecting a sample based on the flux ratio of galaxies (as in \citealt{Lopez-Sanjuan15}), we perform a selection based on the stellar mass ratio, as it was done in \citet{Mundy17}. Although we mainly follow the selection method described in these two works with some minor modifications, we describe this method in full detail below. See Table~\ref{tab:merger_studies} for examples of close-pair selection methods and criteria applied by other studies.

\subsection{Close-pair selection criteria}
\label{sec:close-pair selection criteria}

All studies usually define three main selection criteria to find major galaxy mergers. As the name `close-pair' suggests, galaxies have to be within some defined physical separation $r_{\mathrm{max}}$ projected on the plane of the sky, which translates to an observed angular separation $\theta_{\mathrm{max}}$. To avoid confusing star-forming clumps within the same galaxy with mergers and ensure clearly deblended sources, many studies define a minimum separation $r_{\mathrm{min}}$ as well (and corresponding $\theta_{\mathrm{min}}$). The second criterion for two galaxies to be in a close pair is close proximity in the radial direction. In spectroscopic studies, this is usually defined as $\Delta v_{\mathrm{max}} \leq 500 \/\ \mathrm{km \/\ s^{-1}}$ in velocity space \citep[e.g.,][]{Patton_2000}, while in photometric surveys this translates to close proximity in photometric redshift space \citep[e.g., $\Delta z < 0.1$;][]{Man16}. For finding major mergers, the third requirement is that the stellar mass ratio between the secondary and the primary galaxy is $\mu = M_2 / M_1 \geq 0.25$ \citep[e.g.,][]{Lotz_2011}. For minor mergers, this value is usually defined as $0.1 \leq \mu \leq 0.25$. Once the number of close pairs is determined, the pair fraction $f_{\mathrm{pair}}$ can be easily computed
\begin{equation}
 f_{\mathrm{pair}} = \frac{N_{\mathrm{pair}}}{N_{1}},
 \label{eq:f_pair}
\end{equation}
where $N_{\mathrm{pair}}$ is the number of close-pairs, and $N_1$ is the number of primary galaxies within some pre-defined initial sample, e.g. a volume-limited sample of galaxies within a specified mass bin.

To summarise these three main selection criteria for finding close pairs of major galaxy mergers:
\begin{enumerate}[(i)]
    \item Close projected physical separation $r_{\mathrm{min}} \leq r \leq r_{\mathrm{max}}$ (with our choice of $r_{\mathrm{min}}=5 \ {\rm kpc}$ and $r_{\mathrm{max}} = 30 \ {\rm kpc}$)\footnote{For an analysis using different separation criteria see Appendix~\ref{sec:appendix_rvir}}, translating to close angular separation on the sky $\theta_{\mathrm{min}} \leq \theta \leq \theta_{\mathrm{max}}$,
    \item Close proximity in redshift or velocity space,
    \item Stellar mass ratio above $\mu \geq 0.25$ to find major mergers.
\end{enumerate}

In the following, we describe in detail the motivation and implementation of these selection criteria for a large photometric survey, where we include the $P(z)$ distributions of the estimated photometric redshifts.

\begin{table*}
    \centering
    \caption{Summary of the selection criteria of close-pair studies that are frequently referenced in this work, displaying the redshift and stellar mass ranges probed, the projected separation criteria, and the selection method, i.e. either spectroscopic, photometric (using the peak photometric redshifts) or probabilistic (using the full posterior distributions of the photometric redshifts).}
    \begin{threeparttable}
    \begin{tabular}{lcccc}
        \hline
        Study & Redshift range & Mass range, $\log(M_\star / {\rm M_\odot})$& $r_{\rm sep}$ [kpc] & Selection method \\
        \hline
        \citet{Lopez-Sanjuan15} & [0.4, 1] & $M_{\rm B} \leq -19, -19.5, -20$\tnote{a} & [10, 50] $h^{-1}$ & Probabilistic, based on $P(z)$ \\
        \citet{Man16}           & [0.1, 3] & >10.8 & [10, 30] $h^{-1}$ & $|z_1 - z_2| / (1+z_1) < 0.1$\tnote{b} \\
        \citet{Mundy17}         & [0.005, 3.5] & > 10, 11 & [5, 30] & Probabilistic, based on $P(z)$ \\
        \citet{Ventou_2017}     & [0.2, 6] & $\geq 9.5$, $< 9.5$ & [5, 25] $h^{-1}$ & $\Delta v_{\rm max} < 500$ km/s  \\
        \citet{Mantha18}        & [0, 3] & $> 10.3$ & [5, 50] & $(z_1 - z_2)^2 \leq \sigma_{z,1}^2 + \sigma_{z,2}^2$\tnote{c} \\
        \citet{Duncan19}        & [0.5, 6] & [9.7, 10.3], >10.3 & [5, 30] & Probabilistic, based on $P(z)$ \\
        \citet{Conselice_2022}  & [0, 3] & >11 & [5, 30] & Probabilistic, based on $P(z)$ \\
        \citet{Duan_2024}       & [4.5, 11.5] & [8, 10] & [20, 50] & Probabilistic, based on $P(z)$ \\
        \hline
    \end{tabular}
    \begin{tablenotes}
    \item[a] Sample selected by \textit{B}-band luminosity.
    \item[b] for $z_1<1$, and $<0.2$ for $z_1>1$.
    \item[c] This study used both photomteric and spectroscopic redshifts, and $\sigma$ are the photometric redshift errors.
    \end{tablenotes}
    \end{threeparttable}
    \label{tab:merger_studies}
\end{table*}


\subsection{Initial sample selection}
\label{sec:initial sample selection}

In this section, we describe how we selected the initial galaxy sample for our analysis. First, we limit our sample to the photometric redshift range of $3 \leq z \leq 9$, which is an underexplored redshift range in the study of galaxy mergers. We choose this lower and upper redshift limit, as the $z < 3$ range has been studied extensively in previous research on pair fractions and merger rates, and at $z > 9$, our catalogues contain very few objects after preselection (less than 0.1\% of the $z > 3$ sample), making a meaningful analysis difficult. 

We clean our initial sample by requiring a signal-to-noise ratio in the \texttt{F444W\_KRON\_CONV} photometry (Kron-aperture placed on images that have been convolved to the same resolution) of $\mathrm{SNR} \geq 3$ to filter out background noise and false detections (reduction to $52\%$ of the parent photometric sample of ${\sim}110,000$ objects). We perform an odds cut of $\mathcal{O} \geq 0.3$ as described in Section~\ref{sec:odds quality parameter} to only include galaxies with well-constrained photometric redshifts from \texttt{EAZY} (a further reduction to $63\%$ of the sample from the previous step). Furthermore, we remove any known interlopers, such as 9 brown dwarf candidates \citep{Hainline_2024a}, and falsely identified objects belonging to diffraction spikes of bright stars (see Section~\ref{sec:survey boundaries}), resulting in a further reduction by ${\sim}5\%$ of the remaining sample. Finally, we compute the stellar mass completeness limit for each tier (GS-Deep, GS-Medium, GN-Deep, GN-Medium) as described in Section~\ref{sec:completeness analysis}, and require all primary (and also secondary) objects to be above $M_\star^{\mathrm{comp}}(z)$.

\subsubsection{Redshift and stellar mass bins}
\label{sec:Redshift and stellar mass bins}

Following the initial sample selection, which is uniformly applied for the full analysis for each tier, we define different redshift and stellar mass bins, and we compute the pair fractions within each of these bins. We define the redshift bins equally between $3 < z < 9$ with widths of $\Delta z = 1$, i.e. we obtain results at half redshifts. We then define different constant stellar mass bins which do not depend on redshift. Due to the varying number density of objects having stellar masses above the completeness limit at different redshifts, we define bins with non-equal sizes, centred at $\mathrm{log}(M_\star / \mathrm{M_\odot}) = 8.25 \pm 0.25, 8.75 \pm 0.25, 9.50 \pm 0.50$. These bins were chosen such that they are always above $M_\star^{\mathrm{comp}}(z)$. We select primary galaxies from these stellar mass bins and will search around them for lower-mass satellites in the following parts of the close-pair selection. We can immediately determine the secondary sample of galaxies based on the stellar mass from the primary sample, as the stellar mass ratio is fixed for major mergers at $\mu \geq 0.25$. The primary and secondary sample for a given fixed stellar mass bin is visualised in Figure~\ref{fig:sample} (the number of primary galaxies within each bin can be found in Table~\ref{tab:pair_fraction_merger_rate}).


\subsection{Pair probability function}

In this section, we discuss in detail how we select close pairs from our photometric survey in a probabilistic and statistically robust manner. We build the \textit{pair probability function} (PPF) throughout this section, which mathematically codifies the selection criteria from Section~\ref{sec:close-pair selection criteria}.

\subsubsection{Redshift probability function}

As the first step and most crucial feature of this work, we assess the proximity of potential pairs in redshift space, i.e. closeness in radial distance. To perform this, we incorporate in our analysis the full posterior distributions of the estimated photometric redshifts output from \texttt{EAZY}. If there is an available spectroscopic redshift for a galaxy from our compiled $z_{\mathrm{best}}$ catalogue described in Section~\ref{sec:Spectroscopic redshift catalogue}, we use that instead. To include a $z_{\mathrm{spec}}$ in this probabilistic framework, we fit a narrow Gaussian to this redshift assuming a standard deviation of $\sigma_{z} = 0.01$ and normalisation of 1. 

\begin{figure*}
 \centering
 \subfloat{\includegraphics[width=0.5\linewidth]{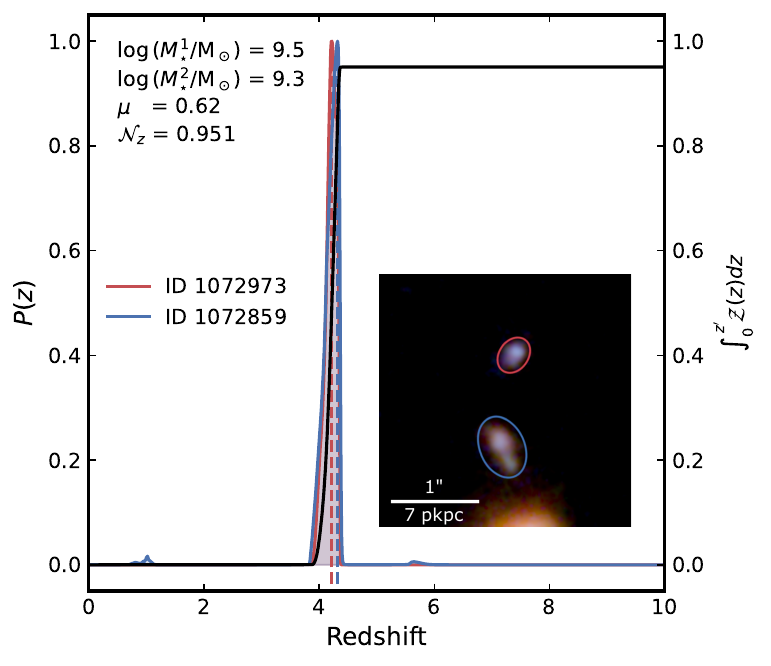}}
 \subfloat{\includegraphics[width=0.5\linewidth]{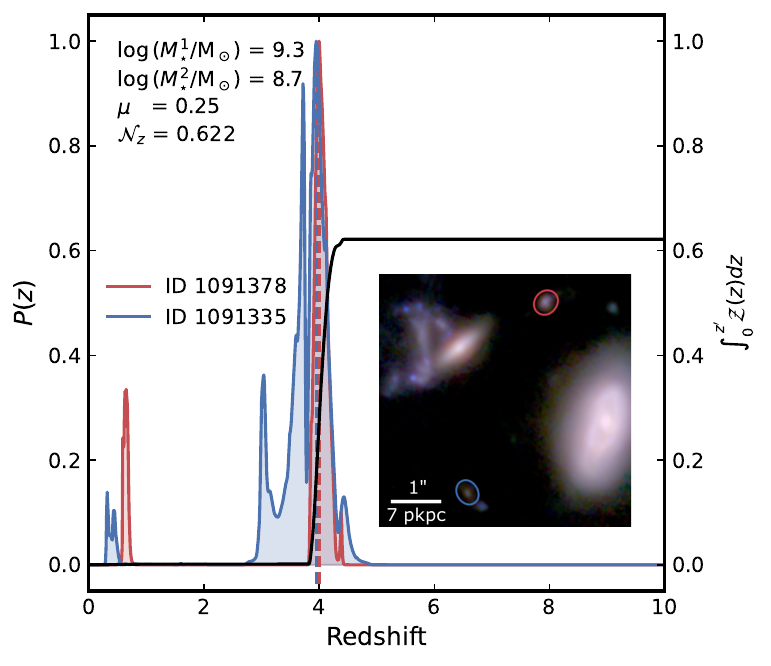}}
 \caption{\textit{Left panel:} example of a candidate close-pair system at $z_1 = 4.22^{+0.08}_{-0.13}$ (primary galaxy), where the red curve indicates the photometric redshift posterior distribution $P_1(z)$ of the primary galaxy, and the blue curve represents $P_2(z)$ for the companion galaxy. Both are well-constrained posteriors with unique narrow peaks, where the dashed vertical lines indicate the location of the peaks. The black continuous curve is the cumulative integral of the redshift probability function $\mathcal{Z}(z)$, resulting in a high number of fractional pairs $\mathcal{N}_z$ indicated on the figure, meaning that they are at the same redshift with a high probability. The two galaxies also satisfy the stellar mass ratio criterion, with values displayed in the upper left corner, where $\mu$ is the mass ratio. The total contribution of the two galaxies to the pair counts is given by $N_{\rm pair} = \int w_2(z) \times {\rm PPF}(z)dz = 1.09$, which is greater than unity due to a boosting by the correction weights. A cutout of two galaxies from the mosaic is displayed on the inset, showing that the system is within a small angular separation. \textit{Right panel:} another example of a close-pair candidate that only passed the close separation criterion by a narrow margin. Although the peak photo-$z$ values are close to each other ($z_a \approx 4.0$), the posteriors are less constrained, and the overlap between them is smaller, resulting in a lower $\mathcal{N}_z$ (lower probability of being a real close-pair). The posteriors also have secondary peaks, suggesting that the estimated photometric redshifts are less reliable.}
 \label{fig:z_pdf}
\end{figure*}

We define the redshift probability function $\mathcal{Z}(z)$ as the convolution of the two individual redshift posteriors $P_1(z)$ and $P_2(z)$ of a given projected close-pair as
\begin{equation}
 \mathcal{Z}(z) = \frac{2 \times P_1(z) \times P_2(z)}{P_1(z) + P_2(z)} = \frac{P_1(z) \times P_2(z)}{N(z)},
\end{equation}
where $N(z)$ is a normalisation defined as 
\begin{equation}
 N(z) = \frac{P_1(z) + P_2(z)}{2},
\end{equation}
which is implicitly constructed such that $\int_0^{\infty} N(z) dz = 1$. Therefore, the meaning of $\mathcal{Z}(z)$ is the number of fractional pairs for a given close-pair considered at redshift $z$. For a redshift bin, $\int_{z_{\mathrm{min}}}^{z_{\mathrm{max}}} \mathcal{Z}(z) dz$ gives the number of fractional close-pairs for a system within that redshift range considered. Therefore, the total number of fractional pairs for a given system of two galaxies can be computed as
\begin{equation}
 \mathcal{N}_z = \int_0^{\infty} \mathcal{Z}(z) dz
\end{equation}
that ranges between 0 and 1. This quantity $\mathcal{N}_z$ essentially relates to the probability of the two galaxies within the system considered being at the same redshift. See Figure~\ref{fig:z_pdf} for examples.

The real power of this analysis is that we can propagate the uncertainty associated with the photo-$z$ estimates throughout our calculations by the redshift probability function $\mathcal{Z}(z)$.

\subsubsection{Angular separation mask}

The second step is to select galaxies from the secondary sample that are in close projected physical distance to galaxies from the primary sample. We start by defining physical separations between the primary galaxy and its companion, as this can be related to the dynamics of the merger. Many previous studies defined a constant physical separation ranging between $r = 5 - 50 \/\ \mathrm{kpc} \/\ h^{-1}$ \citep[e.g.,][]{Lopez-Sanjuan15, Man16, Mundy17, Duncan19, Conselice_2022, Duan_2024}. The choice of maximum separation is crucial for the resulting pair fractions and the comparison to previous results, as we expect a higher number of companions within larger separation limits. Therefore, we choose a constant separation of $r_{\mathrm{max}} = 30 \/\ \mathrm{kpc}$ for easier comparison, where we refrain from further usage of the dimensionless Hubble parameter $h$ \citep{Croton_2013}, as stated before. In the Appendix, we also discuss the physical motivation and effects of a redshift-dependent separation limit. As for the minimum separation, we choose $r_{\mathrm{min}} = 5 \/\ \mathrm{kpc}$, which is motivated by the higher resolving power of JWST, taking care of deblending issues (i.e. single galaxy with clumps or two galaxies) even at $z \sim 9$. This lower separation criterion is further motivated by the typical sizes being ${\sim}0.5-2$ kpc for galaxies at these redshifts \citep{Allen_2024, Miller_2024, Morishita_2024, McClymont_2025b, Westcott_2025}. For a pair fraction analysis using different separation criteria, see Appendix~\ref{sec:appendix_rvir}.

We convert the projected physical separation to an observable angular separation by using the angular diameter distance. Therefore, the angular separation will be redshift-dependent according to
\begin{equation}
\begin{aligned}
 \theta_{\mathrm{max}}(z) = \frac{r_{\mathrm{max}}}{d_A(z)} && \text{and} &&
 \theta_{\mathrm{min}}(z) = \frac{r_{\mathrm{min}}}{d_A(z)},
\end{aligned}
\end{equation}
where $d_A(z)$ is the angular diameter distance that depends on redshift and the assumed cosmology. To calculate $d_A(z)$ we use the \texttt{angular\_diameter\_distance(z)} function from \texttt{astropy.cosmology}.

Finally, using the notation of \citet{Lopez-Sanjuan15}, we define an angular separation mask for each potential close-pair as
\begin{equation}
 \mathcal{M}^\theta(z) = 
 \begin{cases}
  1, & \text{if} \/\ \theta_{\mathrm{min}}(z) \leq \theta(z) \leq \theta_{\mathrm{max}}(z) \\
  0, & \text{otherwise},
 \end{cases}
\end{equation}
which is a binary mask having the same length as the $P(z)$ array from \texttt{EAZY}. For each redshift bin considered, this binary mask will have a window of allowed angular separations.

\subsubsection{Pair selection mask}

The third step is to ensure the correct stellar mass ratio for the pairs considered to find major mergers. We start by defining a mask to select the primary galaxies in a mathematical form corresponding to what is described in Section~\ref{sec:Redshift and stellar mass bins}
\begin{equation}
 \mathcal{S}(z) = 
 \begin{cases}
     1, & \text{if} \/\ M_\star^{\mathrm{lim, 1}}(z) \leq M_{\star, 1}(z) \leq M_\star^{\mathrm{max}} \\
     0, & \text{otherwise}
 \end{cases}
 \label{eq:S(z)}
\end{equation}
where $M_{\star, 1}(z)$ is the redshift-dependent stellar mass of the primary galaxy and
\begin{equation}
 M_\star^{\mathrm{lim, 1}}(z) = \mathrm{max}\{M_\star^{\mathrm{min}}, M_\star^{\mathrm{comp}}(z)\},
\end{equation}
where $M_\star^{\mathrm{min}}$ and $M_\star^{\mathrm{max}}$ are the lower and upper bounds of the stellar mass bin considered, and $M_\star^{\mathrm{comp}}(z)$ is the redshift-dependent stellar mass completeness limit calculated in Section~\ref{sec:completeness analysis}. We note here that in the computation of $M_\star^{\mathrm{lim, 1}}(z)$, we use $M_\star^{\mathrm{comp}}(z_{\mathrm{max}})$ at the upper bound of the redshift bin considered $z_{\mathrm{max}}$ since the stellar mass completeness limit determined monotonically increases with redshift. 

We further note that the stellar mass of the galaxy is expected to evolve with redshift. However, our fitting codes (\texttt{EAZY} or \texttt{Prospector}) do not output the covariance of $M_\star$ with $z$. Therefore, we simply assume that the stellar mass scales as
\begin{equation}
 M_\star(z) = M_\star^{\mathrm{peak}} \left( \frac{1 + z}{1 + z_{\mathrm{peak}}} \right)^2,
\end{equation}
where $M_\star^{\mathrm{peak}}$ corresponds to the stellar mass output from \texttt{eazy-py} or \texttt{Prospector} at the peak of the redshift posterior $z_{\mathrm{peak}}$ (more specifically at $z_a$). This approximation is based on the redshift-luminosity relation by assuming a constant mass-to-light ratio.

Now that we have ensured the correct selection function for the primary galaxy, we have to define a pair selection mask for the secondary galaxy, satisfying the major merger criterion. We define the pair selection mask as
\begin{equation}
 \mathcal{M}^{\mathrm{pair}}(z) =
 \begin{cases}
     1, & \text{if} \/\ M_\star^{\mathrm{lim, 1}}(z) \leq M_{\star, 1}(z) \leq M_\star^{\mathrm{max}} \\
        & \text{and} \/\ M_\star^{\mathrm{lim, 2}}(z) \leq M_{\star, 2}(z) \leq M_\star^{\mathrm{max}} \\
     0, & \text{otherwise},
 \end{cases}
 \label{eq:M_pair}
\end{equation}
where $M_{\star, 2}(z)$ is the redshift-dependent stellar mass of the secondary galaxy and
\begin{equation}
 M_\star^{\mathrm{lim, 2}}(z) = \mathrm{max}\{\mu M_{\star, 1}(z), M_\star^{\mathrm{comp}}(z)\},
\end{equation}
where the criterion for the mass ratio $0.25 \leq \mu \leq 1$ is imposed to search for major mergers.

\subsubsection{Pair probability function}
\label{sec:pair probability function}

We are now equipped with the mathematical forms of the three selection criteria to combine them into a single function that can be used to select close pairs in a probabilistic way, which is the strength of this analysis. We define the pair probability function (PPF) as
\begin{equation}
 \mathrm{PPF}(z) = \mathcal{Z}(z) \times \mathcal{M}^\theta(z) \times \mathcal{M}^{\mathrm{pair}}(z).
 \label{eq:ppf}
\end{equation}

The integral of this function over the full redshift range gives the probability of a system of two galaxies being in a major merger close-pair, $\int_0^{\infty} \mathrm{PPF}(z) dz$. Two examples of close-pair candidates at redshift $z \sim 4$ selected by the above method can be seen in Figure~\ref{fig:z_pdf}.


\subsection{Correction for selection effects}
\label{sec:correction for selection effects}

In this section, we address the different potential selection biases that affect the selection function we defined previously. We assign statistical weights to correct for each of these effects and include this in the final calculation to determine the close-pair fractions.

\subsubsection{Mass incompleteness}
\label{sec:mass incompleteness}

In the case of primary galaxies that are close to the stellar mass completeness limit, the mass range to look for secondary galaxies will be more strongly affected by incompleteness. Mathematically, this happens when $\mu M_{\star, 1}(z) \leq M_\star^{\mathrm{comp}} \leq M_{\star, 1}(z)$. This is due to the requirement in Equation~\ref{eq:M_pair} that the $M_{\star, 2}(z)$ has to be above the completeness limit. Therefore, there is a lower chance of finding satellites around a primary galaxy, and the results will be biased.

To account for this selection bias, we assign statistical weights to the secondary galaxies to boost their number counts, as we might potentially miss some of them due to the incomplete stellar mass search range available. Similarly, we assign a weight to the primary galaxies close to the stellar mass completeness limit to reduce their number counts and, hence, their biasing effect. To compute these weights, we compare the expected and the observable number density of galaxies by using the galaxy stellar mass function (SMF), $\phi(M_\star, z)$. We combine the most recent SMFs available from JWST observations \citep{Harvey_2024, Navarro-Carrera_2024} and extrapolate it to the highest relevant redshifts, as described in Appendix~\ref{sec:appendix_gsmf}. Each secondary galaxy gets the following completeness weight assigned
\begin{equation}
 w_2^{\mathrm{comp}}(z) = \left[ \frac{\displaystyle \int_{M_\star^{\mathrm{lim, 2}}(z)}^{M_{\star, 1}(z)} \phi(M_\star, z) dM_\star}{\displaystyle \int_{\mu M_{\star, 1}(z)}^{M_{\star, 1}(z)} \phi(M_\star, z) dM_\star} \right]^{-1},
 \label{eq:w_2_comp}
\end{equation}

which is the inverse of the ratio between the expected number density of observable galaxies (above the mass completeness limit) and the total number density of galaxies corresponding to the mass search range around the primary galaxy at a given redshift. Similarly, for each primary galaxy, we assign the weight
\begin{equation}
 w_1^{\mathrm{comp}}(z) = \frac{\displaystyle \int_{M_\star^{\mathrm{lim, 1}}(z)}^{M_\star^{\mathrm{max}}} \phi(M_\star, z) dM_\star}{\displaystyle \int_{M_\star^{\mathrm{min}}}^{M_\star^{\mathrm{max}}} \phi(M_\star, z) dM_\star},
 \label{eq:w_1_comp}
\end{equation}
which compares the expected number density of galaxies above the completeness limit to the number density for the full stellar mass bin.

As Figure~\ref{fig:w_comp} shows, these mass completeness weights can drastically increase, approaching the mass completeness limit, with values blowing up at the limit due to the ratio in their definition. To avoid over-correcting our results, we clip the values of $w_2^{\mathrm{comp}}$ at 10.

\subsubsection{Photometric redshift quality}
\label{sec:photometric redshift quality}

The next weighting is incorporated to account for the quality of the photometric redshifts. This is realised through the odds sampling rate (OSR), which was originally introduced in \citet{Lopez-Sanjuan15}. 

The odds quality parameter $\mathcal{O}$ \citep{Benitez_2000, Molino_2014} was already discussed in Section~\ref{sec:odds quality parameter}, which essentially encodes the probability of a galaxy being found within a narrow redshift interval centred on the peak of the photometric redshift posterior distribution. We also require an odds cut of $\mathcal{O} \geq 0.3$ in our pre-selection. The odds sampling rate is simply the ratio between the number of galaxies above this criterion and the total number of galaxies at magnitude $m$, defined as
\begin{equation}
    \mathrm{OSR}(m) = \frac{\displaystyle \sum N(\mathcal{O} \geq 0.3)}{\displaystyle \sum N(\mathcal{O} \geq 0)}.
\end{equation}

We calculate the OSR in magnitude bins of width $\Delta m = 0.25$ and fit a sigmoid curve to the data points to find a continuous function for $\mathrm{OSR}(m)$ at each value of $m$. For this purpose, we convert the fluxes of each galaxy measured in the F444W filter to AB magnitudes and get an $\mathrm{OSR}(m)$ curve such as in Figure~\ref{fig:osr}.

Finally, to compute the photometric redshift quality weight corresponding to each galaxy, we take the inverse of the OSR value as defined below
\begin{equation}
    w^{\mathrm{OSR}} = \frac{1}{\mathrm{OSR}(m)},
\end{equation}
where $m$ is the AB magnitude of the galaxy in the F444W filter.

\subsubsection{Survey boundaries}
\label{sec:survey boundaries}

Finally, we assign weights to correct for incomplete survey areas. Primary galaxies that lie close to the survey boundaries or masked regions (where artefacts were masked out during the data reduction) will have a reduced area around them to look for potential secondary galaxies. This effect does not significantly alter the final results, nevertheless, we correct for it.

Furthermore, we note here that additional masks were created for bright stars and their diffraction spikes. During the initial runs of the merger selection code, it was found that a large number of artificial sources along the diffraction spikes of bright stars (due to an erroneous `shredded' source identification by the automatic source extraction) significantly boosted the close-pair fraction at certain redshifts. Therefore, we generously masked these artefacts, which further reduced the size of the search area and altered its shape.

We calculate the area weight by performing photometry on the annulus surrounding each primary galaxy at inner and outer radii corresponding to $r_\mathrm{min}$ and $r_\mathrm{max}$. In practice, since $\theta_\mathrm{min}(z)$ and $\theta_\mathrm{max}(z)$ are redshift-dependent through the angular diameter distance $d_\mathrm{A}(z)$, the fraction of available area, $f_\mathrm{area}(z)$ will be also redshift-dependent and has to be calculated at each redshift-step. We use the \texttt{photutils} python package \citep{Bradley_2024} to compute the sum over the area within a fixed annulus around a source (see for an example Figure~\ref{fig:area}), containing pixels with non-zero values and the sum over the same area including pixels with zero flux which is equivalent to the masked area. The ratio of these two values gives the area fraction $f_\mathrm{area}(z)$. To correct for the unavailable area and potentially missed secondary galaxies, we assign an area weight to each primary galaxy within our sample
\begin{equation}
    w^{\mathrm{area}}(z) = \frac{1}{f_\mathrm{area}(z)}.
\end{equation}

In practice, we find that the overall effect of these weights is small, on average being $1-2\%$ \citep[see also e.g.,][]{Duncan19}. Therefore, to make our merger selection code more efficient, we only evaluate these area weights at the best available redshift (see Section~\ref{sec:Spectroscopic redshift catalogue}) of each primary source. Hence, we assume that $w^{\mathrm{area}}(z) = w^{\mathrm{area}}$ for the rest of the analysis.

\begin{figure*}
 \includegraphics[width=\linewidth]{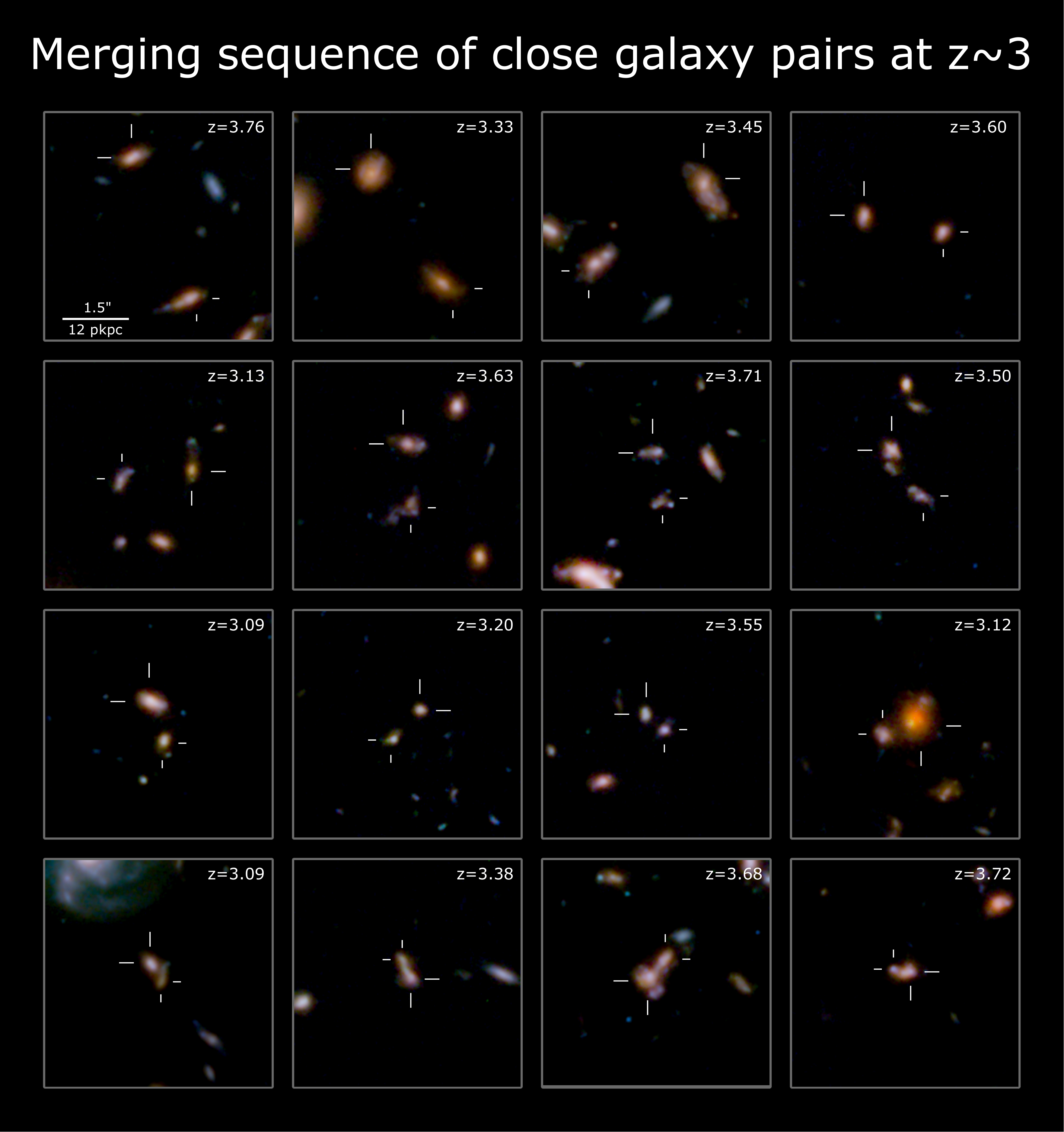}
 \caption{Close galaxy pairs with a high probability of being actual mergers ($\mathcal{N}_z > 0.8$) at $3 < z < 4$. The RGB cutouts (with blue, green, and red assigned to the F090W, F200W, and F444W bands, respectively) are ordered by decreasing projected separation between the close pairs (from the top-left to the bottom-right) in an attempt to build a Toomre sequence \citep{Toomre72} of merging galaxies at $z >3$. The primary galaxies are indicated by the longer half-crosshairs, and the secondary galaxies by the shorter crosshairs. The redshifts of the primary galaxies are shown on the top-right corner of each cutout (with the redshift of the secondary galaxies being at the same or similar values). The angular and proper physical scale (calculated at $z=3.5$) is displayed on the top-left cutout, and all the other cutouts have the same scale (with some minor differences in the proper physical scale due to the deviation from $z=3.5$). The order of these close pairs in the Toomre sequence is both determined by the projected separation and by visual inspection, where one might see tidal features and bridges forming on the bottom row of the cutouts.}
 \label{fig:pairs}
\end{figure*}

\begin{figure*}
 \includegraphics[width=\linewidth]{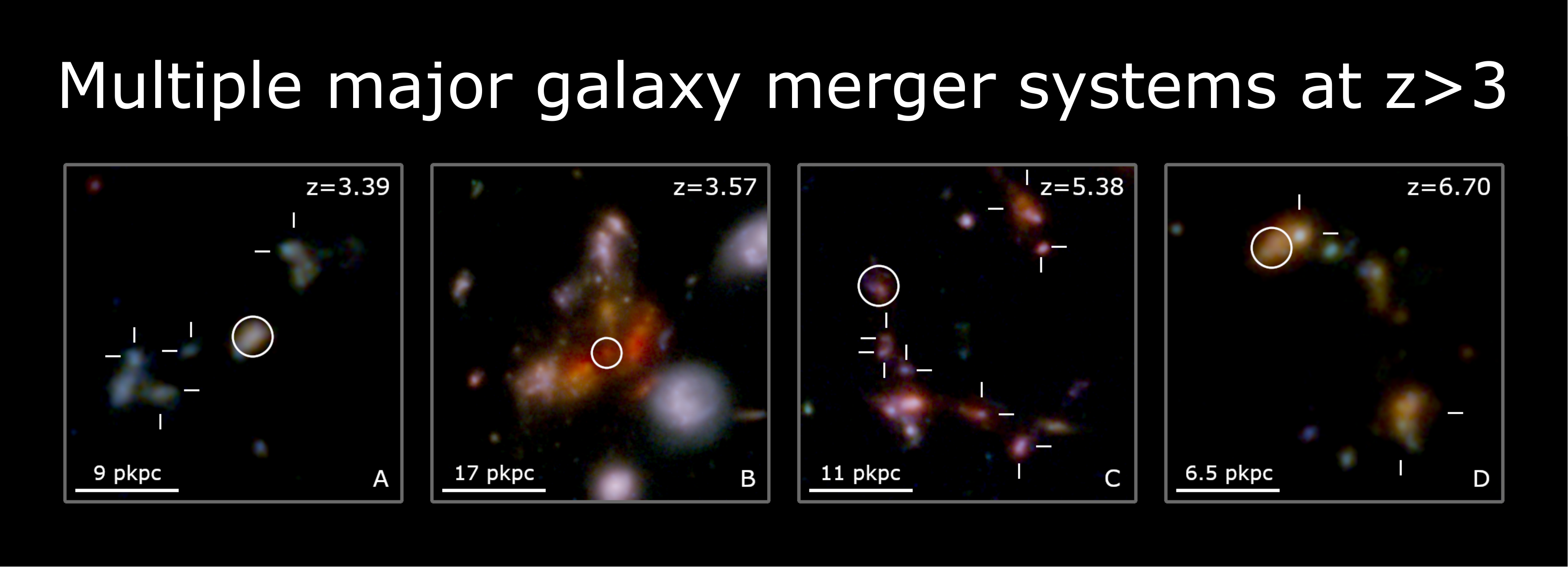}
 \caption{Multi-merger systems at $z>3$ with a high probability ($\mathcal{N}_z > 0.75$), showing an abundant variation of morphologies and multiplicities, reflecting the complexity of these major mergers at high redshift and the difficulty of finding them. The primary galaxy (the most massive galaxy considering all pairs in the system) is indicated by a white circle on each RGB cutout, with half-crosshairs pointing to the secondary galaxies. The redshifts of the primaries are displayed in the top-right corner, as well as the proper physical scale at that redshift for each cutout. \textit{Systems A, C,} and \textit{D} have multiple satellites around each primary galaxy. However, they likely have even more companions but with a lower probability than our selection threshold (and hence not displayed) due to less reliable photometric redshifts that spectroscopic observations could confirm in the future. \textit{System C} falls within the FRESCO footprint, and therefore, all candidates have spectroscopic redshifts from grism observations. The bright galaxy at the lower-left of the cutout is likely to be also a member, but it was not selected by the code as having a different spec-$z$ from higher priority MUSE Wide observations (see Section~\ref{sec:Spectroscopic redshift catalogue} for the spec-$z$ classification). \textit{System B} is a peculiar case where we only display the primary galaxy, as the companions identified by the close-pair selection code are likely clumps associated with the same structure. This extended clumpy morphology causes deblending to fail, which in turn translates to likely false close-pairs. These few peculiar systems (likely only ${\sim}3\%$ of the sample) would require further morphological analysis for confirmation, which is beyond the scope of this paper. 
 }
 \label{fig:multi_merger}
\end{figure*}

\subsubsection{Combined weightings}

Ultimately, we combine all the weightings mentioned above into single weights assigned to each primary and secondary galaxy. We then include these weights, accounting for different selection biases, in integrating the pair probability functions to calculate close-pair fractions.

In the case of each primary galaxy, two of the aforementioned weights are combined, and the total weight is given by
\begin{equation}
    w_{1}(z) = w_1^{\mathrm{comp}}(z) \times w_1^{\mathrm{OSR}}.
    \label{eq:w_1}
\end{equation}
For each secondary galaxy around a primary galaxy, we need to convolve the total weight of the primary with the weights corresponding to the secondary (as its selection is dependent on the former), getting a total combined pair weight of
\begin{equation}
    w_{2}(z) = w_1^\mathrm{area} \times w_1^{\mathrm{comp}}(z) \times w_2^{\mathrm{comp}}(z) \times w_1^{\mathrm{OSR}} \times w_2^{\mathrm{OSR}},
    \label{eq:w_2}
\end{equation}
where we include $w_1^\mathrm{area}$ only for the secondary galaxy as it accounts for potentially missing satellites around a primary source, with the subscript `1' further stressing this point.

We further note that the redshift-dependence of $w_{1}(z)$ and $w_{2}(z)$ arises due to the redshift-dependence of the mass completeness weights, which further depends on the previously calculated mass completeness limit. We calculate this on the same redshift grid as the posterior output of \texttt{EAZY} for consistency. The most significant contribution usually comes from $w_2^{\mathrm{comp}}(z)$. Hence, a clipping is necessary to avoid over-boosting with this weight, as described previously.

These weights are combined with the pair probability function for each potential pair, allowing us to measure volume-limited pair fractions.


\subsection{Final integrated pair fractions}

Finally, using the results from all previous steps, the total integrated pair fractions can be calculated as follows. For each primary galaxy \textit{i} from the initial sample, the number of associated close-pairs in the redshift range $z_\mathrm{min} \leq z \leq z_\mathrm{max}$ (within each redshift bin) can be computed by
\begin{equation}
    N_\mathrm{pair}^i = \displaystyle \sum_j  \int_{z_\mathrm{min}}^{z_\mathrm{max}} w_{2}^j(z) \times \mathrm{PPF}_{ij}(z) dz,
\end{equation}
where \textit{j} are the indices of each potential secondary galaxy around the primary that satisfies the pre-selection criteria, $\mathrm{PPF}_{ij}(z)$ are the pair probability functions for each pair considered given by Equation~\ref{eq:ppf} (see Section~\ref{sec:pair probability function}), and $w_{2}^j(z)$ are the pair weights associated with each secondary galaxy given by Equation~\ref{eq:w_2}. 

Similarly, the contribution to the number of primary galaxies for each system \textit{i} within a $z$-bin corresponding to $N_\mathrm{pair}^i$, is given by\footnote{We note that many previous publications include a $\sum_i$ in front of this equation. We consider this to be a typo as it leads to mathematically erroneous results—in the final form of the pair fraction, there would be a double summation over \textit{i} for $N_\mathrm{1}^i$.}
\begin{equation}
    N_\mathrm{1}^i = \displaystyle  \int_{z_\mathrm{min}}^{z_\mathrm{max}} w_{1}^i(z) \times \mathrm{P}_i(z) \times \mathcal{S}_\mathrm{1}(z) dz,
    \label{eq:N_1_i}
\end{equation}
where $\mathrm{P}_i(z)$ is the normalised redshift posterior distribution function of the primary galaxy (see Section~\ref{sec:photometric redshifts}), $w_{1}^i(z)$ is the corresponding weight given by Equation~\ref{eq:w_1}, and $\mathcal{S}_\mathrm{1}(z)$ is a selection function for the primary galaxy sample. This selection function incorporates the initial selection criteria described in Section~\ref{sec:initial sample selection}, and its mathematical form is given by Equation~\ref{eq:S(z)} (note that we added an extra `1' index to emphasise that it is the primary galaxy selection function).

Finally, we have derived all the necessary quantities to redefine and calculate the galaxy close-pair fraction (given by Equation~\ref{eq:f_pair}). The pair fraction for a redshift-bin of $z_\mathrm{min} \leq z \leq z_\mathrm{max}$, using the probabilistic methodology described above, has the following form
\begin{equation}
    f_\mathrm{pair} = \frac{\sum_i N_\mathrm{pair}^i}{\sum_i N_\mathrm{1}^i},
\end{equation}
and for the sake of completeness, the full form of the pair fraction is given by
\begin{equation}
    f_\mathrm{pair} = \frac{\sum_i \sum_j \int_{z_\mathrm{min}}^{z_\mathrm{max}} w_{2}^j(z) \times \mathrm{PPF}_{ij}(z) dz}{\sum_i \int_{z_\mathrm{min}}^{z_\mathrm{max}} w_{1}^i(z) \times \mathrm{P}_i(z) \times \mathcal{S}_\mathrm{1}(z) dz}.
\end{equation}

In the case of different subfields (i.e., GS-Medium, GS-Deep, GN-Medium, GN-Deep) the total pair fraction can be calculated by
\begin{equation}
    f_\mathrm{pair} = \frac{\sum_k \sum_i N_\mathrm{pair}^{i, k}}{\sum_k \sum_i N_\mathrm{1}^{i, k}},
    \label{eq:f_pair_subfields}
\end{equation}
where the index \textit{k} corresponds to the subfield, and the stellar mass completeness limit $M_\star^{\mathrm{comp}}(z)$ is calculated separately for each subfield.


\subsection{Bootstrapping analysis}

We estimate the uncertainties on the $f_\mathrm{pair}$ values using the common bootstrapping technique \citep{Efron_1979, Efron_1981}. The merger selection code estimates the pair fractions and their errors for each stellar mass–redshift bin by randomly drawing a primary sample of galaxies with replacement from the initial sample and performing for $N$ independent realisations. We set $N=100$ for most of our analysis as we find that the uncertainties relatively quickly converge after a few resamples.

The standard error on the pair fraction in each bin from the bootstrapping analysis is given by
\begin{equation}
    \sigma_{f_\mathrm{pair}} = \sqrt{\frac{\sum_i (f_\mathrm{pair}^i - \langle f_\mathrm{pair} \rangle)^2}{N - 1}},
\end{equation}
where $\langle f_\mathrm{pair} \rangle$ is the average pair fraction of the $N$ independent realisations given by
\begin{equation}
    \langle f_\mathrm{pair} \rangle = \frac{\sum_i f_\mathrm{pair}^i}{N}.
\end{equation}

All the above definitions and calculations are incorporated in the merger selection \texttt{python} code. We measure the pair fraction in this probabilistic approach for a large sample of ${\sim} 110,000$ galaxies (size of the original photometric catalogue before sample cleaning) for both GOODS-South and GOODS-North. For a few examples of close-pairs found by the code, see Figure~\ref{fig:pairs}, which shows close-pairs with a high probability of being actual mergers ordered in a Toomre sequence, and Figure~\ref{fig:multi_merger} for examples of peculiar multi-merger systems. Figure~\ref{fig:multi_merger} \textit{System B} is an extended clumpy object where the failed deblending results in identifying a spurious close-pair system with a high multiplicity. A visual inspection of 500 randomly selected close-pairs from our final catalogue shows that approximately ${\sim}3\%$ of the sample may be affected by deblending issues. This percentage is likely even lower regarding the number of galaxies affected since these pairs often have a high multiplicity. However, these problems mainly occur at lower redshifts in extended objects, where the number counts are sufficiently high to avoid significant boosting by these spurious pairs.

\section{Pair fraction}
\label{sec:pair fraction}

In this section, we discuss the results of the probabilistic close-pair fraction analysis. We present our fiducial results for the previously defined data set and selection criteria and compare them to the existing literature. We also investigate the effect on the resulting pair fractions using different data sets and varying some parameters in the selection criteria. In the next section, we convert the pair fractions to the merger rate by assuming a merger timescale.

\subsection{Major merger pair fraction evolution with redshift}

In our fiducial analysis, we use stellar masses obtained from \texttt{Prospector} and a redshift-independent separation criterion. See Appendix~\ref{sec:appendix_eazy} for an alternative analysis using stellar masses from \texttt{eazy-py}, and Appendix~\ref{sec:appendix_rvir} for a redshift-dependent separation selection criterion.

\begin{figure}
 \includegraphics[width=\columnwidth]{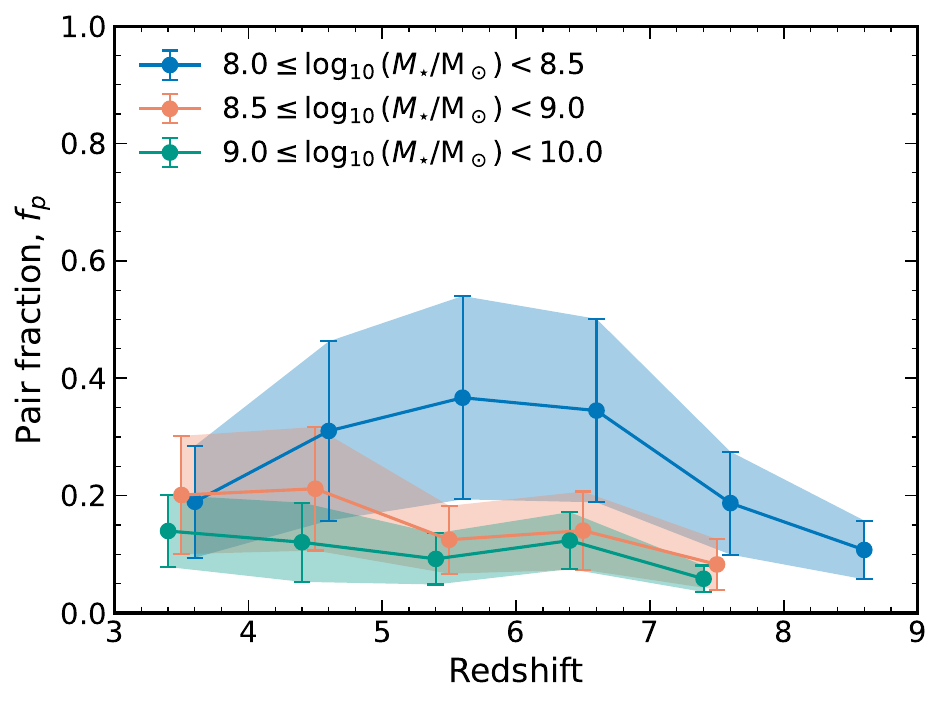}
 \caption{Redshift-evolution of the major galaxy close-pair fractions for three different stellar mass bins. Uncertainties are obtained by bootstrapping with 100 resamples. The lowest stellar mass bin shows the most clear evolution by increasing up to $z \sim 6$ and then turning over at higher redshifts. Pair fractions at higher stellar mass bins show little to slightly decreasing evolution with redshift overall. Each data point represents the pair fraction in a given redshift and stellar mass bin for the overall sample, including all subfields (e.g., GS and GN, deep and medium tier).}
 \label{fig:pair_fraction_summary}
\end{figure}

\begin{figure*}
 \centering
 \includegraphics[width=\linewidth]{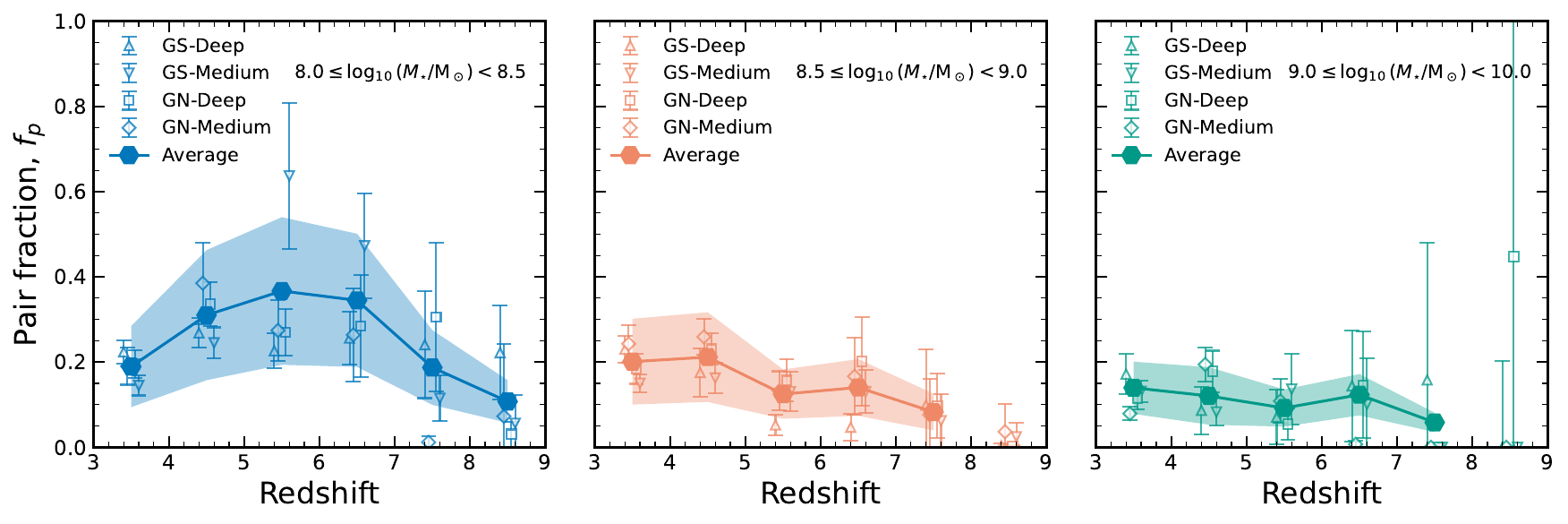}
 \caption{Redshift-evolution of the close-pair fractions for three different stellar mass bins, represented by the three subplots. Each subplot displays the field-to-field variation within each bin for the four subfields (e.g., GS and GN, deep and medium tier), with the lowest stellar mass bin showing the most variations. Overall, cosmic variance does not seem to significantly affect the resulting pair fractions. The averages here represent the overall pair fractions as calculated from the union of the four subfields (Equation~\ref{eq:f_pair_subfields}) instead of just calculating the mean. For the two highest stellar mass bins we do not show the averages at the highest redshift bin at $z \sim 8-9$ as the size of the initial sample is not sufficiently large to allow us measuring pair fractions robustly.}
 \label{fig:pair_fraction_bins}
\end{figure*}

Pair fractions are calculated in the redshift range $z=3-9$ and stellar mass range ${\rm log}(M_\star / {\rm M_\odot}) = 8.0-10.0$ and physical projected separation between $r = 5-30$ kpc, as defined in Section~\ref{sec:Redshift and stellar mass bins}. Results are plotted in Figure~\ref{fig:pair_fraction_summary} for the three stellar mass bins, and values can be found in Table~\ref{tab:pair fraction and merger rate}. Pair fractions in the lowest stellar mass bin ($8.0 \leq {\rm log}(M_\star / {\rm M_\odot}) < 8.5$) show the most significant evolution. There is an increase up to $z\sim6$, where 37\% of the galaxies have major companions, and it subsequently turns over to a declining trend. Pair fractions in the higher stellar mass bins show little evolution with redshift, with potentially a slight decrease. We note that data points represent the total pair fractions in a given bin for the overall sample, which is obtained by Equation~\ref{eq:f_pair_subfields} and not by averaging the values obtained by each subfield. Bins at high redshifts (above $z \sim 7-8$) can become unreliable as the number counts of primary (massive) galaxies become very low and are prone to variations, especially at the higher stellar mass bins (see Table~\ref{tab:pair_fraction_merger_rate}). Therefore, we decided to discard the data points at the highest redshifts at $z \sim 8-9$ for the ${\rm log}(M_\star / {\rm M_\odot}) = 8.5-9.0$ and ${\rm log}(M_\star / {\rm M_\odot}) = 9.0-10.0$ stellar mass bins as it can be seen in Figure~\ref{fig:pair_fraction_summary} (see Figure~\ref{fig:pair_fraction_bins} for measurements in the individual tiers for these bins).

We also look at the field-to-field variation of the pair fractions in each stellar mass bin, which can be seen in Figure~\ref{fig:pair_fraction_bins}. The most significant variation can be found in the ${\rm log}(M_\star / {\rm M_\odot}) = 8.0-8.5$ stellar mass bin, with the GS-Medium results differing from the other three the most. We note that, for the ${\rm log}(M_\star / {\rm M_\odot}) = 8.0-8.5$ bin, the GS-Medium results drive the peak of $f_{\rm P}$ to be at $z \sim 5-6$, and without GS-Medium the peak becomes less evident. This higher close-pair count could be driven by the presence of multiple overdensities at $z \sim 5-6$, discovered by \citet{Helton_2024, Helton_2024b}. Since this lowest stellar mass bin is the closest to the mass-completeness limit, we expect that it is still somewhat affected by incompleteness, even after using the appropriate correction weights, defined in Section~\ref{sec:mass incompleteness}. While the statistical uncertainties estimated by bootstrapping do not reflect this, there could still be a systematic uncertainty affecting our completeness estimates due to a higher time-variability of star formation in low-mass and/or high-redshift galaxies \citep[see e.g.,][]{Sun_2023}. In our stellar mass completeness estimates, based on the method of \citet{Pozetti_2010}, we implicitly assume the same colours for the undetected galaxy population below the completeness limit, which might be strongly affected by the higher burstiness in star formation of low-mass galaxies at high-z. Furthermore, based on these plots, we conclude that our samples in the different subfields are not significantly affected by cosmic variance.

\subsection{Comparison to literature}

In this section, we compare our results for the pair fractions with those of the existing literature. This is a difficult task as most other works use different stellar mass bins or separation limits or both (e.g., see Table~\ref{tab:merger_studies}). Nevertheless, we attempt to find results from the literature whose selection criteria are the closest to our choices and compare them to our results. In the Appendix, we discuss multiple variations of selection criteria for the fiducial analysis and their effects on the results.

For comparing results from the existing literature to our findings, we attempt to keep the distinction between the three stellar mass ranges. While this categorisation is not always perfectly clear and well defined in the source papers, we nevertheless try to associate data points with one of the three stellar mass bins (e.g., if a data point is slightly above ${\rm log_{10}}(M_\star / {\rm M_\odot}) = 10$, we will include it in the $9 \leq {\rm log_{10}}(M_\star  / {\rm M_\odot}) < 10$ bin). 

\begin{figure*}
 \includegraphics[width=\linewidth]{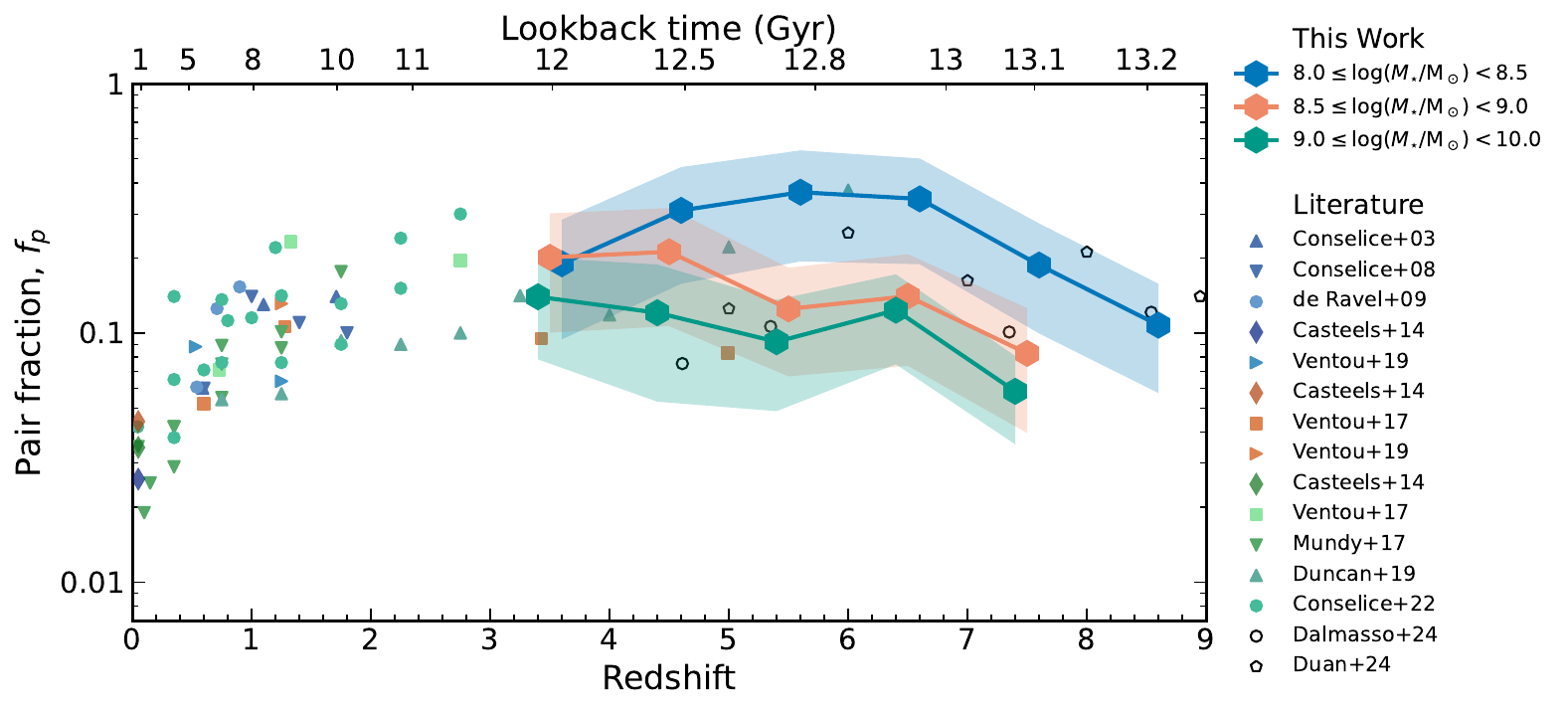}
 \caption{The evolution of major galaxy pair fractions measured in this work and their associated uncertainty bands, colour-coded by stellar mass. Data points from the existing literature \citep{Conselice_2003, Conselice_2008, deRavel_2009, Casteels_2014, Mundy17, Ventou_2017, Duncan19, Ventou_2019, Conselice_2022, Dalmasso_2024, Duan_2024} are plotted with uncertainties omitted to avoid overcrowding the figure, and their specific details are described in the text. Although most of these works use different selection criteria and stellar mass ranges, we attempted to select the closest, most relevant and colour-coded so that they approximately match our stellar mass bins. We find that the pair fractions weakly scale with stellar mass and that pair fractions at higher stellar masses peak at later epochs (i.e. at lower redshifts). We also observe a slight turnover in the pair fraction trends at the highest redshifts.
 }
 \label{fig:pair_fraction_literature}
\end{figure*}

The final results for pair fractions and their comparison to values from the literature can be found in Figure~\ref{fig:pair_fraction_literature}. The values associated with the three stellar mass bins are colour-coded, and uncertainty bands for results from this work are plotted but omitted in the case of literature results to avoid overcrowding the figure. The blue upward-pointing triangles represent data from \citet{Conselice_2003} for morphologically selected $M_\star > 10^8~{\rm M_\odot}$ major mergers at $z \leq 3$, while the blue downward-pointing triangles data from \citet{Conselice_2008} for galaxies selected by a similar method in the $10^8-10^9~{\rm M_\odot}$ range. The blue circles represent results from a spectroscopic close-pair study by \citet{deRavel_2009} for a sample of $M_B \leq -18$ galaxies at $z<1$, that approximately corresponds to $>10^8~{\rm M_\odot}$. The blue, orange, and green diamonds correspond to measurements from \citet{Casteels_2014} that match our stellar mass bins for $z < 0.2$ galaxies, selected by a morphological method. Orange and green squares represent measurements from deep MUSE observations by \citet{Ventou_2017} extending up to $z \approx 6$. The green downward-pointing triangles show results from \citet{Mundy17} for $M_\star > 10^{10}~{\rm M_\odot}$ galaxies with $r=5-30$ kpc separations. The blue and orange rightward-pointing triangles show results from \citet{Ventou_2019} of the MUSE deep fields that approximately match our stellar mass bins. The green upward-pointing triangles represent CANDELS data from \citet{Duncan19} for $10^{9.7} < M_\star / {\rm M_\odot} < 10^{10.3}$, extending up to $z \approx 6$. The green circles show results from \citet{Conselice_2022} for many different datasets for $M_\star > 10^{11}~{\rm M_\odot}$. The open black circles show recent results from \citet{Duan_2024} using JWST data for the same overall mass range as this work (hence no colour coding) and similarly for results from \citet{Dalmasso_2024} that use morphological parameters to select close-pairs.

We fit a power law + exponential function to all the data points from both our measurements and the literature, which is theoretically motivated by the Press–Schechter formalism for merging galaxies \citep{Carlberg_1990, Conselice_2008} and seems to better describe the pair fraction evolution in the literature. This function has the form
\begin{equation}
    f_\mathrm{P}(z, M_\star) = f_0 \times (1+z)^m \times e^{\tau(1+z)},
    \label{eq:pair fraction fit}
\end{equation}
where $f_0$, $m$, and $\tau$ are parameters for which we perform weighted non-linear least squares fitting with uncertainties incorporated as weights. The fitted parameters and their uncertainties can be found in Table~\ref{tab:fit parameters}. We find that the fit is worse in the case of the individual stellar mass bins, and therefore, we omit these from Table~\ref{tab:fit parameters} and do not plot them in Figure~\ref{fig:pair_fraction_literature}.

There is a clear difference between the pair fraction evolution with redshift in the case of the three different stellar mass bins. Based on the measured values (displayed in Figure~\ref{fig:pair_fraction_literature}), the peak of the pair fractions is at lower redshifts for increasing stellar masses of the primary galaxies. The pair fraction trend in the lowest stellar mass bin has the steepest evolution and is dominating at higher redshifts. The opposite can be said about pair fractions in the highest stellar mass bin. There is also a slight turnover in the trends of pair fraction evolution in all three cases.

\section{Major merger rate}
\label{sec:major merger rate}

While the galaxy close-pair fraction is a useful quantity to identify mergers, it is a purely observational one that depends on the exact methodology chosen, such as the selection criteria definitions and the stellar mass range being probed. Therefore, it is hard to compare directly to other studies that use slightly different selection parameters. Conversely, the merger rate is a physically more meaningful quantity that is universal (cf. SFR) and can be directly compared with results from different methodologies (such as derived from close-pair fractions or morphological identification). Converting the observed close-pair fractions to merger rates requires the assumption of a merger observability timescale, which is discussed in the next section.

We define two merger rate quantities that are fundamental to the study of galaxies via mergers. The first is the merger rate per galaxy ($\mathcal{R}_\mathrm{M}$), which traces the number of mergers occurring per unit time per massive galaxy. The second is the merger rate density ($\Gamma_\mathrm{M}$), which is the overall merger rate per comoving volume in ${\rm cMpc^3}$.

\subsection{Merger observability timescale}
\label{sec:merger observability timescale}

Converting the inferred pair fractions to merger rates requires the assumption of a merger observability timescale that corresponds to the duration when the merger can be selected as a close pair.

Typically, the merger rate is defined in the following form
\begin{equation}
    \mathcal{R}_\mathrm{M}(z, M_\star) = \frac{f_\mathrm{merg}(z, M_\star)}{\tau_\mathrm{merg}(z)} = \frac{f_\mathrm{P}(z, M_\star) \times C_\mathrm{merg}}{\tau_\mathrm{merg}(z)},
    \label{eq:r_m_c}
\end{equation}
where $f_\mathrm{merg}$ is the \textit{merger fraction}, i.e. the fraction of galaxies within a sample that will actually merge, and $\tau_\mathrm{merg}$ is the merger timescale (dynamical time) that is redshift dependent. Here, we make a distinction between the \textit{pair fraction} and \textit{merger fraction}, where the former is the fraction of observed close pairs, and the latter is the fraction of galaxies that will actually merge. The conversion factor between the quantities is denoted by $C_\mathrm{merg}$, which is necessary as two nearby galaxies will only have some probability to eventually merge over some timescale and might orbit each other for a significantly longer time \citep[e.g., due to a long dynamical friction timescale;][]{Duncan19}. A typical value of $C_\mathrm{merg} = 0.6$ is assumed, which is derived from simulations for all possible merging scenarios \citep{Lotz_2011, Conselice_2014, Mundy17}. However, this value is approximate and other sources place it in the range $C_\mathrm{merg} \sim 0.4-1.0$.

In this work, we define the merger rate as
\begin{equation}
    \mathcal{R}_\mathrm{M}(z, M_\star) = \frac{f_\mathrm{P}(z, M_\star)}{\tau_\mathrm{P}(z)}~({\rm Gyr^{-1}}),
    \label{eq:r_m}
\end{equation}
where $f_\mathrm{P}$ is the pair fraction that we measured before, and $\tau_\mathrm{P}$ (or $\tau_\mathrm{obs}$) is the \textit{merger observability timescale}, which factors in the effects associated with $C_\mathrm{merg}$ in Equation~\ref{eq:r_m_c}. The observability timescale is the duration a merging pair can be identified in a galaxy catalogue.

The derived merger rate is sensitive to the right choice of the observability timescale (hereafter just merger timescale or timescale for shortness) according to Equation~\ref{eq:r_m}. Various previous studies established that this timescale is roughly constant with time at lower redshifts (typically $z < 3$) ranging between $\tau_\mathrm{P}(z) \sim 0.3 - 1.0$ Gyr \citep{Conselice_2006, Kitzbichler_2008, Lotz_2008b, Lotz_2010}. However, more recent studies using the IllustrisTNG simulations \citep{Genel_2014, Vogelsberger_2014} at higher redshifts discovered that the merger timescale does have a redshift dependence and evolves as a power law \citep{Snyder17, Duncan19}
\begin{equation}
    \tau_\mathrm{P}(z) = 2.4 \times (1+z)^{-2}~{\rm Gyr}.
    \label{eq:t_p}
\end{equation}

A further refined version of the equation above by \citet{Conselice_2022} based on \texttt{TNG300-1} is also dependent on the mass ratio such that
\begin{align}
    \tau_\mathrm{P}(z) = a \times (1+z)^b \notag \quad \text{with} \quad
    a &= -0.65 \pm 0.08 \times \mu + 2.06 \pm 0.01 \\
    b &= -1.60 \pm 0.01,
    \label{eq:t_p_conselice}
\end{align}

\begin{figure}
 \includegraphics[width=\columnwidth]{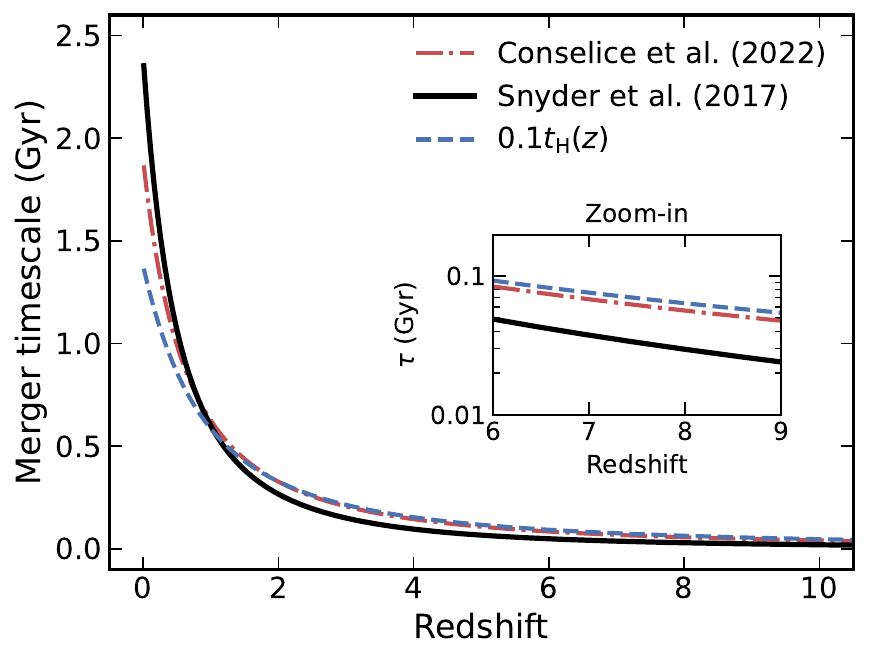}
 \caption{Comparison between the merger observability timescale derived by \citet{Snyder17} that is redshift-dependent, and the timescale of \citet{Conselice_2022} that is also mass-ratio dependent, and 10\% of the Hubble time. They agree well, especially at higher redshifts, with the lowest being the timescale by \citet{Snyder17}, which we choose as our fiducial timescale for deriving the merger rates.}
 \label{fig:timescale}
\end{figure}

We note these quantities do not differ greatly from the typical halo dynamical timescale, $0.1t_\mathrm{H}(z)$, where $t_\mathrm{H}(z)$ is Hubble-time (i.e. age of the Universe) at redshift $z$, as can be seen in Figure~\ref{fig:timescale}. For our fiducial merger rate analysis, we advocate for the timescale derived by \citet{Snyder17}, similarly to \citet{Duncan19}.

\subsection{Merger rate per galaxy}

Using all definitions and results from above, we are now able to calculate the galaxy major merger rate, given by Equation~\ref{eq:r_m}. This quantity essentially measures the average number of mergers per massive (primary) galaxy. We note here that this is equivalent to setting $C_\mathrm{merg}=1$ in Equation~\ref{eq:r_m_c} and just using the appropriate merger timescale as in \citet{Duncan19, Conselice_2022, Duan_2024}.

\begin{figure}
 \includegraphics[width=\columnwidth]{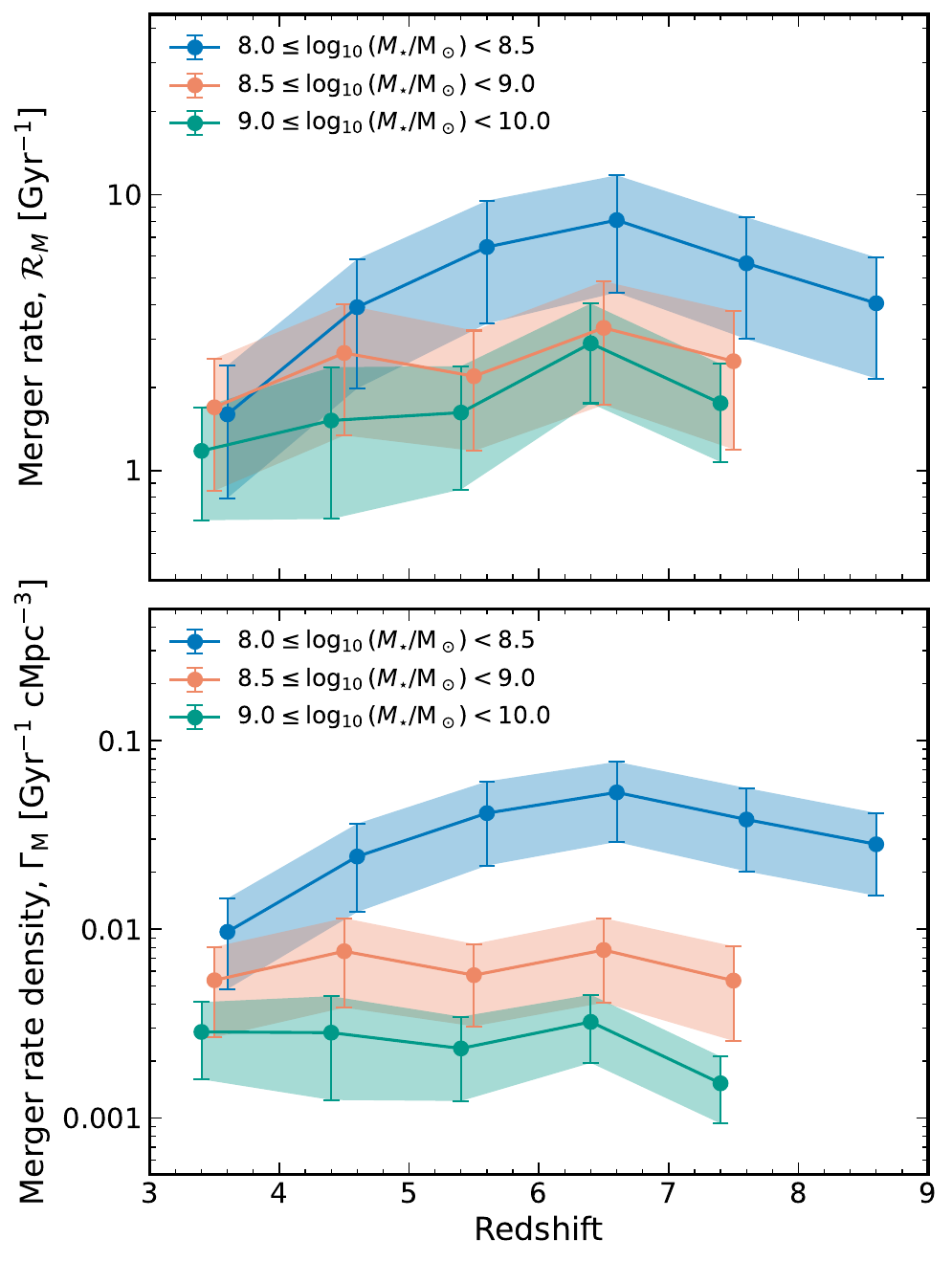}
 \caption{Evolution of the galaxy major merger rate (\textit{top panel}) and merger rate density (\textit{bottom panel}) with redshift for the three stellar mass bins between $z \approx 3-9$. Merger rates steeply increase at intermediate redshifts ($z \approx 3-6$), after which they flatten and become constant at high redshifts ($z \approx 6-9$). At lower stellar masses, potentially there could be a turnover beyond $z \approx 7$. The same can be observed in the case of the redshift evolution of the merger rate densities, which show a stronger scaling with stellar mass.}
 \label{fig:merger_rate}
\end{figure}

As the merger timescale decreases with redshift, the merger rate seen in Figure~\ref{fig:merger_rate} increases at higher redshifts even though pair fractions are declining or constant in that range. Our results show that merger rates at different stellar mass bins steeply increase with redshift at intermediate redshifts ($z \sim 3-6$) and then become constant at high redshifts ($z \sim 6-9$) in the range of $2-8~{\rm Gyr^{-1}}$ per galaxy. At the lowest stellar masses, it seems that there might be a turnover at $z \sim 6-7$. However, this might be affected by the uncertainties in the stellar mass correction weights, especially at high redshifts and low stellar masses, or potentially due to the lower limit (i.e. 5 kpc) adopted for the distance to consider a close pair.

\begin{figure*}
 \includegraphics[width=\linewidth]{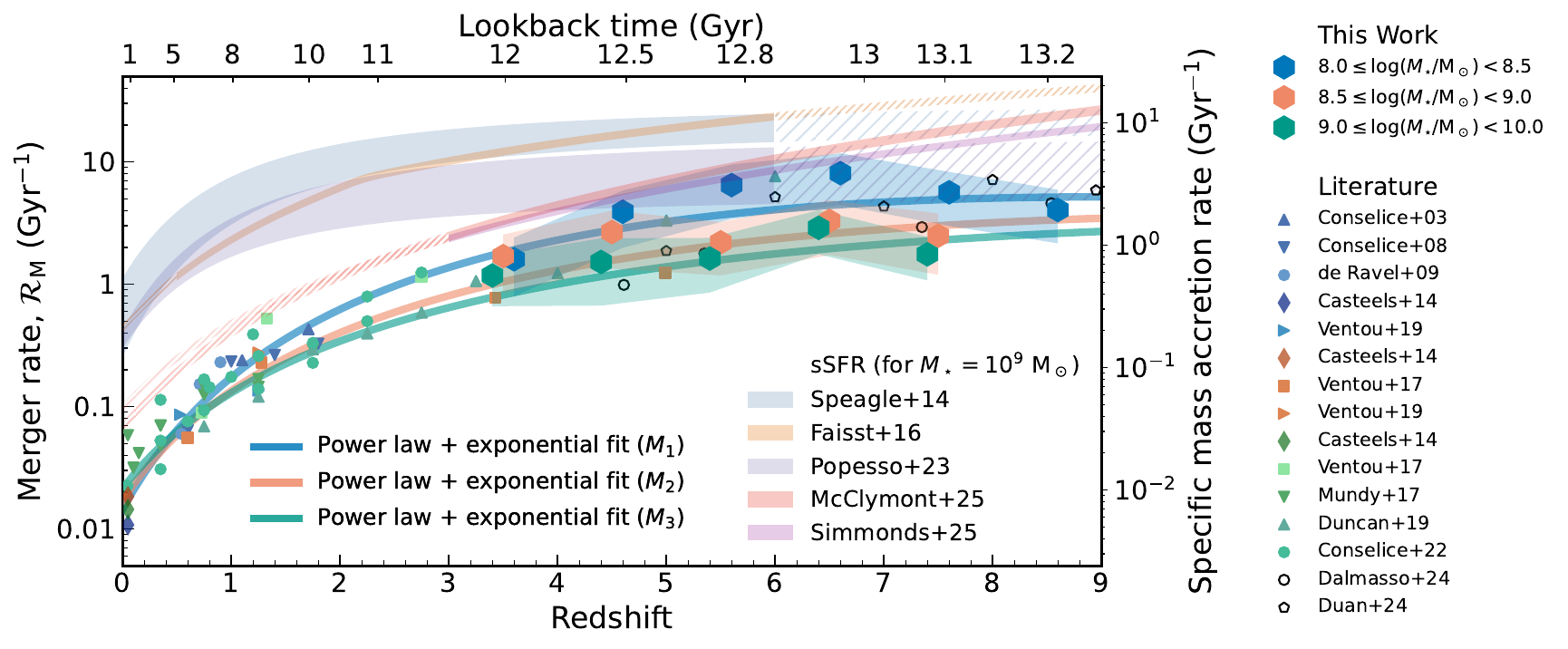}
 \caption{Evolution of the galaxy major merger rate with redshift for the three stellar mass bins between $z \approx 3-9$. Merger rates increase at intermediate redshifts ($z \approx 3-6$), after which they flatten and become constant at high redshifts ($z \approx 6-9$). By fitting a power law + exponential curve to our measured merger rates and the ones available from the literature calculated at different stellar mass bins \citep{Conselice_2003, Conselice_2008, deRavel_2009, Casteels_2014, Mundy17, Ventou_2017, Duncan19, Ventou_2019, Conselice_2022, Dalmasso_2024, Duan_2024}, it becomes evident that the evolutionary trends show a weak mass dependence. Overlaid is the specific star formation rate of the star-forming main sequence at $M_\star =  10^9~{\rm M_\odot}$ for different prescriptions from the literature (\citealt{Speagle_2014, Faisst_2016, Popesso_2023, McClymont_2025}; Simmonds et al. in prep.) which can be directly compared to the specific mass accretion rate (sMAR) as plotted on the right-hand side axis (essentially a rescaling of the merger rate; see Section~\ref{sec:mass accretion rate}). The coloured bands show the range of the sSFRs for the stellar mass range considered in this study, where the hatched sections are extrapolations of each respective relation. The sSFR is comparable to or larger than the sMAR in all cases, indicating that in-situ star formation dominates over stellar mass accretion via major mergers.}
 \label{fig:merger_rate_lit}
\end{figure*}

We show measurements from this work and results from the literature in Figure~\ref{fig:merger_rate_lit}. Our results are plotted with uncertainty bands and are colour-coded by stellar mass, as previously shown in the case of pair fractions. Values of merger rates and associated uncertainties can be found in Table~\ref{tab:pair fraction and merger rate}. We also plot results from the existing literature, as in the case of pair fractions. In this case, however, we use values for merger rates as quoted in the respective papers if they exist, where these are often calculated from pair fractions by a different choice of merger timescale that is more appropriate for the respective dataset. Not all works discuss merger rates, and in these cases, we use the same merger timescale as we used to convert our measured pair fractions (see Section~\ref{sec:merger observability timescale}).

We fit a power law + exponential model to our measured merger rates and results from the literature for each stellar mass bin, as in the case of pair fractions, having the form
\begin{equation}
    \mathcal{R}_\mathrm{M}(z, M_\star) = R_0 \times (1+z)^m \times e^{\tau(1+z)},
    \label{eq:merger rate fit}
\end{equation}
where $R_0$, $m$, and $\tau$ are parameters for which we perform weighted non-linear least squares fitting with uncertainties incorporated as weights. The fitted parameters and their uncertainties can be found in Table~\ref{tab:fit parameters}, and the fitted curves are plotted in Figure~\ref{fig:merger_rate_lit} with the respective colour coding.

\subsection{Merger rate density}

Another useful quantity to measure the merger rate that is frequently used in the literature is the merger rate density or total merger rate. The merger rate density describes how many mergers are occurring per unit time per unit comoving volume at a given redshift. We define the comoving merger rate density as
\begin{equation}
    \Gamma_\mathrm{M}(z, M_\star) = \frac{f_\mathrm{P}(z, M_\star) n_\mathrm{c}(z, M_\star)}{\tau_\mathrm{P}(z)},
    \label{eq:G_m}
\end{equation}
where $n_\mathrm{c}(z, M_\star)$ is the comoving number density of (primary) galaxies within a redshift and stellar mass bin. The comoving number density of galaxies is calculated by directly integrating the SMF (Equation~\ref{eq:SMF}) for a given stellar mass and redshift bin. While the volume-averaged galaxy merger rate $\Gamma_{\rm M}(z)$ is a useful measure, in the following, we will only use the number of mergers per massive galaxy, i.e., the merger rate $\mathcal{R}_{\rm M}(z)$.

\subsection{Mass accretion rates}
\label{sec:mass accretion rate}

We define the specific mass accretion rate (sMAR) as in \citet{Duncan19} in the following way:
\begin{equation}
    \mathrm{sMAR}(z) = \frac{\overset{.}{M}}{M} = \mathcal{R}_{\mathrm{M}}(z) \overline{\mu},
\end{equation}
where $\overline{\mu}$ is the median mass ratio (or merger ratio). This quantity is analogous to the sSFR and measures the rate of mass accumulated by mergers per unit mass. The merger ratio $\overline{\mu}$ can be computed in each stellar mass–redshift bin by finding the median of the cumulative distribution of pair fraction as a function of merger ratio (see Figure~\ref{fig:mu_example} in Appendix~\ref{sec:appendix_mu} for an example). These median merger ratios vary with stellar mass and redshift, and their average is $\langle \overline{\mu} \rangle = 0.485$ (see Figure~\ref{fig:mu_med} in Appendix~\ref{sec:appendix_mu}).

Another way of calculating the specific mass accretion rate is to simply take the mass accretion rate $\rho_{1/4}$ and divide it by the average stellar mass of primary galaxies \citep{Duan_2024} as
\begin{equation}
    \mathrm{sMAR}(z) = \frac{\overset{.}{M}}{M} = \frac{\rho_{1/4}(z)}{\mathcal{M}_{*, 1}(z)} = \mathcal{R}_{\mathrm{M}}(z) \frac{\mathcal{M}_{*,2}(z)}{\mathcal{M}_{*,1}(z)},
    \label{eq:sMAR}
\end{equation}
where the mass accretion rate (MAR; analogous to SFR), which is the average amount of mass added through mergers per galaxy per unit time, is defined as 
\begin{equation}
    \rho_{1/4} = \mathcal{R}_{\mathrm{M}}(z) \times \mathcal{M}_{*,2}(z),
    \label{eq:MAR}
\end{equation}
where the subscript `$1/4$' denotes that this quantity is for major mergers. The average stellar mass of the primary and secondary galaxies in each redshift bin, respectively are defined using the galaxy SMF $\phi (M_\star, z)$ as
\begin{align}
    \mathcal{M}_{*, 1}(z) &= \frac{\displaystyle \int_{M_\star^{\mathrm{min}}}^{M_{\star}^{\mathrm{max}}} \phi(M_\star, z) M_{\star} dM_\star}{\displaystyle \int_{M_\star^{\mathrm{min}}}^{M_{\star}^{\mathrm{max}}} \phi(M_\star, z) dM_\star} \\
    \mathcal{M}_{*, 2}(z) &= \frac{\displaystyle \int_{\mu \mathcal{M}_{\star, 1}}^{ \mathcal{M}_{\star, 1}} \phi(M_\star, z) M_{\star} dM_\star}{\displaystyle \int_{\mu \mathcal{M}_{\star, 1}}^{\mathcal{M}_{\star, 1}} \phi(M_\star, z) dM_\star}.
    \label{eq:mu_estimation_2}
\end{align}
Now the ratio of the average stellar mass of the secondary and primary galaxies will be essentially the same as $\langle \overline{\mu} \rangle$, and we find that $\langle \mathcal{M}_{*,2} / \mathcal{M}_{*,1} \rangle = 0.474$ (see Figure~\ref{fig:mu_med}). This average value is close to the median merger ratio found by the previous method, and although it has a stellar mass–redshift variation, we make the assumption that $\overline \mu = 0.48$ for the rest of the calculations. 

Finally, we plot the specific mass accretion rate as a second y-axis on Figure~\ref{fig:merger_rate_lit} since it is just a rescaled version of the merger rate by a factor of $\overline \mu$ that we calculated earlier. This allows us to directly compare the contribution to mass accretion by mergers via sMAR and star formation via sSFR on the same plot. We plot five different SFMS sSFRs for comparison from \citet{Speagle_2014}, \citet{Faisst_2016}, \citet{Popesso_2023}, \citet{McClymont_2025}, and Simmonds et al. (in prep.). The model by \cite{Speagle_2014} was empirically determined from a compilation of 25 studies from the literature, is valid in the $z \sim 0-6$ range, and is well described by a single time-dependent function. The study by \citet{Faisst_2016} finds a SFMS sSFR with a stronger power-law evolution at $z<2.2$ (extrapolation for $z<0.5$) and a weaker power-law evolution at $z>2.2$ (extrapolation for $z>6$) with redshift, using data from the literature and spectroscopic data from the COSMOS field \citep{Scoville_2007} with median stellar mass of $\sim 10^{9.8}~{\rm M_\odot}$. \citet{Popesso_2023} use a large sample of star-forming galaxies at $0 \leq z \leq 6$ with stellar masses $10^{8.5}-10^{11.5}~{\rm M_\odot}$ and find a flattening sSFR above $z \sim 3$. Since all these empirical SFMS prescriptions are extrapolations beyond $z \sim 6$, and they show flattening sSFRs at higher-$z$ that could be affected by selection effects, we opt for the model of \citet{McClymont_2025}. Their sSFR is derived from the \textsc{thesan-zoom} simulation \citep{Kannan_2025} and increases at higher redshifts, being valid for the $z>3$ range. This SFMS choice is further supported by a good agreement with an observational counterpart based on the JADES dataset at $z \sim 3-9$ by Simmonds et al. (in prep.).

\begin{table*}
    \centering
    \caption{Number of primary galaxies ($N_{\rm primary}$), and the measured major merger pair fraction ($f_{\rm p}$) and merger rate ($\mathcal{R}_\mathrm{M}$) values at each respective redshift and stellar mass bin, with their associated uncertainties. Uncertainties for pair fractions are estimated using bootstrapping (using 100 resamples) and carried forward to calculate uncertainties for merger rates.}
    \label{tab:pair_fraction_merger_rate}
    \begin{threeparttable}
    \begin{tabular}{ccccccc}
        \hline
        \hline
        \multicolumn{1}{c}{Redshift} & $[3.0, 4.0]$ & $[4.0, 5.0]$ & $[5.0, 6.0]$ & $[6.0, 7.0]$ & $[7.0, 8.0]$ & $[8.0, 9.0]$\\
        \hline
        Mass range ($\log M_\star/\mathrm{M}_\odot$) & \multicolumn{6}{c}{\textbf{Number of primary galaxies, $N_{\rm primary}$}\tnote{*}} \\
        $[8.0, 8.5]$ & $3655$ & $2842$ & $2170$ & $1523$ & $501$ & $234$\\
        $[8.5, 9.0]$ & $2295$ & $1397$ & $662$ & $350$ & $95$ & $37$ \\
        $[9.0, 10.0]$ & $1613$ & $760$ & $312$ & $104$ & $34$ & $10$  \\
        \hline
        Mass range ($\log M_\star/\mathrm{M}_\odot$) & \multicolumn{6}{c}{\textbf{Pair fraction, $f_{\rm P}$}} \\
        $[8.0, 8.5]$ & $0.19 \pm 0.10$ & $0.31 \pm 0.15$ & $0.37 \pm 0.17$ & $0.34 \pm 0.16$ & $0.19 \pm 0.09$ & $0.11 \pm 0.05$ \\
        $[8.5, 9.0]$ & $0.20 \pm 0.10$ & $0.21 \pm 0.11$ & $0.12 \pm 0.06$ & $0.14 \pm 0.07$ & $0.08 \pm 0.04$ & $0.02 \pm 0.01$\tnote{\textdagger} \\
        $[9.0, 10.0]$ & $0.14 \pm 0.06$ & $0.12 \pm 0.07$ & $0.09 \pm 0.04$ & $0.12 \pm 0.05$ & $0.06 \pm 0.02$ & $0.09 \pm 0.05$\tnote{\textdagger} \\
        \hline
        Mass range ($\log M_\star/\mathrm{M}_\odot$) & \multicolumn{6}{c}{\textbf{Merger rate, $\mathcal{R}_\mathrm{M}$ (Gyr$^{-1}$)}} \\
        $[8.0, 8.5]$ & $1.60 \pm 0.80$ & $3.91 \pm 1.92$ & $6.46 \pm 3.05$ & $8.08 \pm 3.66$ & $5.64 \pm 2.64$ & $4.04 \pm 1.88$ \\
        $[8.5, 9.0]$ & $1.70 \pm 0.85$ & $2.67 \pm 1.33$ & $2.20 \pm 1.02$ & $3.29 \pm 1.56$ & $2.49 \pm 1.30$ & $0.57 \pm 0.33$\tnote{\textdagger} \\
        $[9.0, 10.0]$ & $1.18 \pm 0.52$ & $1.52 \pm 0.85$ & $1.62 \pm 0.77$ & $2.89 \pm 1.14$ & $1.75 \pm 0.68$ & $3.38 \pm 2.01$\tnote{\textdagger} \\
        \hline

    \end{tabular}
    \begin{tablenotes}
    \item[*] We note here that these values reflect the number of primary galaxies falling into each redshift-stellar mass bin based on their stellar mass and peak photometric (or spectroscopic) redshift ($z_a$) falling within the bin limits. For the purposes of the pair fraction measurement, the total numerical contribution of the primary galaxies in each bin is given by $\sum_i N_1^i$, where $N_1^i$ is given by Equation~\ref{eq:N_1_i}.
    \item[\textdagger] The values presented in the table above are included for the sake of completeness. However, we do not plot or use them in our further calculations, as the number counts from the initial sample selection in these bins are too low to perform a reliable analysis.
    \end{tablenotes}
    \end{threeparttable}
    \label{tab:pair fraction and merger rate}
\end{table*}

\begin{table*}
    \centering
    \caption{Fit parameters for the pair fraction ($f_p$) and merger rate ($\mathcal{R}_\mathrm{M}$) for three stellar mass bins for an assumed power law + exponential model of the form $C \times (1+z)^m \times e^{\tau(1+z)}$, where $C$ is $f_0$ for the pair fraction, and $R_0$  for the merger rate. The fitting is done using the weighted non-linear least squares method with uncertainties incorporated as weights. We also provide overall fits for the full mass range of $\log (M_\star/\mathrm{M}_\odot) = [8,10]$ considered for all available data assuming the same functional form as before, for both the pair fractions and merger rates. Although this fit is somewhat forced and erases the scaling relation with stellar mass, it is useful and simplifies further calculations.}
    \label{tab:pair_fraction_merger_rate_fit_param}
    \begin{tabular}{cccc}
        \hline
        \hline
        \multicolumn{1}{c}{Fit parameters} & $C \times 10^2$ & $m$ & $\tau$ \\
        \hline
        Mass range ($\log M_\star/\mathrm{M}_\odot$) & \multicolumn{3}{c}{\textbf{Pair fraction, $f_p$}} \\
        All & $3.71 \pm 0.24$ & $2.27 \pm 0.20$ & $-0.48 \pm 0.07$ \\
        \hline
        Mass range ($\log M_\star/\mathrm{M}_\odot$) & \multicolumn{3}{c}{\textbf{Merger rate, $\mathcal{R}_\mathrm{M}$ (Gyr$^{-1}$)}} \\
        $[8.0, 8.5]$ & $2.20 \pm 0.20$ & $4.18 \pm 0.29$ & $-0.42 \pm 0.07$ \\
        $[8.5, 9.0]$ & $2.25 \pm 0.37$ & $3.36 \pm 0.40$ & $-0.27 \pm 0.11$ \\
        $[9.0, 10.0]$ & $2.61 \pm 0.17$ & $2.90 \pm 0.20$ & $-0.20 \pm 0.06$ \\
        All & $2.49 \pm 0.16$ & $2.85 \pm 0.20$ & $-0.16 \pm 0.06$ \\
        \hline
    \end{tabular}
    \label{tab:fit parameters}
\end{table*}

\section{Discussion}
\label{sec:discussion}

In this section, we discuss the potential origins and implications of our measurements of galaxy close-pair fractions and major merger rates in the broader context of galaxy formation and evolution, focusing on the mass assembly of galaxies through cosmic time. We give possible explanations for the evolutionary trends found and compare them to predictions and results from large-scale cosmological simulations. We compare the importance of galaxy mergers and star formation in terms of mass assembly, particularly focusing on the difference between in-situ star formation versus ex-situ stellar mass accretion.

\subsection{Qualitative reasoning behind the evolutionary trends found}

In this work, we utilise the deepest available data from JWST NIRCam photometry through JADES observations, combined with all available spectroscopic data, to measure galaxy close-pair fractions and derive major merger rates in the poorly understood redshift range $z \sim 3-9$. We successfully incorporate the full posterior distributions of the estimated photometric redshifts and propagate them throughout our analysis to find robust results with reliable uncertainties.

The redshift evolution of the galaxy close-pair fraction shows a significant dependence on the stellar mass bin considered for selecting the primary galaxies. In the case of the ${\rm log_{10}}(M_\star/{\rm M_\odot}) = [8.0, 8.5]$ stellar mass bin, there is an increase in pair fractions at redshifts $z \sim 3-5$, peaking at $z \sim 5-6$, that is followed by a turn-over (see Figure~\ref{fig:pair_fraction_bins}). For ${\rm log_{10}}(M_\star/{\rm M_\odot}) = [8.5, 9.0]$ a similar, but weaker peak can be observed at $z \sim 4-5$, followed by a decreasing trend. For the third stellar mass bin at ${\rm log_{10}}(M_\star/{\rm M_\odot}) = [9, 10]$, a flat and slightly decreasing trend can be seen, and if there is a peak as in the previous cases, it is at $z \sim 3-4$ or at lower redshift. Therefore, we conclude that the redshift where the pair fraction turns over increases with decreasing stellar mass. The same behaviour was observed in the theoretical study of \citet{Husko_2022}, which is based on the Planck Millennium cosmological simulation. This can be explained by the stellar mass range of interest passing into the exponentially decreasing regime of the galaxy stellar mass function at increasing redshifts. Based on recent studies \citep[e.g.,][]{Navarro-Carrera_2024, Harvey_2024, Weibel_2024} the characteristic stellar mass ($M^\star$), where the SMF transitions from a power-law decline for massive galaxies to an exponential cut-off, has a weak dependence on redshift. This transition happens at different redshifts for the different stellar mass bins considered (the characteristic stellar mass being in the range $M^\star \approx 10^{10-11}~{\rm M_\odot}$), and we simply expect the pair fractions to trace this sudden absence of massive galaxies at higher redshifts. 

Although the $\sim 200~{\rm arcmin}^2$ area studied in this work is small compared to the entire sky, we conclude that we are not significantly affected by cosmic variance since results from both GOODS-South and GOODS-North agree well within the uncertainties (see Figure~\ref{fig:pair_fraction_bins}). Based on our results, there is more variation between the Deep and Medium regions within each field. This suggests that there is more dependence on the depth of the regions considered and, subsequently, on stellar mass completeness.

We also note here that in the literature, two closely related quantities are reported in general, namely the \textit{close-pair fraction} and the \textit{merger fraction}, highlighted by \citet{Husko_2022}. While either of these can be converted to a merger rate by assuming an appropriate timescale, there is a significant difference between them. The close-pair fraction is measured from dynamical and geometrical selection criteria, that is, close separation projected on the sky and proximity in redshift or line-of-sight velocity. On the other hand, merger fractions are determined by selecting close pairs by morphological parameters that show signs of interactions or an early merging phase. This latter quantity attempts to measure the true fraction of galaxies that will actually merge, whereas the former includes random projections as well, which will not merge. At high redshifts ($z \gtrsim 6$), the close-pair method is more reliable for large samples, as it is currently difficult to select mergers based on morphological parameters (although it has been attempted by e.g., \citealt{Dalmasso_2024}).

We observe similar trends in the pair fractions and merger rates, although, for the latter, the redshift evolution is weaker. Our results suggest a flattening in merger rates beyond $z \sim 6$, which is also highlighted in the trends of the fitted power law + exponential curves for each stellar mass bin. Focusing only on the results from this work, one can also see a slight turnover at the highest redshift bins for the different stellar mass ranges. This can be again explained by entering the exponentially decreasing regime of the SMF at high redshifts, as in the case of close-pair fractions \citep{Husko_2022}. 

We also compare our results with recent studies of pair fractions and merger rates. \citet{Duan_2024} use a similar probabilistic approach to this work (although slightly different selection criteria, e.g. separation limits $r=20-50$ kpc), using data from eight JWST Cycle-1 fields, adding up to an area slightly smaller (${\sim}190~{\rm armin^2}$) than in this work. They use \texttt{Bagpipes} \citep{Carnall_2018} for SED-fitting to obtain stellar masses adopting a log-normal SFH, compared to \texttt{Prospector} with a non-parametric continuity SFH used in this work. They look at the same overall stellar mass range of ${\rm log_{10}}(M_\star/{\rm M_\odot}) = [8, 10]$, but they do not carry out a full analysis on smaller stellar mass bins within this mass range as compared to this current work. However, they find lower pair fractions in subsample selected at higher stellar masses ($\log_{10}(M_\star/{\rm M_\odot})=9-10$) that declines beyond $z \sim 5$ compared to a subsample with lower stellar masses ($\log_{10}(M_\star/{\rm M_\odot})=8-9$), which is qualitatively in agreement with our results. Overall, their data points match our results well, falling within our uncertainty ranges, as seen in Figures~\ref{fig:pair_fraction_literature} and \ref{fig:merger_rate_lit}. Furthermore, \citet{Dalmasso_2024} also investigate close-pair fractions at high redshifts in the same stellar mass range but using morphological parameters to identify mergers. They find a flat evolution with redshift, and we convert their pair fractions using the timescale chosen for our analysis to calculate merger rates. Similarly, the merger rates converted from their pair fractions fall within our reported uncertainty ranges.

\subsection{Star formation rate versus merger rate}
\label{sec:star formation rate versus merger rate}

To further assess the importance of major mergers in a quantitative way, we directly compare them to the other main channel of mass growth--star formation. One direct comparison is based on calculating the mass accreted by mergers and by star formation. 

To calculate the cumulative mass accretion by mergers, we simply integrate the mass accretion rate (MAR; Equation~\ref{eq:MAR}) for the desired cosmic time or redshift range as
\begin{equation}
    \delta M(z) = \displaystyle \int_{z_{0}}^{z} \rho_{1/4}(z) dz,
\end{equation}
where $z_{0}$ is chosen at $z=9$ (which is the upper redshift limit of this study) and $z$ is at each subsequent point, and $\rho_{1/4}(z)$ is the mass accretion rate defined previously. We calculate this cumulative mass accretion from mergers, where for simplicity, we assume an overall fit for the merger rate $\mathcal{R}_{\rm M}(z)$ (given in Table~\ref{tab:fit parameters}) for the calculation of $\rho_{1/4}(z)$. 

To directly compare the mass accreted from major mergers over time, we calculated a similar quantity for the mass accreted by star formation. Here, we use the SFMS fit of \citet{McClymont_2025} to derive the sSFR at ${\rm log_{10}}(M_\star/{\rm M_\odot}) = 9$, the central mass in our total stellar mass range of interest. This SFMS sSFR is derived from the \textsc{thesan-zoom} simulation \citep{Kannan_2025} that shows a rising sSFR at high-$z$. We note here that this simulation-based SFMS is in good agreement with its observational counterpart, which is calculated using the JADES dataset by Simmonds et al. (in prep.; see Figure~\ref{fig:merger_rate_lit}). In the next section, we will discuss the motivation and effect of this choice of sSFR compared to other prescriptions in more detail. 

We integrate this SFR the same way as we did for the MAR of mergers to obtain the cumulative mass accretion by star formation. The ratio of these two quantities gives us an approximate way to measure the ex-situ mass fraction (only due to major mergers) in an average galaxy lying on the SFMS. The resulting ex-situ mass fraction ($f_{\rm ex-situ}$) has a weak evolution in the redshift range of $z \approx 3-9$ between $1-8\%$, with a median value of $\sim 5\%$ (see the purple curve on Figure~\ref{fig:cumulative mass fraction}). Although approximate, this is one direct calculation to quantitatively assess the importance of major mergers from our inferred quantities, which shows that in-situ star formation is the dominant channel for the mass growth of galaxies compared to major mergers.

We note here that these results are in contrast with a similar calculation performed by \citet{Duan_2024}, where they find that the contribution of major mergers to galaxy stellar mass is 71 ($\pm25$)\% (equivalent to the contribution from star formation). As their merger rate values agree well with our results, this discrepancy might arise because they use sSFRs estimated by \texttt{Bagpipes} instead of using a SFMS prescription. To further look into this disagreement, we looked at the median sSFR values of the entire initial sample at 10, 50, and 100 Myr measured by \texttt{Prospector} and found a good agreement with the \citet{McClymont_2025} and Simmonds et al. (in prep.) SFMS sSFR trends. Therefore, we conclude that our ex-situ fraction estimates are consistent between SED-derived and simulation-based measurements.

\subsection{In-situ versus ex-situ star formation}

To look in more detail at the difference between the in-situ versus ex-situ mass fraction of galaxies due to star formation and major mergers from another perspective, we build a simple model for galaxy mass growth. We would like to note that our numerical model only focuses on major mergers, and we discuss the potential effect of minor mergers at the end of this section. Our model includes both gas-poor mergers (also referred to as \textit{dry mergers} in the literature) and gas-rich mergers (also known as \textit{wet mergers}). However, in the latter case, it only considers the stellar component, i.e. it ignores the gas that would be accreted and fuel star formation (as well as the triggering of star formation itself).

In our numerical model, we start with a sample of $50,000$ galaxies with initial stellar masses randomly sampled from a linear distribution of masses with slope $\alpha=-1.5$ between $10^{-1}~{\rm M_\odot}$ and $10^{6}~{\rm M_\odot}$ at $z=16$. We then look at the stellar mass growth evolution of these galaxies over $10,000$ time steps until $z=0.5$. Our choices for these values are reasonable but somewhat arbitrary; nevertheless, they do not significantly affect the outputs we are interested in since these initial conditions are quickly erased by the nearly exponential growth of galaxies at these early times. During the evolution, at each time step, we increase the mass of each galaxy in our initial sample by the amount of mass assembled by star formation, where we use the SFMS from \citet{McClymont_2025} (see Figure~\ref{fig:merger_rate_lit}). We discuss the effects of different choices of sSFRs at the end of this section. We also require that star formation is shut down when the galaxy reaches the quenching threshold mass of $M_{\rm quench}=10^{10.8}~{\rm M_\odot}$, which is the location of the knee of the SMF. We also introduce a burstiness parameter for the star formation, that is, $10^b$ with $b$ being a random number drawn from a Gaussian distribution centred at 0 with $\sigma = 0.3$. We multiply the SFR with this burstiness parameter at each step, and it is independent of the previous step. To first order, this takes into account the dispersion around the SFMS \citep{Tacchella_2020}.

\begin{figure}
 \includegraphics[width=\columnwidth]{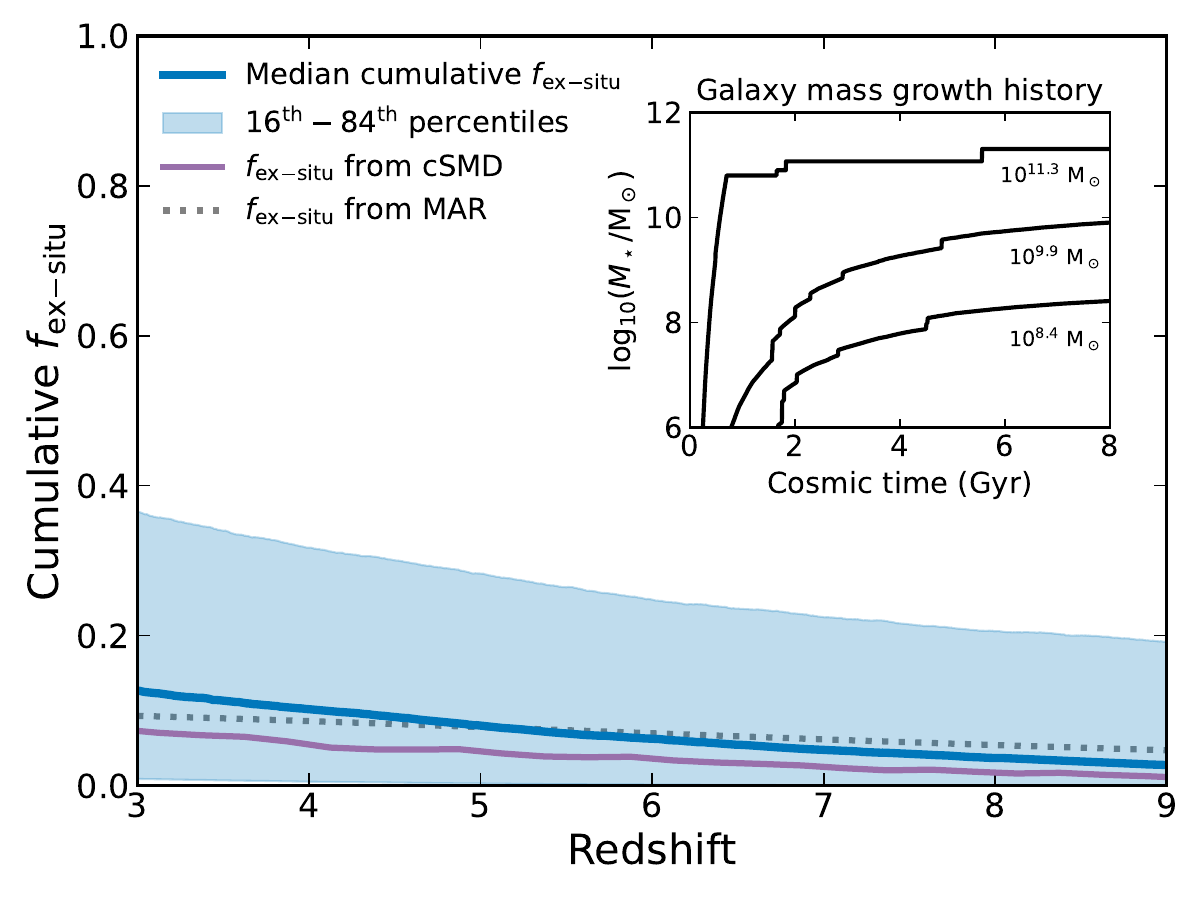}
 \caption{The redshift evolution of the cumulative ex-situ stellar mass fraction due to major mergers from our numerical model that simulates the mass growth of galaxies using the merger rate from our measurements (valid for the $10^8-10^{10}~{\rm M_\odot}$ stellar mass range) and assuming an SFR which always lies on the SFMS of \citet{McClymont_2025}. We plot the median and the corresponding $16^{\rm th}$ and $84^{\rm th}$ percentiles of the distribution of the results from our numerical model of $50,000$ galaxies. The median reaches only up to about $\sim 13\%$ of ex-situ fraction even at low redshifts. In the inset figure, we plot the mass growth history of three example galaxies from our simulation that have undergone several major mergers throughout their evolution, reaching final stellar masses indicated under each respective curve. We also plot the $f_{\rm ex-situ}$ found by weighting it by the SMF (purple curve), and the $f_{\rm ex-situ}$ estimated by taking the ratio between the stellar mass accreted and by mergers by star formation (grey dotted line; Section~\ref{sec:star formation rate versus merger rate}). All three calculations give similar results ranging between $3-13\%$, placing in-situ star formation as the dominant channel for the mass growth of galaxies compared to major mergers.}
 \label{fig:cumulative mass fraction}
\end{figure}

The probability that a merger takes place is given by the merger rate at each time step. For the merger rate, we use a single mass-independent fit from this work for simplicity, that is parametrised in Equation~\ref{eq:merger rate fit} with fitted parameters given in Table~\ref{tab:fit parameters} (see also Figure~\ref{fig:merger_rate_lit}). We assume a Poisson distribution for a merger occurring, as these events are independent and relatively rare over a short time period. If there is a merger in a given time step, we calculate the mass accreted through the merger by multiplying the merger rate $\mathcal{R}_{\rm M}$ at that time (redshift) by the time step and a fraction of the main galaxy's mass. This fraction is the mass ratio that is a random number in the range $\mu = [0.25, 1]$ for a major merger for the purpose of this simulation. In the subsequent time step, there is an independent draw for a merger to take place or not (for examples of galaxy mass growth histories, see the inset in Figure~\ref{fig:cumulative mass fraction}).

This model allows us to calculate the ex-situ mass fraction of galaxies by using empirical values such as merger rates and SFMS SFRs, as well as some simple assumptions. We plot in Figure~\ref{fig:cumulative mass fraction} the cumulative ex-situ mass fraction, $f_{\rm ex{-}situ}$, that is, measuring the redshift-evolution of the mass accreted from mergers compared to the total mass of the galaxy. We plot the median of $f_{\rm ex{-}situ}$ and the $16^{\rm th}$ and $84^{\rm th}$ percentiles of the distribution of the galaxy sample from our simulation. As we might have expected based on Figure~\ref{fig:merger_rate_lit} by comparing the merger rate to the sSFR, the ex-situ mass fraction is slightly lower at higher redshifts ($\sim 3\%$ at $z=9$) and slowly increases with cosmic time (to $\sim 13 \%$ at $z=3$). In other words, based on this calculation, major mergers contributed less to the mass build-up of galaxies at earlier times and subsequently, their relative contribution increased (see the blue curve in Figure~\ref{fig:cumulative mass fraction}). However, even at low-z, the median of $f_{\rm ex{-}situ}$ reaches about $13\%$, and the $84^{\rm th}$ percentile goes slightly below $\sim 40\%$. This is in agreement with our previous estimate of the ex-situ mass fraction in Section~\ref{sec:star formation rate versus merger rate} and places major mergers as a significant but not primary contributor to the mass assembly of galaxies. 

In a third calculation to estimate $f_{\rm ex{-}situ}$, we integrate the SMF (Appendix~\ref{sec:appendix_gsmf}) between $M_{\rm min} = 10^8~{\rm M_\odot}$ and $M_{\rm max} = 10^{12}~{\rm M_\odot}$ to obtain the cosmic stellar mass density (cSMD), and perform the same integration for a SMF convolved with the median $f_{\rm ex{-}situ}$ calculated by the numerical model. This latter calculation should, in principle, give the contribution of major mergers to the cosmic stellar mass density. Finally, we take the ratio between the two quantities at redshifts in the range $z=3-9$ to obtain a SMF weighted $f_{\rm ex{-}situ}$ (see the grey dotted curve in Figure~\ref{fig:cumulative mass fraction}), which shows a similar redshift evolution as the previous fraction at slightly lower values. All three calculations give consistent values of $f_{\rm ex{-}situ}$, ranging between $3-13\%$, and showing a weak redshift evolution.

\begin{figure}
 \includegraphics[width=\columnwidth]{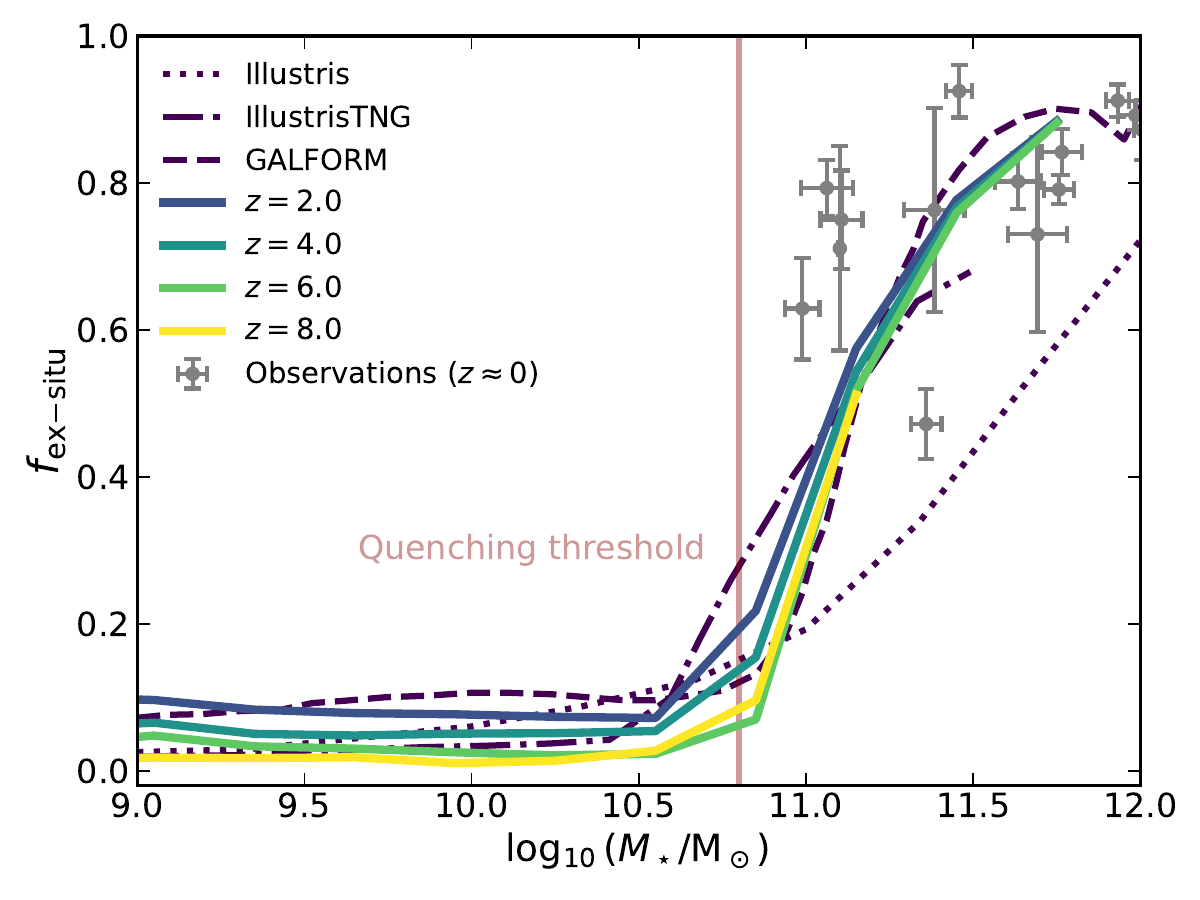}
 \caption{Ex-situ mass fraction $f_{\rm ex{-}situ}$ from our numerical model at different stellar masses and redshifts. We observe that the $f_{\rm ex{-}situ}$ stays approximately constant in the $3-13\%$ range for stellar masses up to $\sim 10^{10.8}~{\rm M_\odot}$, after which it quickly increases due to our mass threshold for quenching star formation. After a galaxy completely quenches its star formation, it can only grow its mass by mergers in this model, which leads to the up-tick in $f_{\rm ex{-}situ}$ above the quenching threshold. We can also observe a dependence on redshift, where $f_{\rm ex{-}situ}$ values are lower for higher redshifts. For comparison, we plot the ex-situ mass fraction trends predicted by the Illustris \citep{Rodriguez-Gomez_2016}, IllustrisTNG \citep{Tacchella_2019}, and \texttt{GALFORM} \citep{Husko_2022} simulations at $z=0$, and find that they are in good agreement with our results. Ex-situ mass fractions measured from observations of local galaxies at $z \approx 0$ \citep{Seigar_2007, Spavone_2017, Spavone_2018, Cattapan_2019, Spavone_2020, Iodice_2020} also seem to match well our model predictions.}
 \label{fig:f_ex-situ}
\end{figure}

We also look at the ex-situ mass fractions as a function of the stellar mass of the galaxy. We plot in Figure~\ref{fig:f_ex-situ} the median $f_{\rm ex{-}situ}$ values of the distribution of the galaxies from our simulation at different redshifts. Notably, we observe a scaling of ex-situ mass fractions with redshift, where $f_{\rm ex{-}situ}$ is higher for lower redshifts, as expected from Figure~\ref{fig:cumulative mass fraction}. There is a sudden steep increase in $f_{\rm ex{-}situ}$ above a mass of $10^{10.8}~{\rm M_\odot}$, which is not surprising, as this is our chosen threshold mass above which a galaxy quenches its star formation, and the only remaining way to accrete more mass is through major mergers in this model. This is in agreement with predictions from \citet{Behroozi_2013} and \citet{Behroozi_2015} (c.f. Figure 2 in \citealt{Behroozi_2015}). These results are also in agreement with the findings of \citet{Rodriguez-Gomez_2016}, \citet{Tacchella_2019}, and \citet{Husko_2022} based on the Illustris, IllustrisTNG, and \texttt{GALFROM} simulations at $z=0$. We also observe the same scaling with redshift as the one reported in \citet{Tacchella_2019}. Furthermore, these results seem to be consistent with observations of quiescent galaxies at lower redshifts at $z \sim 2$ \citep{Ji_2022, Ji_2024a, Ji_2024b}. Since these galaxies are already quenched by the time of observation, their star formation must have stopped in earlier times. Based on detailed analysis of morphology and kinematics of these galaxies, they conclude that quiescent galaxies with $\log(M_\star / {\rm M_\odot})>11$ should have a significant ex-situ mass contribution (e.g. mergers) to alter their structure after quenching, while the ex-situ contribution is much smaller for lower mass $\log(M_\star / {\rm M_\odot})<11$ quiescent galaxies.

We note here that our results for the ex-situ stellar mass fractions depend on the choice of the star-forming main sequence prescription. We use the SFMS$_{10}$ fit of \citet{McClymont_2025}, which is derived from the \textsc{thesan-zoom} simulation \citep{Kannan_2025}. This SFMS is in good agreement with observational measurements based on JADES data (Simmonds et al. in prep.; see Figure~\ref{fig:merger_rate_lit}), as well as \texttt{Prospector}-derived sSFR values of our sample. Other SFMS fits from the literature, which are based directly on observational measurements of stellar mass and SFR \citep[e.g.,][]{Speagle_2014, Faisst_2016, Popesso_2023} agree well with each other in the $z<3$ range (see Figure~\ref{fig:merger_rate_lit}), however, they diverge at the higher redshifts (beyond $z \approx 6$) and lower masses ($\log(M_\star/{\rm M_\odot})<9$) which we consider in this work. Additionally, the flattening of the SFMS evolution at high-$z$, which is seen in observational SFMS prescriptions, can be mostly explained by selection effects \citep{McClymont_2025}. We therefore opt for a simulation-derived prescription, which is not subject to these effects and instead shows a rising sSFR at high-$z$. Nevertheless, we run our numerical model using an sSFR from \citet{Speagle_2014, Faisst_2016, Popesso_2023}, and find ex-situ mass fractions in a similar range (below $\sim 16\%$) and mostly even lower due to the higher sSFR values in the different prescriptions. However, in these cases, the weak redshift evolution of $f_{\rm ex{-}situ}$ is reversed, and it is in general higher at higher redshifts and subsequently slowly decreases with decreasing redshift (as it is expected based on Figure~\ref{fig:merger_rate_lit}). As an additional check, we also run the numerical model for the observational SFMS by Simmonds et al. (in prep.) and find similar but slightly higher $f_{\rm ex{-}situ}$ values (5-12\%) than our fiducial calculation. Therefore, while the numerical values and the weak redshift scaling slightly change based on the assumed SFMS, our overall conclusion about dominant in-situ star formation compared to the effect of major mergers still does hold overall, independent of the choice of the SFMS prescription.

We further note here that our findings regarding the in-situ versus ex-situ mass growth of galaxies are based only on major mergers and do not take into account any contributions from minor mergers, i.e. mergers with mass ratios of $\leq 1/4$. According to the hierarchical growth model of galaxies, minor mergers are expected to be more common than major mergers \citep[e.g.,][]{Fakhouri_2010}. Previous models predicted that quiescent galaxies mainly grow via gas-poor minor mergers rather than major mergers \citep[e.g.,][]{Bezanson_2009}. Studies by HST found that the size growth of massive quiescent galaxies is caused by a significant number of minor mergers \citep[see e.g.,][]{Newman_2012, Belli_2015}. A recent study using JWST by \citet{Suess_2023} found a large population of low-mass companions (mostly having mass ratios $< 1/10$) around massive quiescent galaxies ($\log(M_\star/{\rm M_\odot})>10.5$). Their results suggest that in the redshift range $0.5 \leq z \leq 3$, minor mergers are potentially the main drivers of quiescent galaxy size growth and can also introduce radially decreasing colour gradients. However, the study of minor mergers is beyond the scope of this paper and will require a further independent study to constrain their rate of occurrence.

\subsection{Comparison to theoretical models}
\label{sec:comparison to large scale cosmological simulations}

In this section, we compare our results to findings from large-scale cosmological simulations and semi-analytical models. The existing results from the literature can be divided approximately into two main categories, namely: studies that claim that merger rates flatten or turn over at high redshifts depending on the mass selection \citep[e.g.,][]{Husko_2022}; and sources that advocate for the monotonic increase of merger rates with redshift \citep[e.g.,][]{Hopkins_2010a, Rodriguez-Gomez15, O'Leary_2021}.

Our results agree well qualitatively with the predictions of \citet{Husko_2022} using the Planck Millennium cosmological simulation and the \texttt{GALFORM} semi-analytical model of galaxy formation \citep{Baugh_2019}. Interestingly, most models show an increase in merger rates at all redshifts, unlike \texttt{GALFORM}, which predicts a plateau and a turnover depending on the stellar mass of interest. Although, \citet{Husko_2022} calculated the merger rate at  ${\rm log_{10}}(M_\star/{\rm M_\odot})=[9, 10.3, 11]$, all these results show similar trends to our findings, where they rise and then stabilise at intermediate redshifts, and subsequently turn over (see Figure 6 from the cited work). They explain this by the exponential drop-off in the galaxy SMF. They find similar behaviour in the close-pair fractions (see Figure 7 in their work), where the turnover at higher masses moves toward lower masses with increasing redshift, reflecting the pattern seen in the merger rate. Again, this shift occurs because fewer galaxies populate the steeply declining tail of the SMF. This trend is more prominent in our observational measurements, as can be seen in Figures~\ref{fig:pair_fraction_summary} and \ref{fig:pair_fraction_literature}. This is further supported by not only the decreasing ratio of massive and satellite galaxies (i.e. pair fractions) but also the absolute number counts of primary galaxies declining at high redshifts, as expected.

Notably, \citet{Husko_2022} highlights an interesting difference between physical pairs (with actual 3D separations matching their projected distances) and projected pairs (with significantly larger line-of-sight separations than their projected distances, i.e. the galaxies will not merge), which cannot be disentangled in observational studies as opposed to simulations. Their results reveal that physical pair fractions, analogous to merger fractions, remain nearly constant with stellar mass, while projected pair fractions decline with increasing mass, consistent with the decreasing trend of the SMF. Notably, even at high stellar masses, about half of all pairs are due to projection (reducing by a similar amount the merger rate). For these massive systems, projected pairs are predominantly found within the same dark matter halo.

Other theoretical models predict qualitatively similar trends for both pair fractions and merger rates. The empirical model \texttt{EMERGE} \citep{O'Leary_2021} predicts a flattening in pair fractions and a weak turn over at high redshifts, including a scaling with the stellar mass selection, as found by this work. Similarly, the merger rates calculated from their mock observations can be best described by a power-law exponential evolution. Moreover, \citet{Endsley_2020} use the \texttt{UniverseMachine} semi-empirical model to produce mock observations for JWST and find similar results as from \texttt{GALFORM} and \texttt{EMERGE}. Findings from Illustris by \citet{Snyder17} also agree with these results, especially at the lower redshift regime.

On the contrary, some simulation-based predictions claim an ever-rising merger rate with redshift, even for the high stellar mass regime (e.g., $M_\star > 10^{11}~{\rm M_\odot}$). It has been well established in the past that the dark matter halo-halo merger rate rises with redshift \citep[e.g.,][]{Fakhouri_2010, Genel_2008, Genel_2010}. Therefore, assuming a direct correlation between the halo-halo and galaxy-galaxy merger rates, it would be expected that the latter also increases with redshifts. Using the Illustris simulation, \citet{Rodriguez-Gomez15} claims that the major merger rate increases as a power law with redshift, in agreement with the semi-empirical models of \citet{Stewart_2009} and \citet{Hopkins_2010a}, having a similar redshift dependence as the halo merger rate. As discussed in \citet{Husko_2022}, these discrepancies could arise for several reasons, such as differences in definitions and methodology. The \citet{Stewart_2009} and \citet{Hopkins_2010a} models, like \texttt{EMERGE} and \texttt{GALFORM}, assign merger times when a halo becomes a subhalo but do not track subhalo evolution within the primary halo. Additionally, their results rely on abundance matching up to z = 2, making predictions at higher redshift extrapolations.

Our results relating to $f_{\rm ex-situ}$ are consistent with the findings of \citet{Baker_2025}, where they look at in-situ versus ex-situ mass build-up for quiescent galaxies in the FLAMINGO simulations \citep{Schaye_2023} and conclude that on average high-z quiescent galaxies have not undergone multiple major mergers (${\sim}0.6$ major+minor mergers per galaxy per Gyr using our definition) and that their mass growth was dominated by in-situ star formation.

\subsection{Caveats and further improvements}
\label{sec:caveats and further improvements}

Here, we discuss the main uncertainties related to our analysis of merger rates. The three main sources that could bias our calculations are incompleteness, photometric redshift uncertainties, and the assumed merger timescale. 

We conduct a detailed assessment of the stellar mass completeness, as presented in Section~\ref{sec:completeness analysis}, and define our primary mass bins to be above the completeness limit. We assign weights to account for cases where the mass approaches the completeness limit or when the secondary mass bin partially enters the incomplete regime, as described in Section~\ref{sec:mass incompleteness}. These weights depend on the assumed SMF (see Appendix~\ref{sec:appendix_gsmf}) and might fully account for the incompleteness. Therefore, we might still be affected by incompleteness at high redshifts. Another systematic source of uncertainty simply lies within the nature of photometric redshifts estimated by \texttt{EAZY}. We perform an odds quality parameter cut (see Section~\ref{sec:odds quality parameter}) to keep the sources with well-defined photo-$z$'s and also assign a weight accounting for less constrained redshifts in the case of fainter objects (see Section~\ref{sec:photometric redshift quality}). We also compare the photometric redshifts to available spectroscopic redshifts and find a good agreement (see Section~\ref{sec:Spectroscopic redshift catalogue}). However, considering all this, our sample could still be affected by uncertainties due to projection effects and increased pair probability functions due to the larger overlap between wider photometric redshift posteriors. Furthermore, another selection effect could be related to the distance limits adopted in the definition of close-pairs, and in particular, the lower limit of $r_{\rm min} = 5~{\rm kpc}$. This could affect the inferred merger rates in the higher-$z$ samples, for which the merger timescale is shorter.  

Finally, the assumption of a merger observability timescale has a significant impact on the conversion of pair fractions to merger rates, which should be seriously considered. As we discuss in Section~\ref{sec:merger observability timescale}, the choice of timescale is crucial for estimating merger rates, and various parametrisations exist in the literature. We choose the observability timescale given by \citet{Snyder17}, which is obtained from the Illustris simulation \citep{Genel_2014, Vogelsberger_2014} in order to reconcile the merger rates with the pair counts directly measured from the simulation. It has a redshift dependence proportional to $\sim (1+z)^{-2}$, which leads to shorter timescales at higher redshifts, which in turn raises the declining tail of pair fractions. Merger timescales unfortunately remain impossible to measure directly at high redshifts, and the current theoretical estimates differ by factors of 2-3 \citep{Conselice_2006, Kitzbichler_2008, Lotz_2008b, Lotz_2010, Hopkins_2010b, Snyder17, Husko_2022, Conselice_2022}, therefore by choosing one particular timescale we are inevitably biased to that assumed theoretical model.

\section{Summary \& Conclusions}
\label{sec:conclusion}

This study presents a comprehensive analysis of the major merger history of galaxies in the redshift range $z \approx 3-9$, leveraging the unprecedented depth of the JADES dataset. By combining advanced probabilistic methodologies for incorporating photometric redshift posteriors and robust corrections for selection effects, we provide novel insights into the role of major mergers in galaxy evolution during the early Universe.

We apply a probabilistic method that takes into account the full photometric redshift posterior distributions and available spectroscopic redshifts to select major merger close-pairs with mass ratios $> 1/4$ within projected separations of $r_{\rm p} = 5-30$ kpc for stellar masses in the range $\log_{10}(M_\star / {\rm M_\odot})=[8, 10]$. We perform the analysis in sub-tiers of the GOODS-South and GOODS-North samples from JADES. We also correct for potential biases and selection effects when compiling our close-pair sample. Our main results can be listed as follows:

\noindent\textbf{I. Close-pair Fraction Evolution:}  
Our analysis reveals that the evolution of galaxy pair fractions with redshift follows a distinct pattern depending on the stellar mass range considered. In the lowest stellar mass bin ($8.0 \leq \log(M_\star/M_\odot) < 8.5$), the pair fractions increase up to $z \sim 5-6$ and exhibit a turn-over at higher redshifts. In the case of the two higher-mass bins, there is a flattening and weak decline with increasing redshift. This turnover suggests a reduced abundance of close pairs at the earliest epochs, possibly due to fewer galaxies populating the steeply declining tail of the SMF. At higher stellar masses, the pair fractions show less variation with redshift, pointing to a more stable merger history for massive galaxies. Our results agree well with those of previous close-pair studies.

\noindent\textbf{II. Major Merger Rate Evolution:}  
We convert the pair fractions to major merger rates by assuming a merger timescale, and we observe an apparent flattening in the merger rate evolution beyond $z \sim 6$, stabilising in the range of $2-8~{\rm Gyr^{-1}}$ per galaxy, depending on the stellar mass range considered. This behaviour reflects the interplay between the evolution pair fractions and the assumed merger observability timescale and the influence of the observational selection criteria. The impact of stellar mass on inferred merger rates is weak, nevertheless evident across the studied redshift range. Our inferred merger rates match well with previous studies, and an exponential-power law functional form can well describe their overall redshift evolution.

\noindent\textbf{III. Merger Contribution to Mass Growth:}  
A comparison between the cumulative stellar mass accretion from major mergers and the mass assembled through star formation through a numerical model and, similarly, integrating the SMF convolved with the ex-situ fraction consistently reveal that major mergers (in the mass range $\log_{10}(M_\star/{\rm M_\odot})=8-10$) contribute approximately $3-13\%$ to the total mass growth over $z \approx 3-9$. This contribution is substantial but secondary to in-situ star formation, which remains the primary driver of stellar mass build-up during this epoch in comparison with major mergers. These findings are consistent with predictions from cosmological simulations, which similarly estimate major mergers as a significant but not dominant channel for mass assembly in the early Universe. These results provide a critical constraint on the merger-driven growth of galaxies in the high-redshift Universe.

\noindent\textbf{IV. Methodological Advances:}  
The probabilistic framework further refined and applied in this study ensures robust results by incorporating full photometric redshift probability distributions and applying correction weights for mass incompleteness, photometric redshift quality, and survey boundaries. This methodology overcomes key limitations of previous works and paves the way for future studies of galaxy interactions using JWST data. Our approach also highlights the importance of using mass-selected samples for deriving meaningful close-pair fractions and merger rates, particularly at high redshifts and low stellar masses probed.

This work establishes that major mergers, while an important aspect of hierarchical galaxy formation, play a supplementary role in galaxy mass growth during the early Universe. The trends in pair fractions and merger rates support the view that mergers are a secondary process, acting in concert with star formation to shape the evolution of galaxies. Nevertheless, mergers could still drive star formation within galaxies, and hence, the physical properties of these mergers should be studied next. By constraining the merger history of galaxies at $z \approx 3-9$, this study provides a benchmark for theoretical models and lays the groundwork for future observational efforts to explore galaxy assembly at even higher redshifts with JWST and beyond.

\section*{Acknowledgements}

DP acknowledges support by the Huo Family Foundation through a P.C. Ho PhD Studentship. DP and ST acknowledge support by the Royal Society Research Grant G125142. CS, FDE, GCJ, and RM acknowledge support by the Science and Technology Facilities Council (STFC), by the ERC through Advanced Grant 695671 ``QUENCH'', and by the UKRI Frontier Research grant RISEandFALL. RM also acknowledges funding from a research professorship from the Royal Society. SA acknowledges support from the JWST Mid-Infrared Instrument (MIRI) Science Team Lead, grant 80NSSC18K0555, from NASA Goddard Space Flight Center to the University of Arizona. SA acknowledges grant PID2021-127718NB-I00 funded by the Spanish Ministry of Science and Innovation/State Agency of Research (MICIN/AEI/ 10.13039/501100011033). WMB acknowledges support by a research grant (VIL54489) from VILLUM FONDEN. AJB acknowledges funding from the ``FirstGalaxies'' Advanced Grant from the European Research Council (ERC) under the European Union’s Horizon 2020 research and innovation programme (Grant Agreement No. 789056). SC acknowledges support by the European Union's HE ERC Starting Grant No. 101040227 - WINGS. DJE, ZJ, BDJ, MR, BER, and CNAW acknowledge support by a JWST/NIRCam contract to the University of Arizona, NAS5-02015. DJE is also supported as a Simons Investigator. BER also acknowledges support from the JWST Program 3215. WM thanks the Science and Technology Facilities Council (STFC) Center for Doctoral Training (CDT) in Data Intensive Science at the University of Cambridge (STFC grant number 2742968) for a PhD studentship. H\"U acknowledges funding by the European Union (ERC APEX, 101164796). Views and opinions expressed are however those of the authors only and do not necessarily reflect those of the European Union or the European Research Council Executive Agency. Neither the European Union nor the granting authority can be held responsible for them. The research of CCW is supported by NOIRLab, which is managed by the Association of Universities for Research in Astronomy (AURA) under a cooperative agreement with the National Science Foundation. JW gratefully acknowledges support from the Cosmic Dawn Center through the DAWN Fellowship. The Cosmic Dawn Center (DAWN) is funded by the Danish National Research Foundation under grant No. 140.

This work is based [in part] on observations made with the NASA/ESA/CSA James Webb Space Telescope. The data were obtained from the Mikulski Archive for Space Telescopes at the Space Telescope Science Institute, which is operated by the Association of Universities for Research in Astronomy, Inc., under NASA contract NAS 5-03127 for JWST. These observations are associated with PIDs 1180, 1181, 1210, 1286, 1895, 1963, and 3215. The authors acknowledge the teams led by PIs Daniel Eisenstein (PID 3215), Christina Williams, Sandro Tacchella, Michael Maseda (JEMS; PID 1963), and Pascal Oesch (FRESCO; PID 1895) for developing their observing program with a zero-exclusive-access period.

\section*{Data Availability}

The data underlying this article will be shared at a reasonable request by the corresponding author. Fully reduced NIRCam images and NIRSpec spectra are publicly available on MAST (\url{https://archive.stsci.edu/hlsp/jades}), with \doi{10.17909/8tdj-8n28}, \doi{10.17909/z2gw-mk31}, and \doi{10.17909/fsc4-dt61} \citep{Rieke_2023, Eisenstein_2023b, Bunker_2024, D'Eugenio_2025}.



\bibliographystyle{mnras}
\bibliography{references.bib} 



\appendix

\clearpage
\section{Correction weights}
\label{sec:appendix_weights}

In this section, we discuss in more detail the various correction weights (introduced in Section~\ref{sec:correction for selection effects}) applied to the pair probability function (PPF) for finding major galaxy close-pairs.

\begin{figure}
 \centering
 \includegraphics[width=\columnwidth]{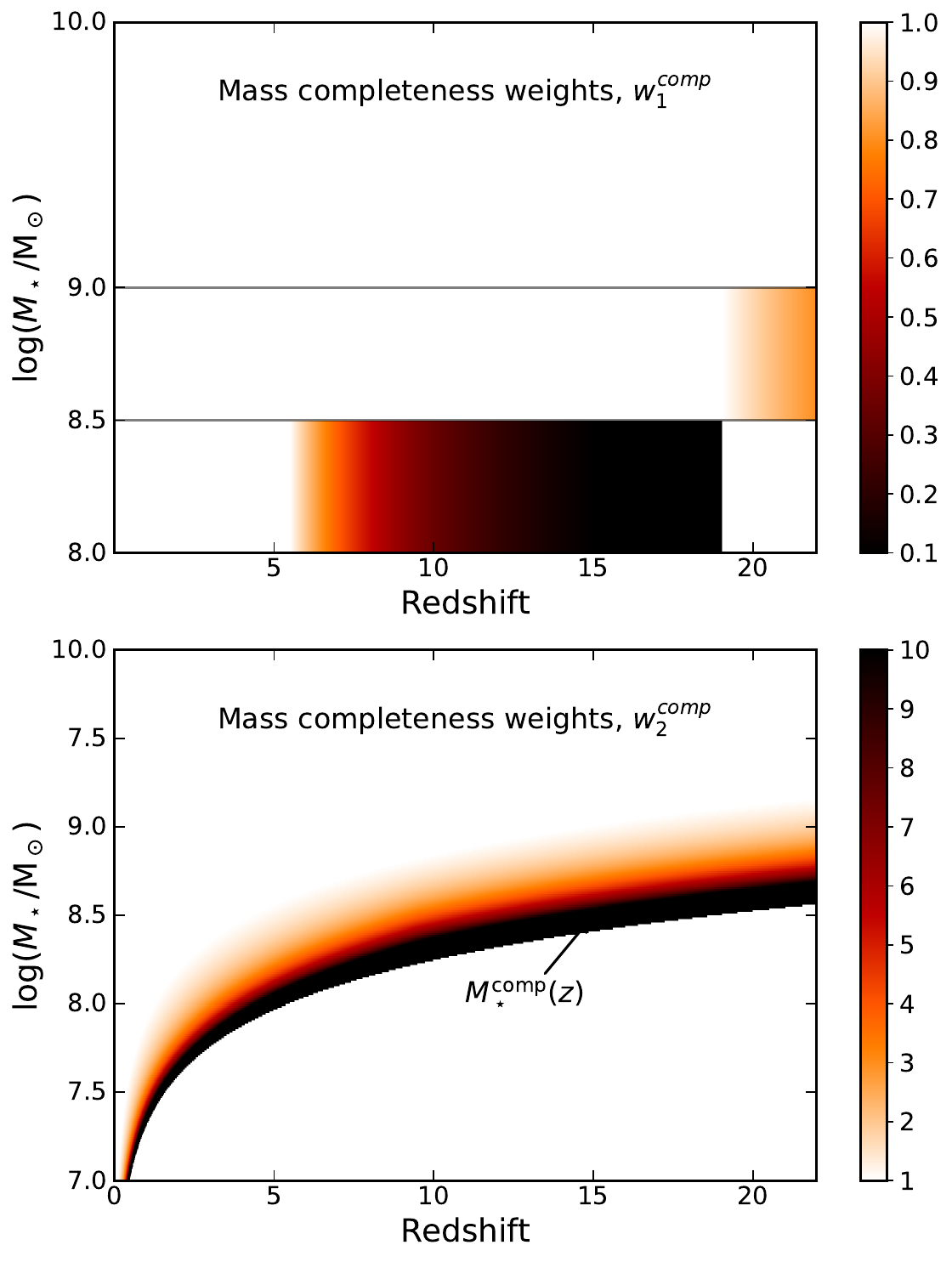}
 \caption{The magnitude of the mass completeness weights on the stellar mass–redshift plane, which are statistical weights incorporated in the pair probability function to account for potentially missing galaxies due to reduced search range in the proximity of the mass completeness limit. As the top panel for $w_1^{\mathrm{comp}}$ and the bottom panel for $w_2^{\mathrm{comp}}$ show, the strength of these weights exponentially increases, approaching the mass completeness limit.}
 \label{fig:w_comp}
\end{figure}

The first weight applied is a correction for stellar mass incompleteness (see Section~\ref{sec:mass incompleteness}). Figure~\ref{fig:w_comp} shows the strength of the mass completeness weights as they approach the mass completeness limit $M_\star^{\rm comp}(z)$. Each subfield (i.e. GS and GN, deep and medium tiers) will have its own corresponding mass completeness weights, as these depend on $M_\star^{\rm comp}(z)$, which in turn directly depends on the 5$\sigma$ point-source limit of the field. These weights are calculated from the ratios of expected number densities of galaxies using the stellar mass function for a given mass bin by Equations~\ref{eq:w_2_comp} and \ref{eq:w_1_comp}. In the case of the primary galaxy's mass completeness weights ($w_1^{\mathrm{comp}}$; top panel of Figure~\ref{fig:w_comp}) there are just three mass bins, and they are not heavily affected by incompleteness (as they were chosen to be above $M_\star^{\rm comp}(z)$). On the other hand, weights corresponding to secondary galaxies have to be calculated for each possible stellar mass ($w_2^{\mathrm{comp}}$; bottom panel of Figure~\ref{fig:w_comp}), and usually significantly affect the calculations (as mass bins of secondary galaxies can potentially enter the incomplete regime at higher redshifts). As these weights exponentially increase in magnitude when approaching the mass completeness limit and also depend on the assumed SMF, we balance their contributions by clipping $w_1^{\mathrm{comp}}$ at the $[0.1, 1.0]$ and $w_2^{\mathrm{comp}}$ at the $[1, 10]$ range.

\begin{figure}
 \includegraphics[width=\columnwidth]{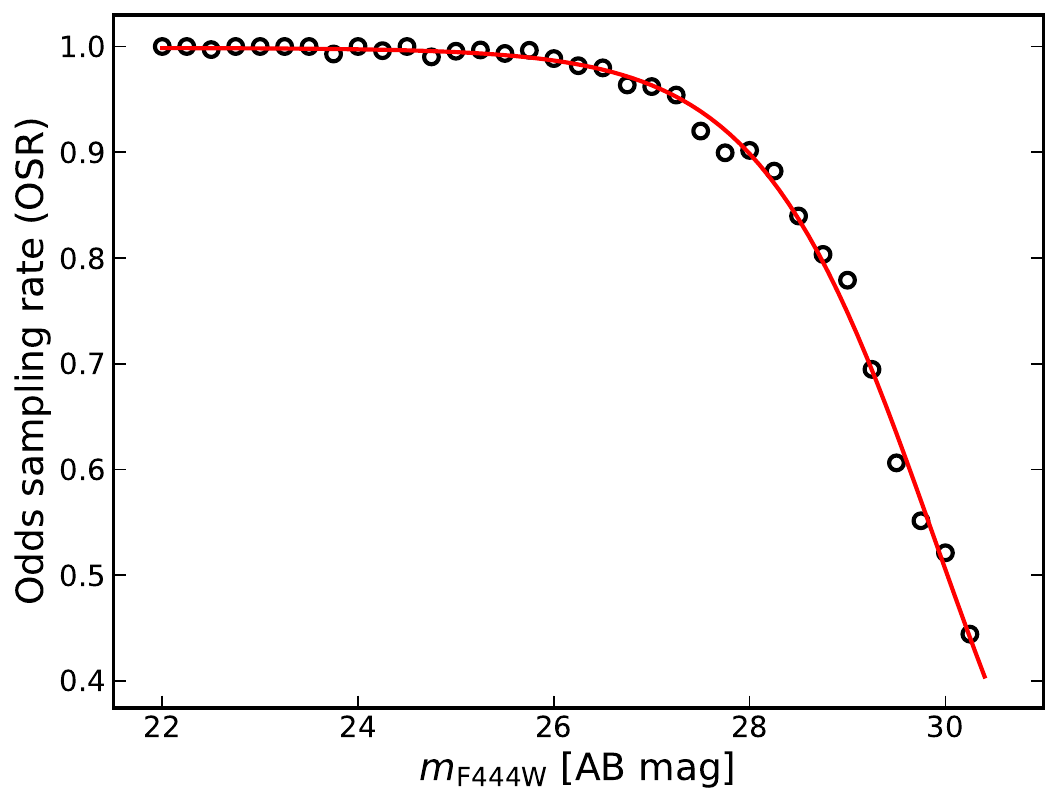}
 \caption{Dependence of the odds sampling rate (OSR) on the AB magnitude of the galaxies (calculated from fluxes measured in F444W) for the GS-Medium field. As the objects get fainter, the OSR significantly drops, meaning they have poorly constrained photometric redshifts. We fit a sigmoid curve to these data points to obtain a continuous ${\rm OSR}(m)$ relation.}
 \label{fig:osr}
\end{figure}

The second weight applied to the PPF accounts for the quality of photometric redshifts for galaxies by their brightness. The odds quality parameter $\mathcal{O}$ effectively measures how well defined is the primary peak of photometric redshifts (as discussed in Sections~\ref{sec:odds quality parameter} and \ref{sec:photometric redshift quality}). The odds sampling rate (OSR) is the ratio between the number of objects above an odds parameter threshold value ($\mathcal{O} \geq 0.3$) and the total number of objects within a magnitude bin of width $\Delta m = 0.25$, which translates to the quality of the photometric redshifts versus the brightness of the source, or how peaked are the photo-$z$ posteriors. Figure~\ref{fig:osr} shows the odds sampling rate curve for the GS-Medium field. The deeper we probe objects at higher magnitudes, the worse the photometric redshift quality becomes; hence, the OSR decreases. According to this, we are potentially missing multiple primary and secondary galaxies due to their poorly constrained photo-$z$'s. Therefore, we account for these potentially missing close-pair candidates by multiplying the pair probability functions of faint objects by $w^{\rm OSR}$ to boost their number counts. All magnitudes are calculated from F444W fluxes, as this filter has the most extensive coverage in our fields, and all objects should be detected in this band, even at high-z.

\begin{figure}
 \centering
 \includegraphics[width=0.8\columnwidth]{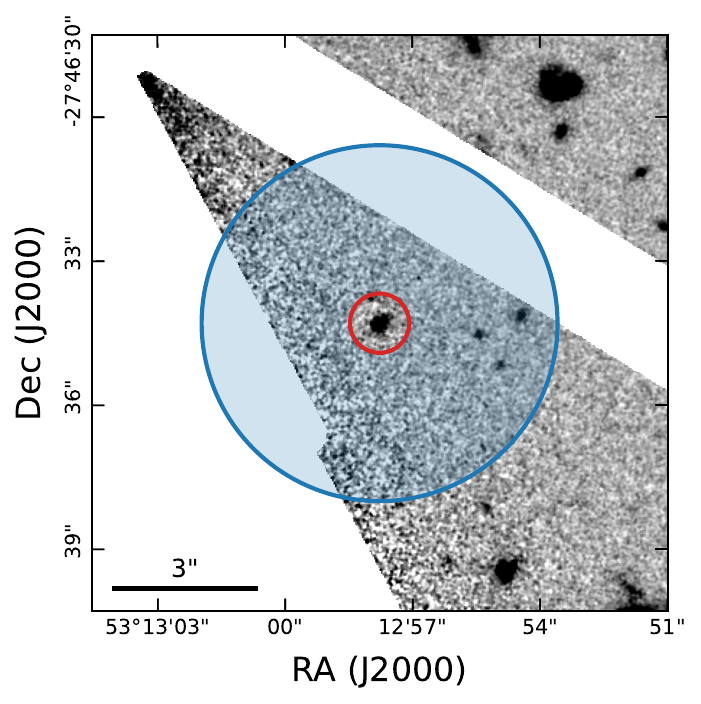}
 \caption{Example of missing search area around a galaxy at $z_{\mathrm{phot}} = 2.73$ in the primary sample in GS-Medium. The inner circle (red) corresponds to $r_{\mathrm{min}}$ and the outer circle to $r_\mathrm{max}$. We are interested in the available search area in the annulus between the two circles. In this particular case, there is a missing area (represented by white) to the left of the source due to the survey boundary and to the top right due to a masked diffraction spike of a nearby bright star. Hence, we compute $f_\mathrm{area} = 0.66$ for this primary galaxy.}
 \label{fig:area}
\end{figure}

The third weight we apply is to account for potentially missing secondary galaxies due to incomplete search annuli around primary galaxies (see Section~\ref{sec:survey boundaries}). This could be caused by two main reasons: a primary galaxy being close to the survey boundary (in most cases, less than $50\%$ of the search area missing) or due to masked regions (see Figure~\ref{fig:area} for an example). As mentioned earlier, we mask out bright stars and strong diffraction spikes, as they can introduce artificial objects and, hence, many close-pairs that are, in fact, artefacts. These area weights have the weakest effect on our results. Nevertheless, we implement them for completeness. We calculate the area weights for each field separately before using them for the close-pair finding algorithm.

\section{Galaxy stellar mass function}
\label{sec:appendix_gsmf}

Here, we present the galaxy stellar mass function (SMF) assumed for this work based on two recent high redshift JWST studies by \citet{Navarro-Carrera_2024} and \citet{Harvey_2024}. 

The first work, by \citet{Navarro-Carrera_2024}, uses JWST data from the Hubble Ultra Deep Field (HUDF) and UKIDSS Ultra Deep Survey field, complemented with ancillary data from CANDELS to measure the galaxy stellar mass function (SMF) for redshifts at $3.5 < z < 8.5$ probing down to stellar masses of $\simeq 10^8~{\rm M_\odot}$. \citet{Harvey_2024} extends measurements to the redshift range of $6.5 < z < 13.5$ and uses deep JWST NIRCam observations from the PEARLS and other publicly available surveys (CEERS, GLASS, JADES GOODS-South, NGDEEP, and SMACS0723) to compile a sample of high-z galaxy candidates and measure the SMF.

We adopt a Schechter function \citep{Schechter_1976} as in both cited works and combine their results for the three fit parameters. This function has a power-law shape with an exponential fall-off and is characterised by three parameters: $\alpha$, $M^\star$, and $\Phi^\star$. The SMF has the following logarithmic form
\begin{equation}
   \begin{split}
    \Phi(\log M) = \ln(10) \Phi^\star {\rm e}^{-10^{\log M - \log M^\star}} \Bigl(10^{\log M - \log M^\star} \Bigr) ^{\alpha + 1} {\rm d} \log M
    \end{split}
    \label{eq:SMF}
\end{equation}
with $\alpha$ being the slope of the low-mass end, $M^\star$ is the characteristic stellar mass at which the function turns from a power-law to an exponential fall-off, and $\Phi^\star$ controls the overall scaling of the SMF.

\begin{figure*}
 \centering
 \includegraphics[width=\linewidth]{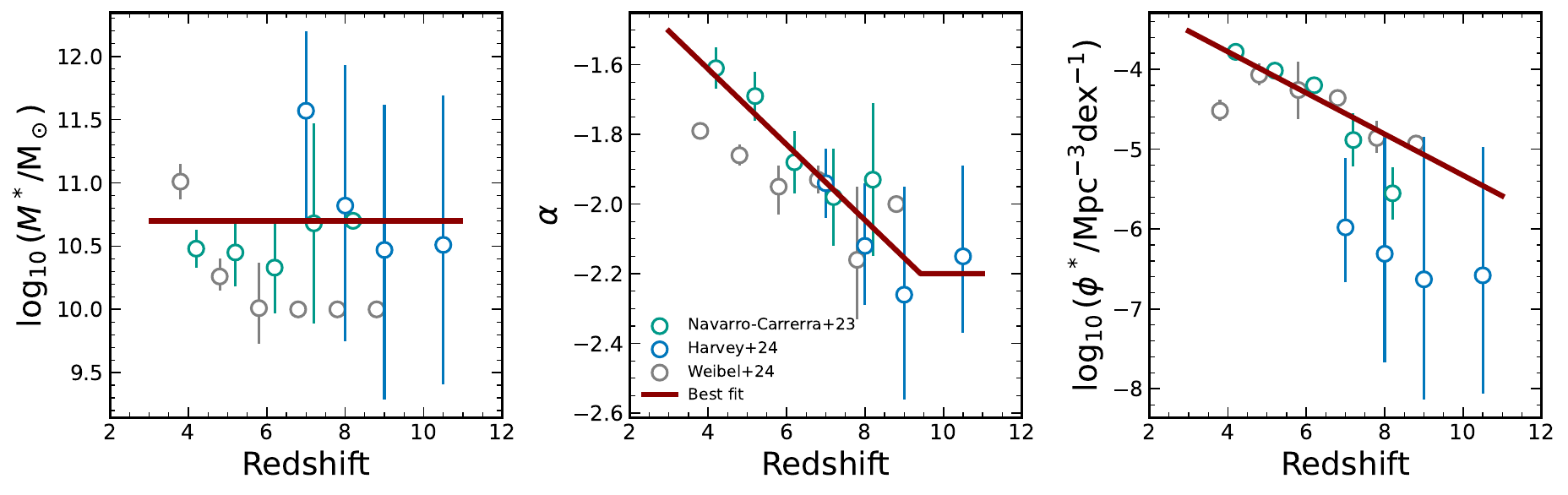}
 \caption{Redshift evolution of the parameters of the galaxy SMF used in this work based on recent JWST studies by \citet{Navarro-Carrera_2024} and \citet{Harvey_2024}. For comparison, we also plot results from \citet{Weibel_2024}, and we find a good agreement overall. We fix the characteristic stellar mass at $\log_{10}(M^\star / {\rm M_\odot})= 10.7$, assuming redshift independence. We fit a line for $\alpha$ and cap it at the lowest allowed value of $\alpha = -2.2$, while we fit a line to $\log_{10}\Phi^\star$ using least-squares fitting with uncertainties.}
 \label{fig:GSMF}
\end{figure*}

For these three parameters, we take the results from the two papers above and fit lines using leas-square fitting, including measurement uncertainties (see Figure~\ref{fig:GSMF}). In the case of the characteristic stellar mass, we fix it at a value of $\log_{10}(M^\star / {\rm M_\odot})= 10.7$, which is motivated mainly by previous studies and the seemingly weak dependence on redshift. For $\alpha$, we fit a line but cap it to a constant value of $\alpha = -2.2$ at $z > 9.5$. For $\log_{10}\Phi^\star$, we simply fit a line. We use these combined parameters in our SMF to cover intermediate and high redshifts with deep JWST data.

\clearpage
\section{Results with \texttt{eazy-py}}
\label{sec:appendix_eazy}

In this section, we discuss a variation of our fiducial analysis of major galaxy close-pairs, using stellar masses computed with \texttt{eazy-py}. 

\begin{figure}
 \centering
 \includegraphics[width=\columnwidth]{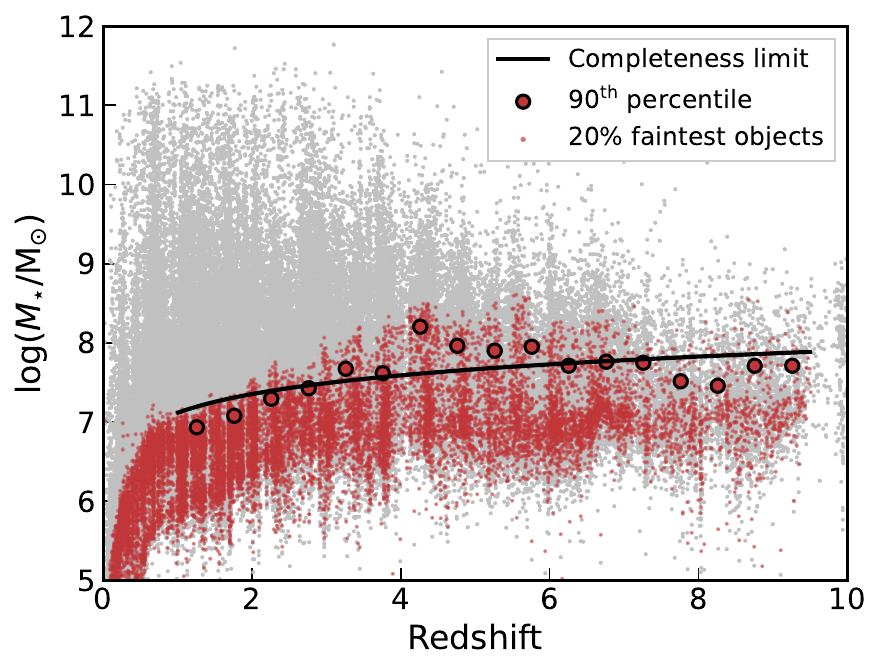}
 \caption{Stellar mass distribution and completeness calculation for the total GOODS-South sample using masses calculated by \texttt{eazy-py}. A significant scatter can be observed in the 90\textsuperscript{th} percentile completeness limits between the different redshift bins, which is caused by the lower accuracy of \texttt{eazy-py} compared to \texttt{Prospector}. While the overall sample size is larger compared to the \texttt{Prospector} fits, as each photometric source will have a stellar mass estimate by \texttt{eazy-py}, the initial sample after pre-selection on which we perform our analysis has a similar size.}
 \label{fig:completeness_eazy}
\end{figure}

Running \texttt{eazy-py} takes significantly less computational resources than \texttt{Prospector}; however, it results in less reliable stellar masses and other estimated physical parameters. Therefore, we obtain stellar masses for each photometric source, resulting in a significantly larger overall sample. For consistency with the fiducial analysis, we choose the same redshift limit, SNR and odds selection criteria, resulting in an initial sample with a size comparable to the \texttt{Prospector} analysis. Figure~\ref{fig:completeness_eazy} shows the stellar mass distribution of our sample against redshift. We perform the same completeness analysis as in Section~\ref{sec:completeness analysis} by \citet{Pozetti_2010}, and we observe a significantly larger scatter in the 90\textsuperscript{th} percentile completeness limits in the redshift bins, especially at $4 < z < 5$.

\begin{figure}
 \centering
 \includegraphics[width=\columnwidth]{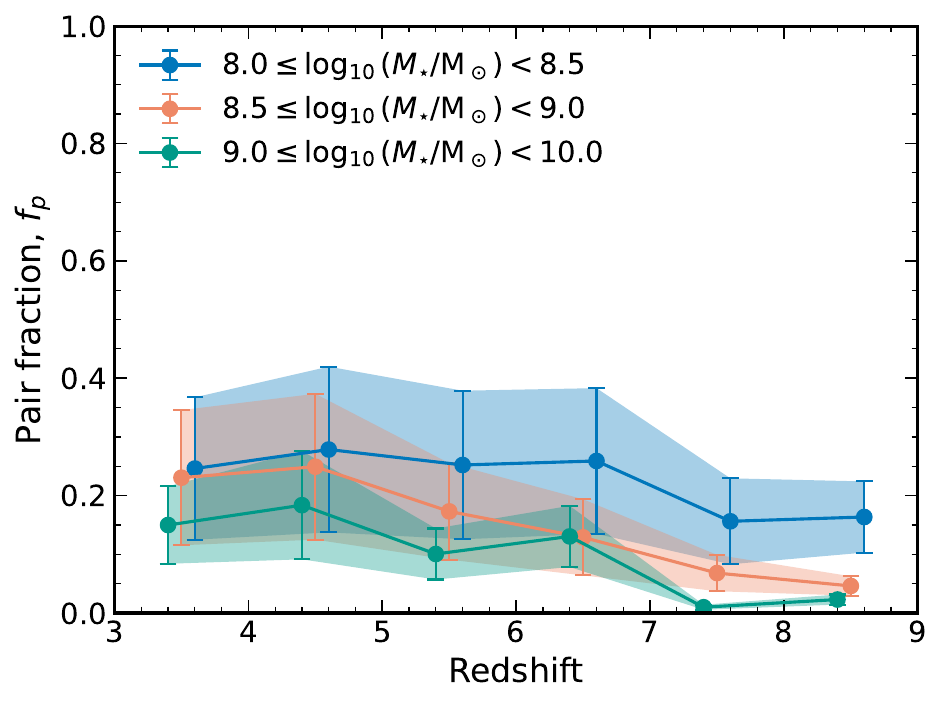}
 \caption{Evolution of the major merger close-pair fractions with redshift stellar masses estimated by \texttt{eazy-py}, instead of \texttt{Prospector} as in the fiducial analysis. We observe an almost constant and slightly declining trend (beyond $z \approx 5$) with increasing redshifts, where the decline is more pronounced at the higher stellar mass bins. There is a significant contrast between the results from the fiducial analysis and this one, mainly in the lowest stellar mass bin, that might be caused by the lower accuracy of \texttt{eazy-py}.}
 \label{fig:pair_fractions_eazy}
\end{figure}

We perform the same close-pair selection as described in Section~\ref{sec:close-pair methodology}, with the exception of using a data set with stellar masses estimated by \texttt{eazy-py} instead of \texttt{Prospector}. We present our results in Figure~\ref{fig:pair_fractions_eazy}, where we find an almost constant evolution on pair fractions with redshifts in all three stellar mass bins. There is a weak decline after $z \approx 5$, which is more pronounced in the higher stellar mass bins. It is interesting to compare this result with Figure~\ref{fig:pair_fraction_summary}, where the most striking difference is between the trends found in the $8.0 \leq \log_{10}(M_\star / {\rm M_\odot}) < 8.5$ mass bin. In our fiducial model, there is a pronounced increase, a strong peak at $z \approx 6$, followed by a decline at high-$z$, whereas here, we find an almost flat evolution. Although we conduct the same analysis in this case, we are more confident in the results from the fiducial analysis, as we better trust the stellar masses estimated by \texttt{Prospector} than the ones from the fast template-fitting code \texttt{eazy-py}. Nevertheless, this additional analysis shows the effect of the adopted SED-fitting codes that can significantly alter the results. Ultimately, large spectroscopic surveys at high redshifts could give precise and accurate close-pair measurements that are currently not yet available for a large enough sample.

\clearpage
\section{Pair fractions at redshift-dependent separations}
\label{sec:appendix_rvir}

In this section, we discuss another variation of the fiducial close-pair analysis, where we use a redshift-dependent separation criterion instead of a constant one. 

\begin{figure}
 \centering
 \includegraphics[width=\columnwidth]{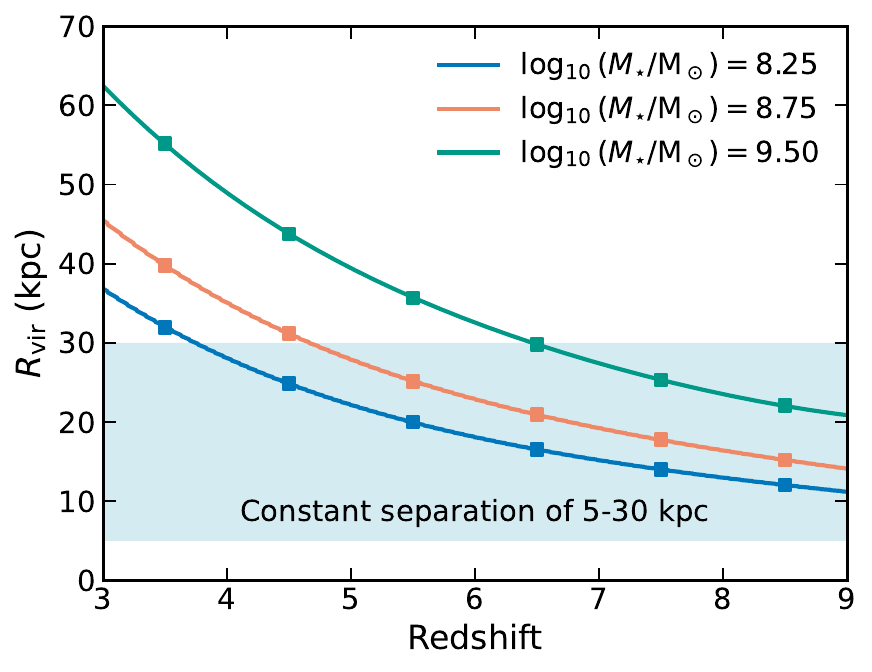}
 \caption{Redshift dependence of the virial radii of dark matter haloes corresponding to the primary galaxies that we adopt as the redshift-dependent close-pair selection criterion. The square markers represent the values at each stellar mass–redshift bin. The constant separation criterion of $r_{\rm sep} = [5, 30]~{\rm kpc}$ from the fiducial close-pair analysis is highlighted as a light blue rectangle for better comparison with the redshift evolution of the virial radii.}
 \label{fig:rvir}
\end{figure}

The separation criterion for this analysis is based on the virial radius of the dark matter halo hosting each primary galaxy at each redshift. To estimate this radius, we use the stellar mass - halo mass (SMHM) relation from \citet{Behroozi_2019} using the \texttt{UniverseMachine} code. From this parametrisation, we obtain the peak halo mass $M_{\rm halo}$ corresponding to the stellar mass $M_\star$ in the centre of each stellar mass bin (at 8.25, 8.75, and 9.5 $\log{\rm M_\odot}$ respectively). In the case of $\log_{10}(M_\star / {\rm M_\odot}) = 9.5$ the parametrisation of \citet{Behroozi_2019} does not go beyond $z=8.6$ and we extrapolate the trend up to $z = 9$, which gives reasonable results (see Figure~\ref{fig:rvir}). We can compute the virial radius corresponding to the peak halo masses using the equation
\begin{equation}
    R_{\text{vir}} = \left( \frac{2 G M_{\text{halo}}}{\Delta_c H^2(z)} \right)^{1/3},
\end{equation}
where $H(z)$ is the Hubble parameter, $G$ is the gravitational constant, and we assume $\Delta_c = 200$ for the density contrast. Using this equation, we get redshift-dependent virial radii for the DM haloes hosting the average primary galaxy in each redshift bin (see Figure~\ref{fig:rvir}).

We adopt these redshift-dependent virial radii as the new separation criteria that we determine for each stellar mass - redshift bin (as indicated by the square markers on Figure~\ref{fig:rvir}). Here, we make the assumption that if a secondary galaxy is closer to the primary than the corresponding virial radius of the DM halo hosting it, then the system forms a close pair and will eventually merge. This is a simple assumption that is physically motivated. 

\begin{figure}
 \centering
 \includegraphics[width=\columnwidth]{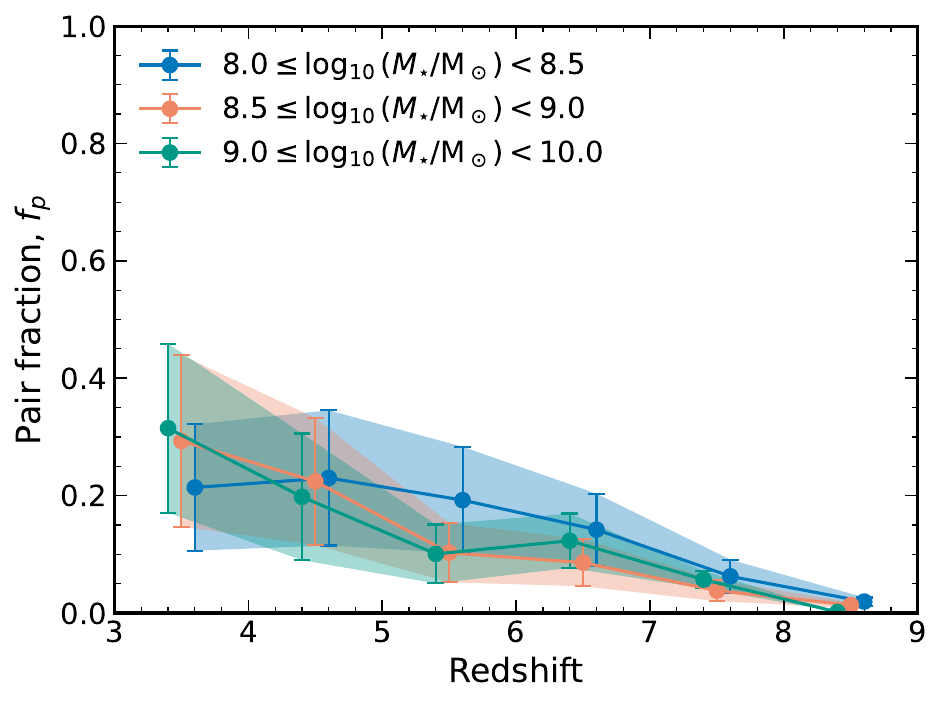}
 \caption{Close-pair fraction evolution using stellar masses from \texttt{Prospector}, but adopting a redshift-dependent separation criterion instead of a constant one, based on the virial radii of the dark matter haloes hosting the primary galaxies. We find that the pair fractions are smoothly declining for all stellar mass bins, in contrast to the results found in the fiducial analysis.}
 \label{fig:pair_fractions_rvir}
\end{figure}

The resulting pair fractions are plotted in Figure~\ref{fig:pair_fractions_rvir}, which show somewhat different behaviour than the fiducial analysis and the results from \texttt{eazy-py}. For this analysis, we are using stellar masses from \texttt{Prospector} as in the fiducial analysis. We find that the pair fraction declines with redshift in all stellar mass bins considered. The decline is smoother and goes to lower values than in the fiducial analysis (c.f. Figure~\ref{fig:pair_fraction_summary}). In comparison, this is simply caused by the smaller search radius around high-$z$ primary galaxies according to the relation from Figure~\ref{fig:rvir}. However, this might be physically more motivated than looking at a constant separation criterion, as the dark matter haloes at high-$z$ were significantly smaller than the ones at lower-$z$, and therefore, gravitational interaction and mergers took place at smaller physical scales.

This analysis shows that choosing a different separation limit significantly affects the resulting close-pair measurements. We also note here that most previous works adopted a constant separation limit, whereas a redshift-dependent one is physically more motivated.

\clearpage
\section{Median mass ratio}
\label{sec:appendix_mu}

Here, we discuss in more detail the analysis of the merger ratio distribution of our sample and the calculation of the average merger ratio. This step is essential to convert the merger rate $\mathcal{R}_{\rm M}(z)$ to the specific mass accretion rate (sMAR) (see Equation~\ref{eq:sMAR}).

\begin{figure}
 \centering
 \includegraphics[width=\columnwidth]{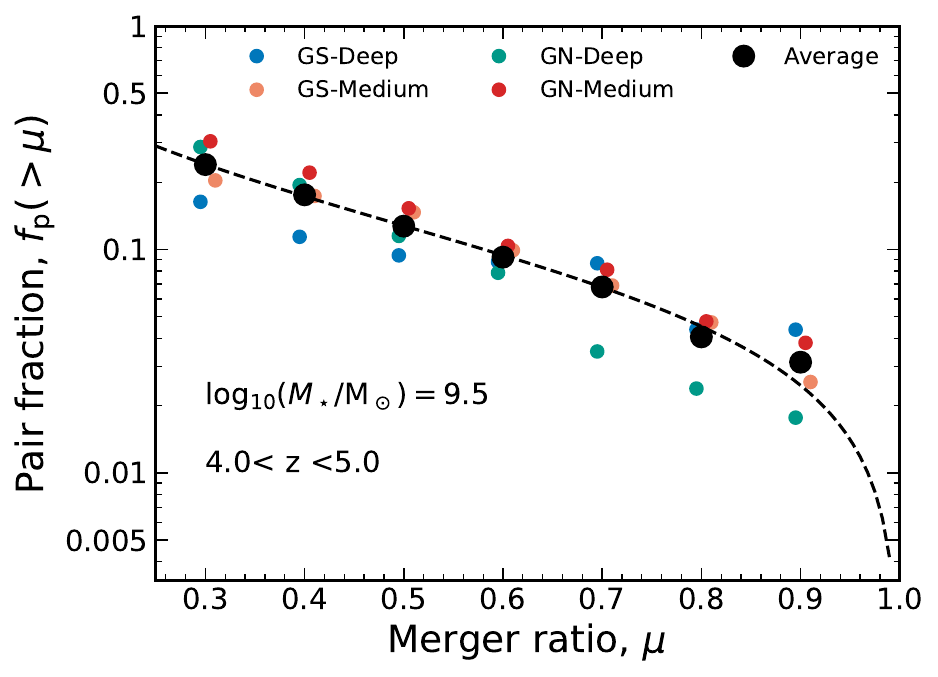}
 \caption{Example distribution of close-pair fractions with increasing mass ratio cuts for the stellar mass-redshift bin indicated on the plot for the four tiers and their averages. This is a cumulative distribution as pair fractions with lower $\mu$ values contain close-pairs with higher mass ratios than that $\mu$ limit (e.g., $f_{\rm P}(> \mu = 0.3)$ also contains close-pairs included in $f_{\rm P}(> \mu = 0.4)$). We calculate the median mass ratio in this bin by calculating the $50^{\rm th}$ percentile of this distribution by extrapolation and also fit a curve described by Equation~\ref{eq:mu_param}.}
 \label{fig:mu_example}
\end{figure}

First, we calculate the median mass ratio (merger ratio) $\overline{\mu}$ for our close-pair sample. This can be done by computing the pair fraction $f_{\rm P}(> \mu)$ in each redshift – stellar mass bin above increasing values of $\mu$ (see Figure~\ref{fig:mu_example} for an example). We then calculate the median value of the mass ratio from this cumulative pair fraction distribution by extrapolation to find the $\mu$ value at which the distribution reaches the $50^{\rm th}$ percentile. We also fit a curve to the average $f_{\rm P}(> \mu)$ values that is described by a parametrisation \citep{Duncan19} of the form
\begin{equation}
    f_{\rm P}(> \mu) = A \times \left( \frac{1}{\mu} - 1\right)^B.
    \label{eq:mu_param}
\end{equation}

\begin{figure}
 \centering
 \includegraphics[width=\columnwidth]{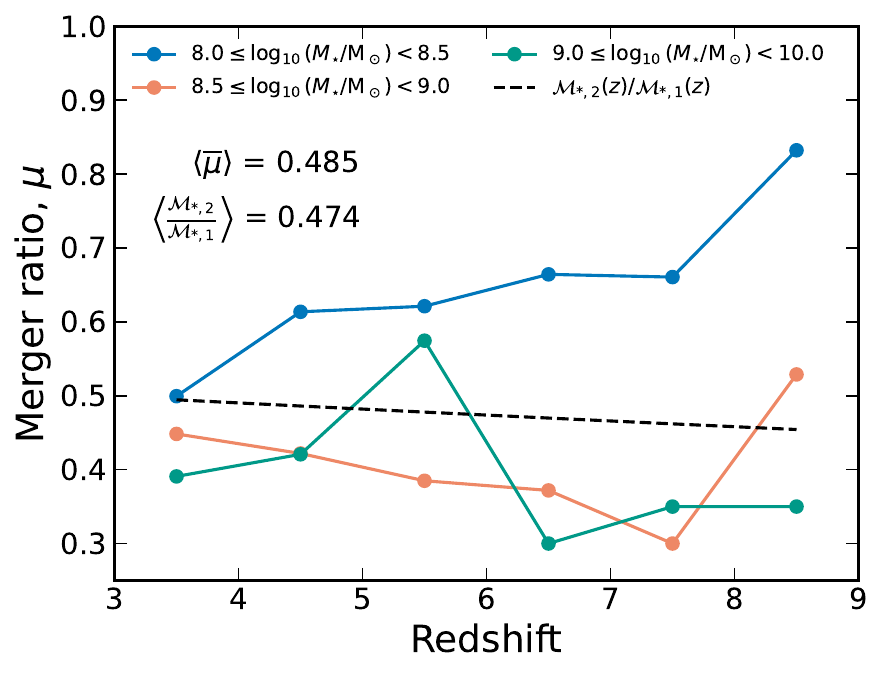}
 \caption{Empirically determined median merger ratios from the $f_{\rm P}(> \mu)$ cumulative distributions in each stellar mass - redshift bin. We find that there is a significant scatter in these values a higher redshifts, which is due to the low number counts of close-pairs that results in a poor resolution of $f_{\rm P}$ in terms of merger ratio, subsequently leading to less reliable $\overline \mu$. We also plot the redshift evolution of the predicted merger ratio of $\langle \mathcal{M}_{*, 2}(z) / \mathcal{M}_{*, 1}(z) \rangle$ using the SMF. The averages of these two quantities are displayed on the figure and agree well with each other.}
 \label{fig:mu_med}
\end{figure}

We then calculate the median mass ratios in each stellar mass - redshift bin and plot them on Figure~\ref{fig:mu_med}. Interestingly, we find that there is a considerable variation of $\overline \mu$ values at different redshifts and stellar masses. This is mainly due to having fewer close pairs in the higher mass bins that are difficult to resolve into smooth $f_{\rm P}(> \mu)$ distributions to infer $\overline \mu$. We simply take the average of all these median mass ratios to compute $\langle \overline \mu \rangle = 0.485$.

We compare the previously computed average median mass ratio $\langle \overline \mu \rangle$ to a predicted mass ratio based on the stellar mass function. This is done by calculating the ratio between the predicted average mass in the secondary stellar mass bins $\mathcal{M}_{*, 2}(z)$, and the primary stellar mass bins $\mathcal{M}_{*, 1}(z)$ defined in Equation~\ref{eq:mu_estimation_2}. We calculate these ratios for the three stellar mass bins and find that they are similar and have a weak dependence on redshift (as seen in Figure~\ref{fig:mu_med}). We then take the average of these ratios and find that $\langle \mathcal{M}_{*, 2}(z) / \mathcal{M}_{*, 1}(z) \rangle = 0.474$. This is in good agreement with our empirically determined value from the $f_{\rm P}(> \mu)$ distributions.


\bsp	
\label{lastpage}
\end{document}